\documentclass[prb,twocolumn,floatfix,showpacs,superscriptaddress]{revtex4-1}

\usepackage{epsfig}
\usepackage{amsmath}
\usepackage{color}

\newcommand{\rv}{\mathbf{r}}
\newcommand{\conc}{x_\mathrm{H}}
\newcommand{\conca}{x_\alpha}
\newcommand{\concb}{x_\beta}
\newcommand{\kB}{k_\mathrm{B}}
\newcommand{\gh}{\hat{g}}
\newcommand{\gch}{\hat{g}_\mathrm{c}}
\newcommand{\goh}{\hat{g}_\mathrm{o}}
\newcommand{\ggradh}{\hat{g}_\mathrm{g}}
\newcommand{\gelh}{\hat{g}_\mathrm{el}}
\newcommand{\omz}{\Omega_0}
\newcommand{\Nz}{N_0}
\newcommand{\muHh}{\hat{\mu}_\mathrm{H}}   
\newcommand{\muH}{\mu_\mathrm{H}}   
\newcommand{\Gibbs}{{\cal G}}    
\newcommand{\Energy}{{\cal E}}    
\newcommand{\lagrangeh}{\hat{\mu}_0}    
\newcommand{\ghbar}{\hat{\bar{g}}}           
\newcommand{\muhbar}{\hat{\bar{\mu}}_\mathrm{H}}           
\newcommand{\muhat}{\bar{\mu}_\mathrm{H}}           
\newcommand{\omegahbar}{\hat{\bar{\omega}}_\mathrm{H}}           
\newcommand{\omegahat}{\bar{\omega}_\mathrm{H}}           
\newcommand{\muel}{\mu_\mathrm{el}}   
\newcommand{\muelh}{\hat{\mu}_\mathrm{el}}   
\newcommand{\muela}{\mu_\mathrm{el}^\alpha}   
\newcommand{\muelb}{\mu_\mathrm{el}^\beta}   
\newcommand{\much}{\hat{\mu}_\mathrm{c}}   
\newcommand{\muoh}{\hat{\mu}_\mathrm{o}}   
\newcommand{\tr}{\mathrm{tr}\,}   
\newcommand{\gc}{g_\mathrm{c}}    
\newcommand{\go}{g_\mathrm{o}}    
\newcommand{\Gel}{{\cal G}_\mathrm{el}}    
\newcommand{\Gc}{{\cal G}_\mathrm{c}}    
\newcommand{\Go}{{\cal G}_\mathrm{o}}    
\newcommand{\NH}{N_\mathrm{H}}   
\newcommand{\NM}{N_\mathrm{M}}   
\newcommand{\xn}{x_0}
\newcommand{\Rhydride}{R_\mathrm{H}}     
\newcommand{\Gelb}{{\cal G}_\mathrm{el}^\mathrm{bulk}}    

\begin{document}

\title{Scale bridging description of coherent phase equilibria in the presence of surfaces and interfaces}

\date{\today}

\author{R. Spatschek}
\affiliation{Institute for Energy and Climate Research, Forschungszentrum J\"ulich GmbH, 52425 J\"ulich, Germany}
\affiliation{Max-Planck-Institut f\"ur Eisenforschung GmbH, 40237 D\"usseldorf, Germany}
\author{G. Gobbi}
\affiliation{Max-Planck-Institut f\"ur Eisenforschung GmbH, 40237 D\"usseldorf, Germany}
\affiliation{Department of Mechanical Engineering, Politecnico di Milano, 20156 Milan, Italy}
\author{C. H\"uter}
\affiliation{Institute for Energy and Climate Research, Forschungszentrum J\"ulich GmbH, 52425 J\"ulich, Germany}
\affiliation{Max-Planck-Institut f\"ur Eisenforschung GmbH, 40237 D\"usseldorf, Germany}
\author{A. Chakrabarty}
\affiliation{Max-Planck-Institut f\"ur Eisenforschung GmbH, 40237 D\"usseldorf, Germany}
\affiliation{Texas A\&M University at Qatar, P.O. Box 23874, Doha, Qatar}
\author{U. Aydin}
\affiliation{Max-Planck-Institut f\"ur Eisenforschung GmbH, 40237 D\"usseldorf, Germany}
\author{S. Brinckmann}
\affiliation{Max-Planck-Institut f\"ur Eisenforschung GmbH, 40237 D\"usseldorf, Germany}
\author{J. Neugebauer}
\affiliation{Max-Planck-Institut f\"ur Eisenforschung GmbH, 40237 D\"usseldorf, Germany}

\begin{abstract}
We investigate phase separation including elastic coherency effects in the bulk and at surfaces and find a reduction of the solubility limit in the presence of free surfaces.
This mechanism favours phase separation near free surfaces even in the absence of external stresses.
We apply the theory to hydride formation in nickel, iron and niobium and obtain a reduction of the solubility limit by up to two orders of magnitude at room temperature in the presence of free surfaces.
These effects are concisely expressed through a solubility modification factor, which transparently expresses the long-ranged elastic effects in a terminology accessible e.g.~to {\em ab initio} calculations.
\end{abstract}


\maketitle

\section{Introduction}

Phase transitions are one of the most important aspects in physics and materials science, and they largely influence our entire environment.
It is one of the cornerstones of thermodynamics and statistical physics that macroscopic phase separation is determined by minimisation of certain thermodynamic potentials, depending on the boundary conditions or constraints of the system.
Classically, this energy minimization is described via the common tangent construction, which has initiated the cartography of phase transitions in all kind of materials, and enables the prediction of phase stability for applied materials, having triggered many Calphad activities\cite{Kaufman:1970aa, Lukas:2007aa}.
Usually, elastic effects are not considered in the prediction of phase coexistence, and despite their enormous  relevance this is still a widely unresolved issue.
Important progress has been made by Cahn\cite{Cahn:1962aa}, who demonstrated that in case of identical elastic constants for both precipitate and matrix and isotropic lattice expansion, the common tangent approach remains valid.
More generally, the coherent phase diagram can be obtained by adding the elastic energy of the system to the free energies and minimizing the resultant energy\cite{Cahn:1984aa, Williams:1984aa, W.C.Johnson:1987aa}. 
This classical work focuses on bulk elastic effects and does not take into account interfacial effects.

Complementary to the bulk and continuum perspective many investigations have been done on the phase formation at surfaces, also in combination with elastic effects.
An important example is heteroepitaxal growth in thin films\cite{Johnson:1988aa}, where stresses can arise due to the mismatch between film and substrate.
Cluster expansion methods, which are powerful methods for predicting alloy phase diagrams, have also been applied to epitaxial films\cite{Wood:1988aa,Liu:2009aa}. 
However, in general such techniques do not take into account the elastic effects, which are induced by surfaces and interfaces, beyond the nano scale for substitutional solid solutions\cite{Ozolins:2002fk, Muller:1999aa}. 
Also related atomic scale approaches have been used to study surface phase formation.
Two-dimensional alloy formation at surfaces has frequently been reported in the literature, exhibiting solubility under conditions, where phase separation would occur in the bulk, see e.g.~Ref.~\onlinecite{Neugebauer:1993aa}.
Tersoff\cite{Tersoff:1995ab,Tersoff:1995aa} inspected them on a generic level under the aspect of elastic effects, which are partially released near the surfaces using atomistic descriptions.
The effective elastic interaction of individual atoms with the surface is found to decrease on the scale of a single atomic layer and strongly competes with surface energy effects, giving rise to various types of surface patterns which are influenced by anisotropy\cite{Alerhand:1988aa}.
Also, it generically leads to a short scale increase of the solubility directly at the surface relative to the bulk.
These microscopic considerations however do not allow to gain insights into phase separation influenced by the presence of surfaces on macroscopic scales, where bulk effects dominate.

The present article offers a complementary perspective on phase separation near surfaces and interfaces under the influence of elastic effects, starting from a mesoscopic continuum approach and connecting it to {\em ab initio} simulations to predict local phase diagrams.
On this scale, the phenomena reported by Tersoff et al.~reduce to an effective surface contribution with negligible thickness, and which is not further considered here -- a complete description will therefore require the superposition of effects on these different scales.
On the present larger scale phase separation occurs and leads to the appearance of coherency stresses due to lattice mismatches between precipitates and matrix phase.
In agreement with the thermodynamic picture these stresses are energetically unfavorable in the bulk and can therefore suppress phase separation.
Near surfaces, but on the scale of the precipitate sizes and therefore significantly larger than the atomic scale, the elastic stresses can partially relax, similarly to the arguments used by Tersoff, and therefore affect phase separation.
Thermodynamically, the incorporation of interstitial or substitutional impurity atoms leads to the formation of solid solution phases for low concentrations.
Continued insertion of impurity atoms leads at some concentration to the formation of precipitates and phase separation, and this concentration is denoted as solubility limit (or terminal solubility).
According to the above arguments, the solubility limit is expected to be decreased in the presence of free surface, in comparison to the (deep) bulk, even in the absence of interfacial effects.
The quantitative discussion of these effects on different scales is the subject of the present paper.
Our focus is on equilibrium phase separation and not their kinetics.

Deeper inside the two-phase regions phase separation can occur in particular via spinodal decomposition, and elastic  coherency also plays a role there.
Coherent spinodal decomposition \cite{Kappus:1977aa, Tang:2012aa} can lead to spontaneous phase separation in a near surface region although bulk spinodal decomposition may still be suppressed.
However, it should be pointed out that spinodal decompositions occurs at impurity concentrations, which are often orders of magnitude higher than the solubility limit. 
Similar to the discussion above, these effects, which are a result of combined bulk elastic effects near boundaries and appear also in thermodynamically large systems, have to be distinguished from finite size effects on phase diagrams, which result from the finite number of degrees of particles and the absence of singularities in the relevant thermodynamic potentials, see the discussion e.g.~in Ref.~\onlinecite{Pohl:2012aa} and references therein.

The concepts, which are developed and inspected in the present paper with different analytical and computational methods on different scales, are fully generic, and we apply them in the later part of the manuscript explicitly to different metal-hydrogen systems.
Throughout the article we therefore use the terminology of metal-hydrogen systems, despite the conceptual generality.
The motivation for this specific application are often low solubility limits in the room temperature regime, which can lead to the formation of hydrides at low hydrogen concentrations \cite{Hydrides1968}.
For example, in the room temperature regime the extrapolated solubility limit of $\alpha$-Zr is of the order $\mathrm{[H]/[Zr]}\sim 10^{-5}$, see Ref.~\onlinecite{Khatamian:1997aa}.
In accordance with the above statements a conclusion of the present analysis will be that the formation of hydrides is therefore more likely near free surfaces which are formed at cracks.
As hydrides are typically brittle phases, this hydride formation at cracks can contribute to hydrogen embrittlement \cite{AIME1980, TMS1990}.
In contrast, high concentrations of hydrogen are important for the use of metals as hydrogen storage materials\cite{Ledovskikh:2016aa}.

Surface effects in combination with elasticity near the critical point have been considered theoretically by Bausch et al.\cite{Bausch:1975aa} for different geometries, based on the description of elastic interaction in coherent metal-hydrogen systems by Wagner and Horner\cite{Wagner:1974aa, Wagner:1978aa}.
Fluctuations play a central role near the critical point and are affected by the elastic effects, in contrast to the low temperature and concentration regime, which we focus on in the present article.
The knowledge of elasticity as driving force has also consequences for the kinetics, and therefore the careful understanding of the thermodynamics is essential.
Application of the theoretical description by Janssen\cite{Janssen:1976aa} has allowed in particular to determine diffusion coefficients of Nb and Ta using the Gorsky effect \cite{Gorsky:1935aa} by Bauer et al.\cite{Bauer:1978aa, Volkl:1979aa}.
As a result, the interaction of bulk elastic and surface effects can lead to substantially different diffusion behavior of hydrogen in wire and foil specimens \cite{Fukai:2005aa}.
Indirect evidence for the role of surface effects stems from high pressure hydrides of iron and its alloys\cite{Antonov:2002aa}.
An anomalous volume increase for high hydrogen concentrations in Ni$_{0.8}$Fe$_{0.2}$H$_x$ leads to an inconsistency between (surface sensitive) X-ray and (volume averaged) neutron measurements of bulk and powder materials, resulting in the conclusion of concentration inhomogeneities and the formation of a high hydrogen concentration surface phase.
An explicit experimental proof of macrosopic modes and their shape dependence through elastic effects on phase separation in Nb-H near the critical point has been given by Zabel and Peisl \cite{Zabel:1979aa} for coherent\cite{Zabel:1980aa} and incoherent cases\cite{Zabel:1979ab}.
Griessen and coworkers studied the destabilization of the Mg-H system and found a strong influence on the plateau pressures for hydride formation in thin magnesium films, which are either stress free or clamped \cite{Baldi:2009aa, Baldi:2009ab}.
Differences between free standing and clamped thin films of Pd-H clearly demonstrate the role of elasticity and boundary conditions \cite{Pivak:2009aa, Gremaud:2009aa}.
For Pd nanoparticles, hydrogen intercalation is consistent with a coherent interface model and differs significantly from bulk behavior \cite{Griessen:2015aa}.
Experimental findings for hydride formation in niobium for different mechanical boundary conditions show drastic variations of the solubility limits, which are not yet understood systematically \cite{Northemann:2008uq,Pundt:2006aa,Northemann:2011aa}.
This open issue will specifically be addressed in the present article.
This may also play a role for precipitation in Nb-H and Nb-D, where enhanced hydride formation near a surface scratch occurs\cite{Whitton:1976aa}, and which may be a signal of the reduced solubility limit near surfaces, as the scratch increases the available surface and allows for stress relaxation.
This exemplary list of experimental and theoretical results demonstrates the relevance of elastic effects in combination with surfaces and interfaces for metal hydrogen systems.

The present article is organised as follows, in order to provide a comprehensive, multi-scale and multi-method approach to phase separation in the presence of surfaces and interfaces.
We start with the analysis of a continuum model, which is conceptually close to phase field descriptions.
This model allows to gain valuable insights into the problem based on the numerical solution of the underlying equations.
In the later sections, we deduce this a priori phenomenological model from an atomistic picture, in particular {\em ab initio} simulations, which quantifies the model in the low temperature and concentration limit.
We point out that for many applications this limit is of major interest, in particular for hydrogen embrittlement, which takes place around room temperatures at minute hydrogen concentrations.
We predict the solubility limit of hydrogen based on the numerical simulations.
It differs significantly from stress free and coherent bulk equilibrium solubility limits and introduces the concept of a surface induced solubility limit.

Before we enter into this matching between continuum and atomistic scales, we discuss the influence of elastic effects on interface equilibria in a sharp interface picture.
This complementary approach is necessary to fully interpret the findings of the continuum model.

As a final step, we establish the matching between continuum and discrete descriptions.
This leads to a closed analytical prediction of the solubility limit using only parameters, which are directly accessible from {\em ab initio} simulations.
Via the detour of bulk coherent phase equilibria we predict an easy-to-use expression for the near-surface solubility limit, which shall be of general relevance for many phase diagram predictions in the low concentration regime.


\section{Continuum modeling}
\label{continuum::section}

We start the description from the continuum perspective, where we inspect a simple model concerning phase separation.
Here we pay special attention to the role of elasticity, assuming coherent interfaces between the phases.
This assumption allows to investigate the differences between phase separation in the bulk, near free surfaces and at interfaces.
For the model we choose a Cahn-Hilliard model, in which the concentration is taken as order parameter.
The model is chosen as simple as possible to illustrate the main physical features.
However, we point out that it gives an accurate description in the low temperature and low concentration regime, where the model parameters can be directly taken from {\em ab initio} simulations, as will be further explained in Section \ref{abinitio::section}.
The focus is on equilibrium phase separation;
kinetic mechanisms, also of spinodal decomposition, are presently of minor interest.
The generic framework can also be used for more complex systems, e.g.~of phase field models for phase separation in alloys, where additional order parameters are introduced to distinguish between the phases\cite{Steinbach:1998aa,Steinbach:2013aa,steinbach,Fleck:2010il,Spatschek11,Spatschek06,Spatschek:2007vn,Monas:2013uq}.

\subsection{Model description}

The purpose of this section is to set up a minimalistic continuum model, which captures the essential physical ingredients to describe the influence of elastic coherency effects on phase equilibria in the bulk and near surfaces.
We use here the terminology of a metal-hydrogen system, although all concepts are fully transferrable to other systems.
The model is based on a Gibbs free energy density $\gh$, where the hat indicates that this energy density has the dimension energy per volume.
It consists of four contributions, $\gh = \goh + \gch + \ggradh + \gelh$, which will be explained in the following.
The integrated Gibbs energy reads
\begin{equation} \label{eq1}
\Gibbs = \int \gh d\rv = \int \left( \goh + \gch + \ggradh + \gelh \right) d\rv,
\end{equation}
where the integration is taken over the volume of the system.
In the spirit of a Lagrangian description of the arising elastic deformations, the reference state is the non-deformed, single phase system. 
We are mainly interested in situations in which the system can freely expand (NPT ensemble, i.e.~fixed particle numbers, pressure and temperature), specifically with vanishing external pressure, $P=0$.
Therefore, the Gibbs energy is the appropriate thermodynamical equilibrium potential.

Before going into detail, we briefly introduce the energy terms.
First, the two contributions $\goh(\conc)$ and $\gch(\conc, T)$ determine the phase diagram via a regular solution model, which has a large miscibility gap in this study.
They depend on the local dimensionless concentration of the impurities, $\conc$, and on the temperature $T$.
$\goh$ describes the enthalpy of mixing and favors the separation into a hydrogen free phase with $\conc=0$ and a hydride phase with $\conc=1$.
$\gch$ represents the configurational entropy of the alloy and is written here for an interstitial solid solution.
Since we focus on low temperatures (below the Debye temperature), we ignore thermal vibrations.
Hence, the dominant temperature dependence comes from the configurational degrees of freedom.
$\ggradh(\nabla\conc)$ is a term which penalizes the presence of interfaces (as they are accompanied by sharp concentration gradients) and therefore introduces an interfacial energy.
For the main application of the model we keep this term negligibly small (hence we use it only to stabilize the numerical convergence), such that the results are dominated by bulk effects.
$\gelh(\conc)$ is the elastic energy which appears due to the widening of the lattice through the presence of inhomogeneous hydrogen distributions.
In particular, this term is responsible for the coherency stresses between the metal and the hydride phase.

In detail, the interstitial hydrogen concentration is defined as the ratio of interstitial hydrogen to the number of host lattice atoms, $\conc = \NH/\NM$.
The mixing enthalpy expression
\begin{equation} \label{gohdef}
\goh(\conc) = \alpha \frac{\Nz}{\omz} \conc(1-\conc),
\end{equation}
has only one energy parameter $\alpha>0$ in the simplified model.
For the proper normalization we introduce $\omz=a^3$ as the volume of a cubic, hydrogen free unit cell with (hydrogen and stress free) lattice constant $a$ of the reference state, and $\Nz$ is the number of metal atoms per unit cell, e.g.~$\Nz=4$ for fcc metals.
We tacitly assume that the concentrations remain in the interval $\conc \in [0, 1]$.
This constraint can e.g.~be achieved by additional infinite energy barriers for concentrations outside this interval, similarly to phase field models\cite{Steinbach:1998aa,Steinbach:2013aa,Bhogireddy:2014aa,steinbach}.
Since this constraint is only required for the case $T=0$, we do not pay further attention to it, as otherwise it is ensured by the entropic Gibbs energy contribution $\gch$.

Let us briefly comment on the physical role of $\goh$.
For low concentrations it is dominated by the linear contribution $\goh \simeq \alpha \conc \Nz/\omz$.
In this regime $\goh$ is proportional to the energy which is needed to inject an isolated hydrogen atom into the metal matrix, i.e.~the formation enthalpy in the dilute limit.
With the chemical potential being proportional to the derivative of the Gibbs energy density with respect to concentration, it therefore only leads to a constant shift of the chemical potential relative to an (arbitrary) reference potential.
For a fixed total number of hydrogen atoms in the system (in accordance with the desired NPT ensemble) this linear term does therefore not influence the phase separation behavior.
For higher concentrations, deviations appear due to the quadratic term $-\alpha \conc^2 \Nz/\omz$.
This contribution is central, as it is responsible for the phase separation into a hydrogen poor metal (in the following also denoted as $\alpha$ phase) and a hydrogen rich hydride phase ($\beta$ phase).
It captures an effective attractive hydrogen-hydrogen interaction (for $\alpha>0$), as this energy contribution lowers the energy relative to the energy of isolated hydrogen atoms, which comes from the aforementioned linear term.
In the context of the metal-hydrogen system, this interaction is often considered to be at least partially of elastic origin, due to the lattice expansion\cite{Fukai:2005aa, Alefeld:1972aa, Wagner:1978aa}.
To avoid confusion, we point out that this ``elastic term'' is conceptually different from the explicit elastic term $\gelh$, which is discussed below.
The latter is related to a mesoscopic change in the hydrogen concentration, which occurs in particular at the boundary between a hydride precipitate and the metal.
This term captures the energetic cost of coherency stresses between the phases with different equilibrium lattice constants.
For a homogeneous system, which can freely expand, the explicit elastic energy $\gelh$ vanishes.
This is different for the quadratic contribution contained in $\goh$, which lowers the energy even for spatially homogeneous concentrations and is responsible for the phase separation.
The distinction between microscopic elastic effects in $\goh$ and mesoscopic contributions in $\gelh$ will be further elucidated later in the text, see Section \ref{abinitio::section} and Appendix \ref{micromesoelastic::appendix}.

The configurational contribution $\gch$ is given by
\begin{equation} \label{gchdef}
\gch(\conc, T) = \frac{\kB T\Nz}{\omz} \left[ \conc\log\conc + (1-\conc)\log(1-\conc) \right],
\end{equation}
which is a standard expression in statistical physics. 
It is based on the assumption that the hydrogen atoms occupy all interstitial sites with equal probability, neglecting interactions between them.
The expression is accurate in the limiting cases $\conc\ll 1$ and $1-\conc\ll 1$, when only a small amount of hydrogen atoms or vacancies is present.
Deviations can be expected inside the  concentration interval $0<\conc<1$.
With the present focus on the low concentration regime $\conc\ll 1$ the expression (\ref{gchdef}) is therefore a suitable description.
Here, we implicitly assumed that the only relevant phases in the system are a dilute $\alpha$ phase with $\conc\ll 1$ and the hydride ($\beta$ phase) with $1-\conc\ll 1$.

The gradient energy term,
\begin{equation}
\ggradh = \frac{\gamma_0}{2}(\nabla \conc)^2
\end{equation}
penalises interfaces between $\alpha$ and $\beta$ via a contribution $\gamma_0$.
From a phase field perspective, it also establishes together with $\goh$ a smooth order parameter profile between the two phases and leads to the appearance of an interfacial energy.
Since our focus is on bulk terms only, we intentionally set $\gamma_0=0$ throughout the entire article.
For numerical purposes, it can sometimes be beneficial to use a small value of $\gamma_0$ to stabilise interfaces and to introduce a finite interface thickness.

The elastic contribution $\gelh$ is based on the isotropic linear theory of elasticity.
In the usual way we define the strain tensor as $\epsilon_{ij}=(\partial_i u_j + \partial_j u_i)/2$ using the displacement field $u_i(\rv)$.
Moreover, we take into account that hydrogen widens the host lattice, which leads to a diagonal eigenstrain $\epsilon_{ij}^0=\chi \conc \delta_{ij}$ with the Vegard coefficient $\chi$.
For a wide range of metals the assumption of Vegard's law, i.e.~the linear dependence of the eigenstrain on the local concentrations, is well satisfied.
We exclude here the appearance of e.g.~tetragonal distortions, as we aim to establish a fully isotropic theory.
Tetragonal distortions can e.g.~play a role for carbon as interstitial element, which can occupy different sub-lattices of bcc metals.
For hydrogen in fcc metals, tetragonal distortions are negligible, which justifies our assumption.
With this, the isotropic elastic energy density reads
\begin{equation} \label{eq3}
\gelh = G (\epsilon_{ij} - \chi \conc\delta_{ij})^2 + \frac{1}{2}\lambda (\epsilon_{kk} - 3\chi \conc)^2,
\end{equation}
with $G$ and $\lambda$ being shear modulus and Lam\'e coefficient respectively, which we assume to be concentration and phase independent, for simplicity.
As usual, the Einstein sum convention is applied for repeated indices.
For a validation of this energy expression, also in comparison to {\em ab initio} simulations, we refer to the extended discussion in Appendix \ref{micromesoelastic::appendix}.
We note that this description is the isotropic variant of the continuum model by Wagner and Horner \cite{Wagner:1974aa}, where for the occupation of hydrogen on fcc octahedral sites the force-dipole tensor is diagonal, containing the widening of the lattice through hydrogen\cite{Alefeld:1969aa, Alefeld:1972aa}.

The integration in Eq.~(\ref{eq1}) is taken over the volume $V$ of the body. 
Since we are interested in the influence of free boundaries, we assume boundary conditions $\sigma_{in}=0$ there, with $n$ being the surface normal direction.
As mentioned before, this is in line with the use of the Gibbs energy instead of a Helmholtz free energy as generating functional for fixed displacement (i.e.~volume) boundary conditions.
Notice that due to the traction-free boundary conditions ($P=0$), Gibbs and Helmholtz free energy actually coincide.

For the Gibbs energy minimization with respect to concentration one can use different versions of conserved dynamics.
In the present article, we focus on equilibrium properties, and therefore the path to the energetic minimum does not matter.
For details, we refer to Appendix \ref{section::FEM}.

Additionally, the elastic equilibrium conditions read
\begin{equation}
\frac{\delta\Gibbs}{\delta u_i} = 0,
\end{equation}
which translates into the usual static elastic equilibrium condition
\begin{equation}
\frac{\partial\sigma_{ij}}{\partial x_j} = 0
\end{equation}
with the stress tensor
\begin{equation} \label{isostress}
\sigma_{ij} = 2G (\epsilon_{ij} - \delta_{ij}\chi \conc) + \lambda \delta_{ij} (\epsilon_{kk} - 3\chi \conc).
\end{equation}

The above equations fully describe the system, and the material parameters are summarised in Table \ref{table1}.
They are extracted from {\em ab initio} simulations in Section \ref{abinitio::section}.

\subsection{The phase diagram without elasticity}

\begin{figure}
\begin{center}
\includegraphics[width=8.5cm]{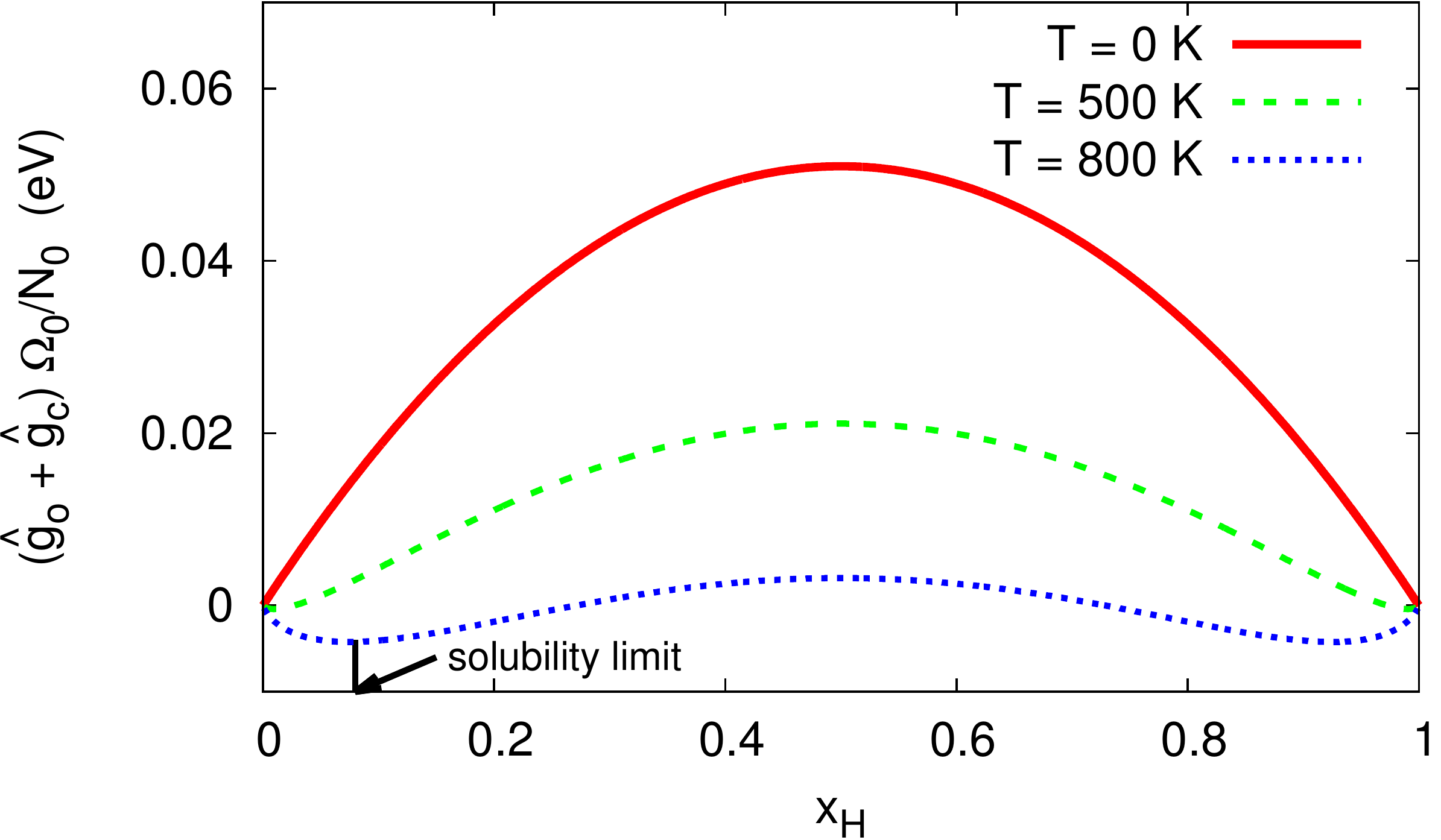}
\caption{(Color online) Concentration dependent energy $\goh(\conc) + \gch(\conc,T)$ for the Ni-H system.
For $T=0\,\mathrm{K}$ the configuration term vanishes, and only $\goh$ remains.
The temperature dependent solubility limits are given by the positions of the local minima of the Gibbs energy.
}
\label{fig1}
\end{center}
\end{figure}

Let us for the moment ignore the elastic contribution to the Gibbs energy, which therefore consists only of the mixing enthalpy $\goh$ and configurational term $\gch$ as bulk terms.
Also, the interfacial term $\ggradh$ does not appear in the thermodynamic limit.
The energy density is sketched in Fig.~\ref{fig1} for different temperatures.
Here we use parameters adjusted to the Ni-H system.
\begin{table}
\caption{Parameters for Ni-H in isotropic appoximation.}
\label{table1}
\begin{tabular}{c|c|c}
parameter & meaning & Ni-H value \\
\hline
$N_0$ & no.~of metal atoms per unit cell & 4 \\
$\omz$ & unit cell volume & $40.953\,\mathrm{\AA^3}$ \\
$\alpha$ & obstacle potential parameter & $0.2039\,\mathrm{eV}$ \\
$G$ & bulk modulus & $1.05\,\mathrm{GPa}$ \\
$\lambda$ & Lam\'e coefficient & $2.88\,\mathrm{GPa}$ \\
$\chi$ & Vegard coefficient & 0.0623
\end{tabular}
\end{table}

\begin{figure}
\begin{center}
\includegraphics[width=8.5cm]{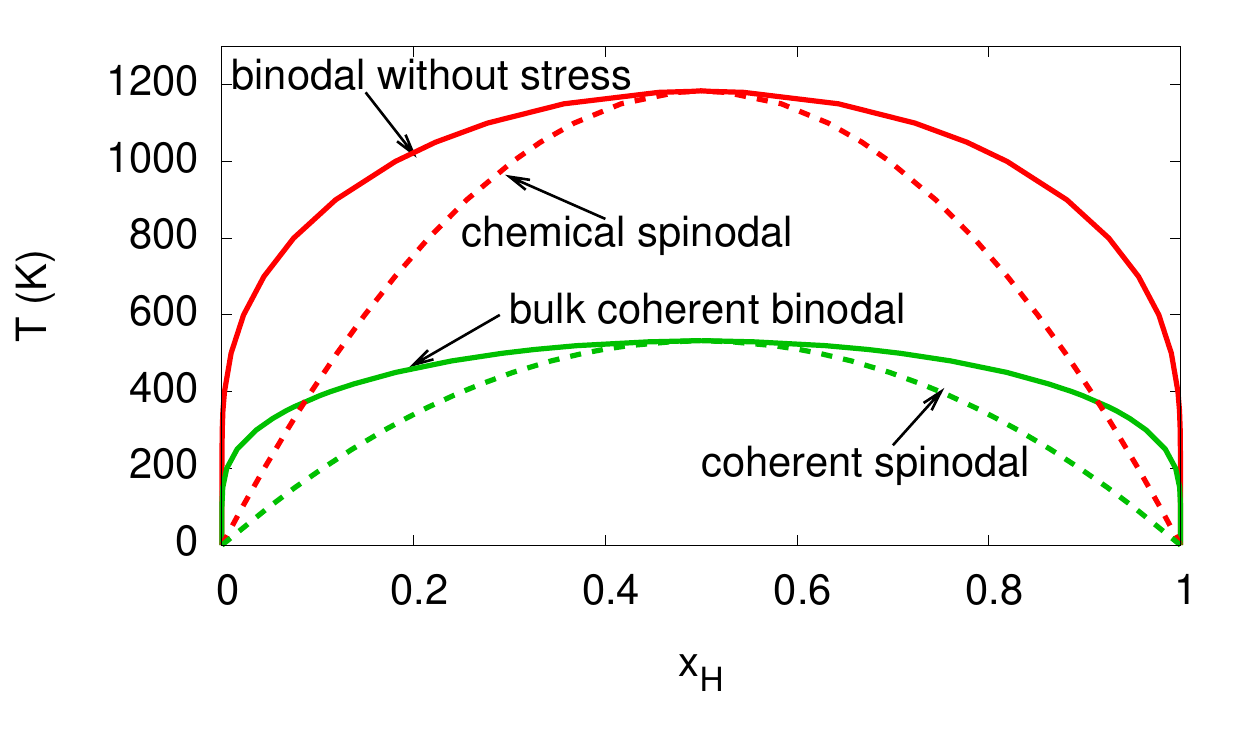}
\caption{(Color online) Phase diagram for Ni-H using the simplified model system, without stress effects (solid red curve).
The critical point is at $T_c=1184\,\mathrm{K}$.
Additionally, the solid green curve shows the coherent bulk phase diagram (see Section \ref{section::coherentphasediagram}), and the dashed curves are the spinodal lines (see Section \ref{section::spinodaldecomposition}). 
}
\label{fig2}
\end{center}
\end{figure}

In equilibrium, the concentrations are constant in each phase, and the common tangent construction describes the equilibrium concentrations.
This construction is particularly simple for our symmetrical potential, where it appears as a horizontal line, which connects the two minima of the potential $\goh(\conc) + \gch(\conc,T)$.
Hence the equilibrium concentrations are here determined simply by $[\goh(\conc) + \gch(\conc,T)]'=0$.
The resulting phase diagram is shown in Fig.~\ref{fig2}.
%
We mention once more that the present model system is not intended to give a quantitative description in the entire concentration and temperature domain, but focus on the low temperature and concentration regime.
In particular, at higher temperatures vibrational contributions are not properly accounted for, and therefore the phase diagram becomes inaccurate in this temperature regime.
This property is reflected in particular by a far too high critical point compared to experimental data\cite{Shizuku:2002aa,Fukai:2004aa} ($T_c\approx 360^\circ\,\mathrm{C}$ at $x_c\approx 0.5$ for a pressure of $p_c\approx 1.4\,\mathrm{GPa}$).

\subsection{The coherent phase diagram}
\label{section::coherentphasediagram}

If we additionally take into account the elastic term, the two-phase region becomes smaller (lower critical temperature and reduced region of phase separation) and is bound by the coherent spinodal.
The intuitive explanation for this result is -- in contrast to the single phase situations, which are stress free as the material can freely expand -- that coherency stresses arise if the two phases coexist.
This raises the total energy and therefore makes the transition to a phase separated phase energetically unfavourable.
As a consequence, the miscibility gap becomes smaller, or, in other words, the solubility limits of the two phases are increased.
This means that the $\alpha$ phase can take more interstitial hydrogen atoms, and conversely more hydrogen vacancies can exist in the homogeneous $\beta$ phase.

In general, the concept of a common tangent construction does not apply in the presence of long-ranged elasticity, which induces nonlocal effects.
However, the isotropic system is an exception.
It was suggested by Cahn that coherent bulk phase equilibria with isotropic elasticity, equal elastic constants and a purely dilatational eigenstrain can still be described by a common tangent construction, which is based on a generalised Gibbs energy\cite{Cahn:1962aa},
\begin{equation} \label{CahnModifiedPotentialJustUsed}
\ghbar = \goh + \gch + \frac{E \chi^2 \conc^2}{1-\nu}.
\end{equation}
Here, $E=G(3\lambda+2G)/(\lambda+G)$ and $\nu=\lambda/[2(\lambda+G)]$ are Young's modulus and Poisson ratio. 
With this, the phase equilibrium conditions are given by two conditions.
The first is the generalized equality of chemical potentials, $\muhbar(\conca) = \muhbar(\concb)$, where the chemical potential is defined as 
\begin{equation}
\muhbar(\conc) = \frac{\partial \ghbar}{\partial\conc}.
\end{equation}
We mention here in passing that this chemical potential has dimension energy per volume, and it is related to the usual chemical potential, which has dimension energy per particle as $\mu_\mathrm{H} = \hat{\mu}_\mathrm{H}\omz/\Nz$.
The second condition is the equality of grand potentials, $\omegahbar(\conca) = \omegahbar(\concb)$ with
\begin{equation}
\omegahbar(\conc) = \ghbar - \muhbar\conc.
\end{equation}

A derivation for the above equilibrium conditions is given in Section \ref{coherent::section}.
Here we point out that the equilibrium conditions are based on the assumption that the concentrations in the two phases are spatially constant.
This is the case for phase separation in the bulk, as is demonstrated explicitly in Appendix \ref{Eshelby::section} for the special case of the Eshelby problem.

If we apply these equilibrium conditions similar to the stress free case in the previous section, but now using the modified Gibbs potential, we obtain the coherent phase diagram, which is shown also in Fig.~\ref{fig2}.
As expected, the two phase region is here smaller and fully inside the corresponding region for the stress free case.

\begin{figure}
\begin{center}
\includegraphics[width=8.5cm]{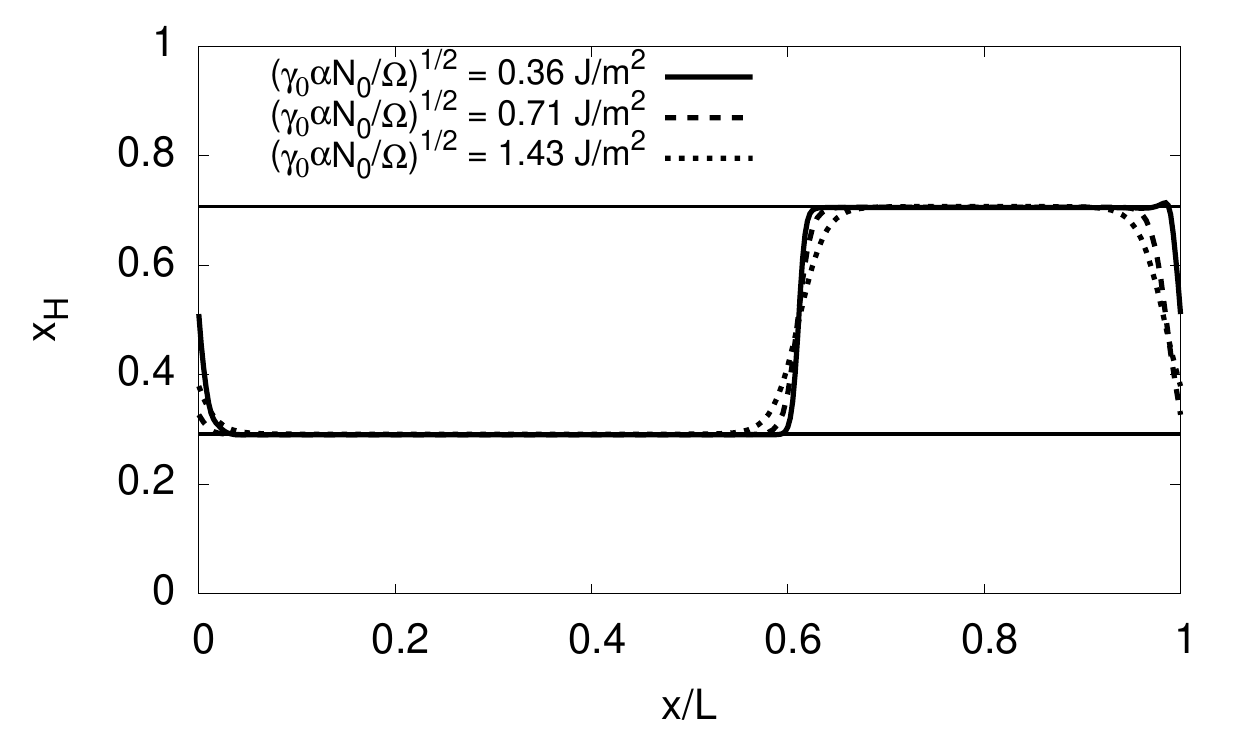}
\caption{Phase separation with elastic effects at $T=500\,\mathrm{K}$.
The concentration profiles result from continuum simulations using different values of the interfacial energy parameter $\gamma_0$.
The (dimensionless) system sizes range from $L(\alpha\Nz/\gamma_0\omz)^{1/2}\approx 223$ to $893$, with $L$ being the system length.
For the parameters in the legend and Table \ref{table1} this corresponds to $L=10^{-7}\,\mathrm{m}$.
The solid horizontal lines are the bulk solubility limits according to the theoretical coherent binodal.
}
\label{fig4aaa}
\end{center}
\end{figure}

To demonstrate that also the continuum model reproduces the coherent phase diagram in the bulk, we consider for example a quasi one-dimensional system.
This is not a limitation, as the elastic energy does not depend on the shape of the inclusion but only on its volume fraction due to the Bitter-Crum theorem\cite{Fratzl:1999aa}, and therefore the one-dimensional system is equivalent to a full three dimensional bulk case.
In this system, we initialise a strong concentration gradient in one direction.
By that, the concentration profile remains translational invariant in the other spatial directions if we use periodic boundary conditions for a three dimensional box.
The initial concentration gradient triggers the phase separation also in absence of thermal noise, which would otherwise be needed to overcome nucleation barriers.
As a result, we obtain phase separation in equilibrium, see Fig.~\ref{fig4aaa}.
First, we note that the phase concentrations are spatially constant in each phase, and they match the values predicted in the coherent phase diagram.
This demonstrates that the model indeed reproduces the thermodynamic bulk equilibrium.
Second, we use different gradient energy terms.
As expected, a larger value of $\gamma_0$ leads to a wider interface.
For large systems the interfacial contribution does not affect the phase separation.

\subsection{Phase separation near free surfaces}
\label{CahnHilliardSurface::section}

\begin{figure}
\begin{center}
\includegraphics[trim=5cm 3cm 7.5cm 1cm, clip=true, width=6cm]{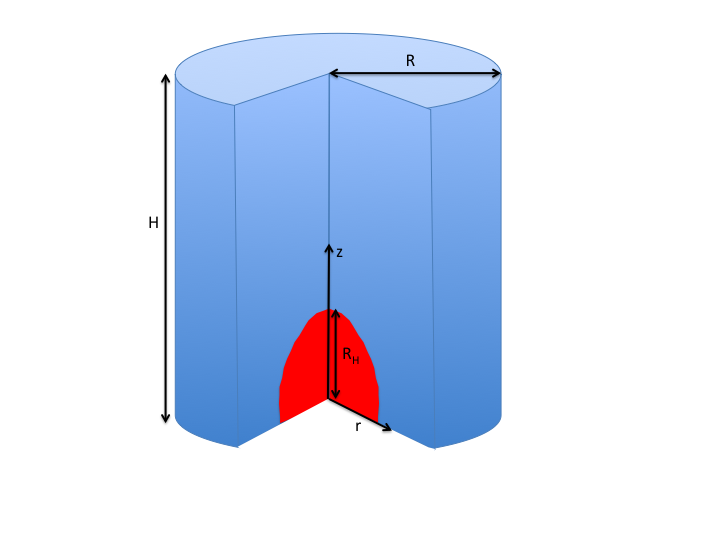}
\caption{(Color online) Sketch of the geometry for the finite element simulations of the continuum model. 
The entire sample has a cylindrical shape, which is here cut to allow a view into the interior. 
All surfaces, top, bottom and mantle are stress free. 
A hydride is nucleated near the lower surface, shown here in red.
Energy minimisation then determines the shape and size of the equilibrium hydride precipitate.
The system has a cylindrical symmetry, which allows to reduce the problem to a two-dimensional description.
}
\label{fig5}
\end{center}
\end{figure}

Since the main topic of the present article is to study phase separation in the presence of surfaces, we consider now a system with traction free boundaries, i.e.~a system that is not confined by external stresses ($\sigma_{in}=0$), using a three-dimensional description.
In order to reduce the numerical efforts, we assume without loss of generality a cylindrical symmetry of our problem, as sketched in Fig.~\ref{fig5}.
This means that our sample has a cylindrical shape, and both the mantle as well as top and bottom are assumed to be traction free.
Top and bottom are therefore equivalent locations for the precipitate, and its equilibrium location is selected by the initial conditions for the energy minimisation.
In the simulations we ensure that the cylinder radius $R$ is sufficiently large, such that the mantle boundaries do not play a role, and the system can be considered as infinitely extended, $R\to \infty$.
In a full three dimensional description all positions at the bottom (or top) surface are energetically equivalent and we can place the hydride on the axis of symmetry;
in the two-dimensional representation with cylindrical symmetry off-centred positions would correspond to an unfavourable toroidal precipitate shape.

\begin{figure}
\begin{center}
\includegraphics[width=4.2cm]{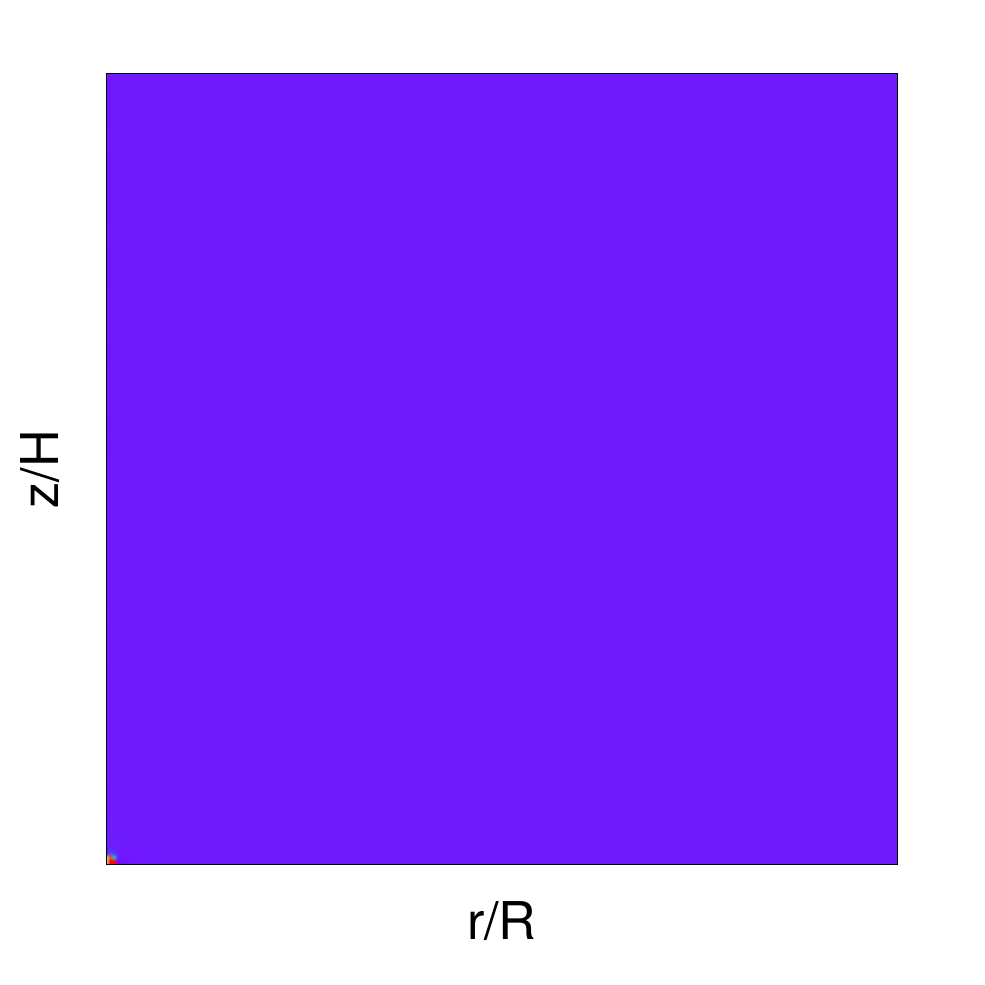}
\includegraphics[width=4.2cm]{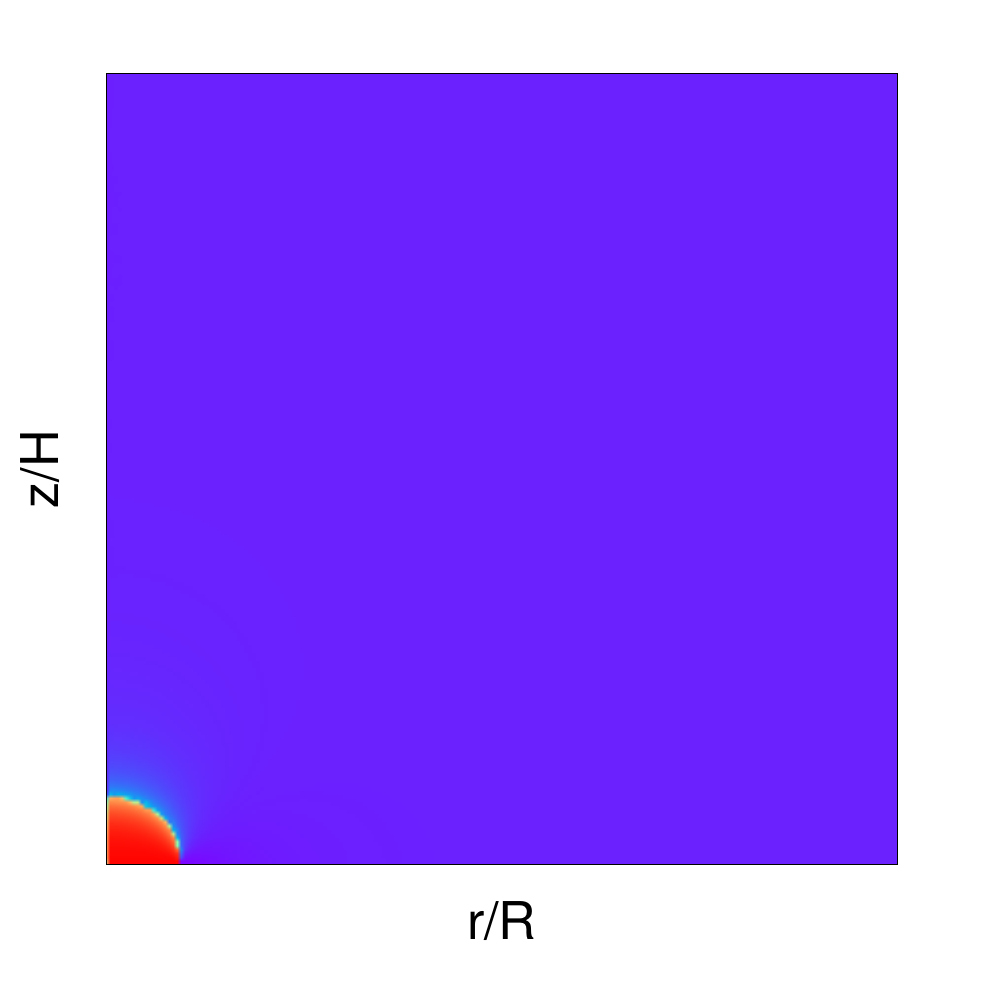}
\includegraphics[width=4.2cm]{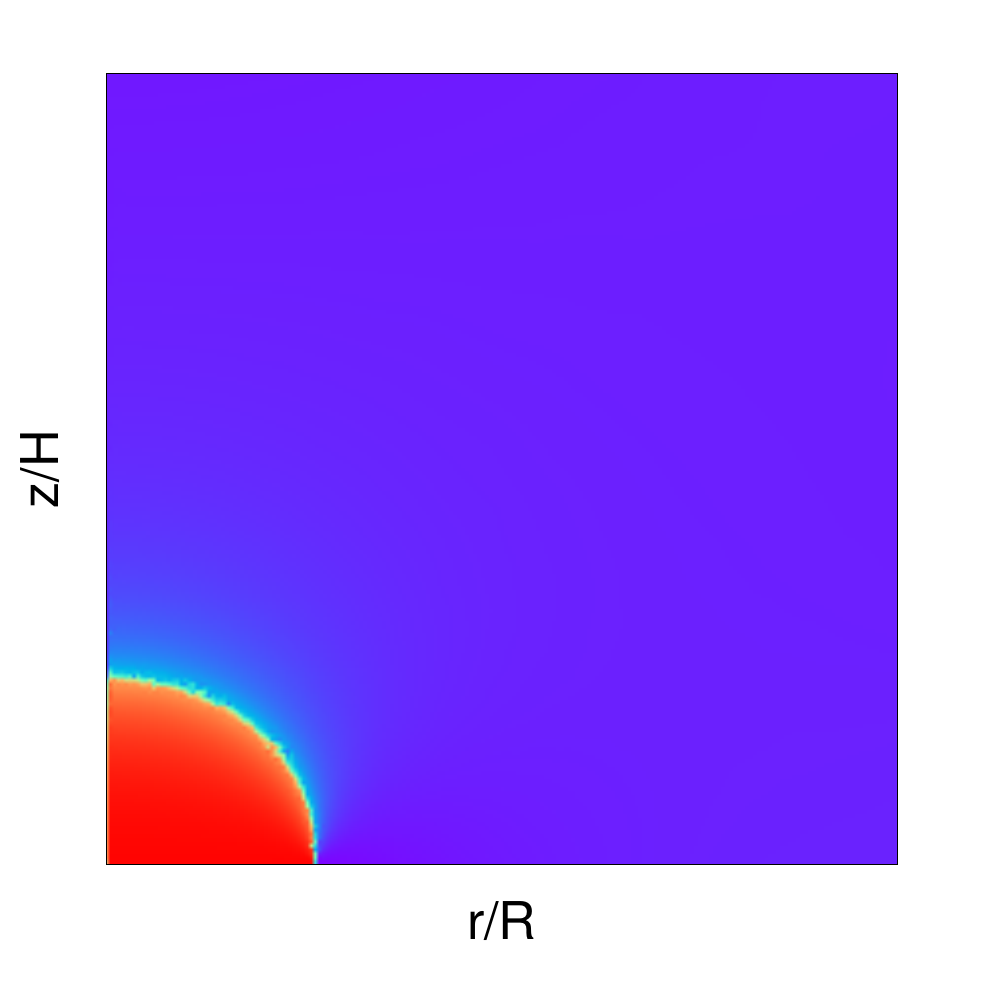}
\includegraphics[width=4.2cm]{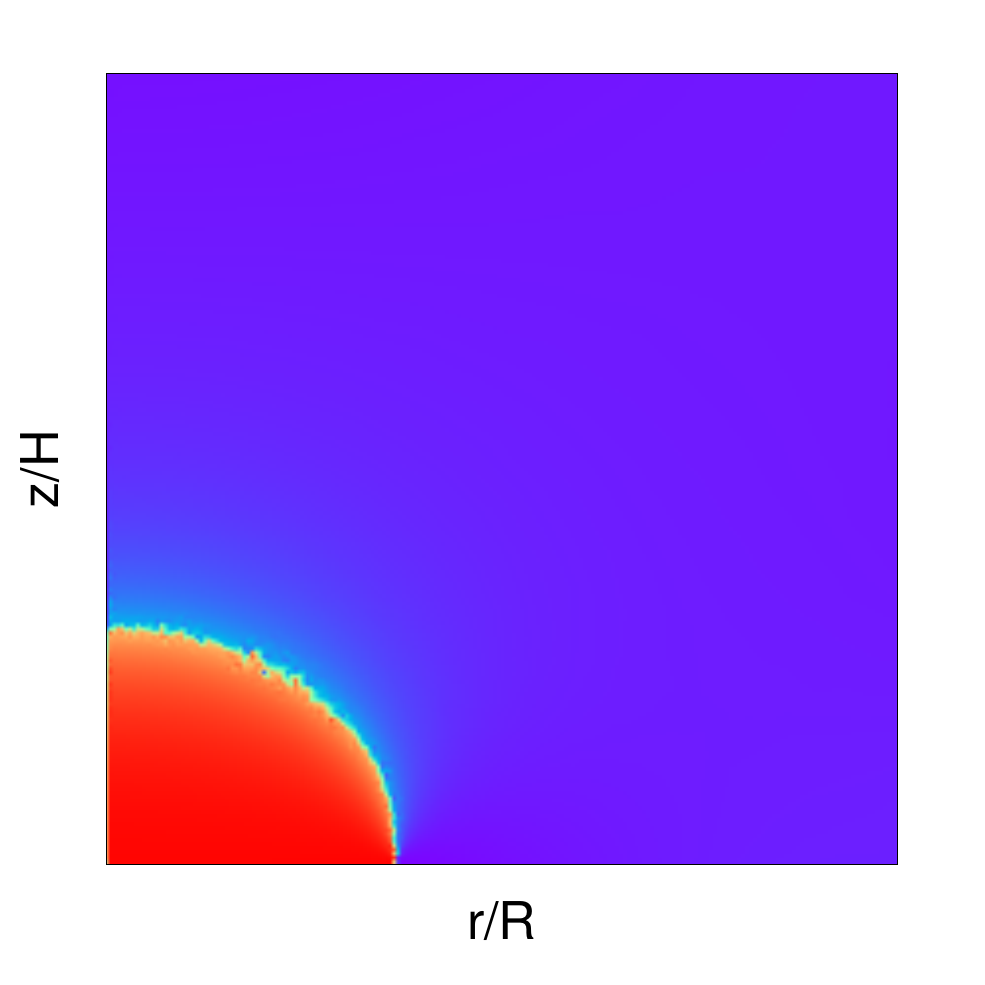}
\caption{(Color online) Color plot of the local hydrogen concentration at $T=500\mathrm{K}$, for different average hydrogen concentrations $\conc$.
Blue corresponds to a low hydrogen concentration ($\alpha$ phase), red a high concentration, i.e.~the hydride ($\beta$ phase).
The $z$ is the axis of rotational symmetry.
The aspect ratio is $H/R=1$.
For the lowest concentration (top left) at $\conc=0.031$ the system is in equilibrium in a single phase state.
For $\conc=0.041$ (top right) a nucleus forms, which is getting larger for higher concentration, $\conc=0.051$ (bottom left) and $\conc=0.061$ (bottom right).
}
\label{fig6}
\end{center}
\end{figure}

The elastic Cahn-Hilliard model is solved using a finite element description (see Appendix \ref{section::FEM} for details) and the FreeFEM package\cite{Hecht:2012aa}.
Initially, we place a spherical hydride inside the volume, in proximity to either the top or bottom surface.
This initial hydrogen distribution determines the average concentration.
The evolution is then modelled according to Eq.~(\ref{eq10}) until the system has reached an equilibrium configuration.
Energy reduction leads to an effective attractive interaction between the inclusion and the free surfaces, and therefore the equilibrium precipitate forms an almost semi-spherical inclusion at this surface, see Fig.~\ref{fig6}.

\begin{figure}
\begin{center}
\includegraphics[width=8.5cm]{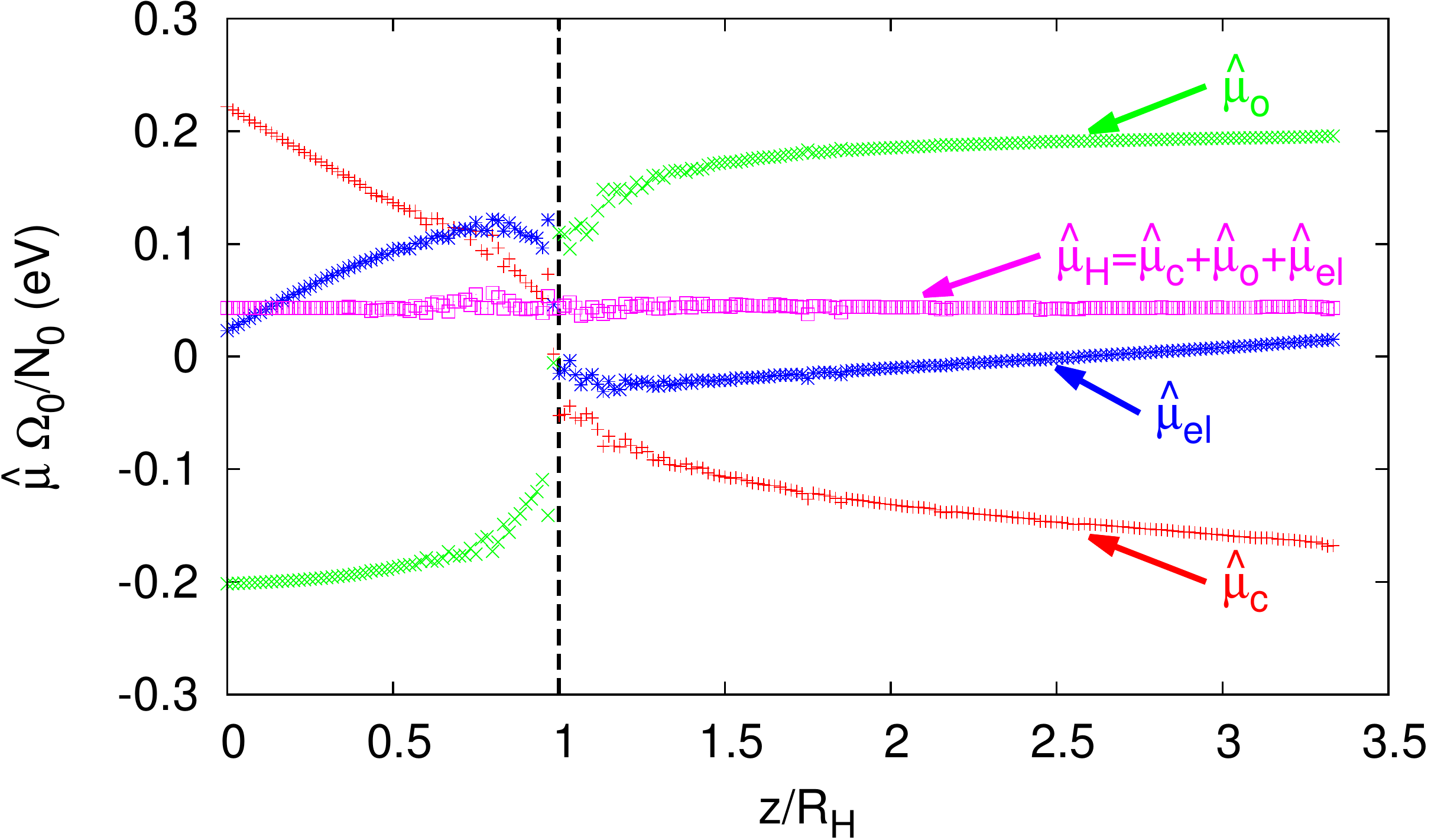}
\caption{(Color online) The contributions to the chemical potential along the axis of symmetry, $r=0$. 
Whereas the total chemical potential is spatially constant, the individual contributions are varying inside the individual phases and have a jump at the interface $z=\Rhydride$.
The data stems from an equilibrium configuration for $T=500\,\mathrm{K}$ and average concentration of $\conc=0.061$ (see also Fig.~\ref{fig6}).
The numerical noise results from data interpolation in between the nodes of the FEM implementation.
}
\label{fig7}
\end{center}
\end{figure}

In equilibrium, the chemical potential is spatially constant in the system.
This is shown in Fig.~\ref{fig7}, along the axis of symmetry, $r=0$.
The contributions to the chemical potential are defined as $\hat{\mu}=\gh'(\conc)$ for each of the terms in Eq.~(\ref{eq1}).
In contrast to the total chemical potential $\muHh=\much+\muoh+\muelh$ the individual contributions are not spatially constant inside the phases and have a jump at the interface.
Notice that the interfacial contribution vanishes for $\gamma_0=0$.

The most important result is that we find phase separation in regimes where we would not have expected it from the bulk phase diagram.
In the example in Fig.~\ref{fig6} we see the formation of a hydride at $\conc=0.041$ at $T=500\,\mathrm{K}$, whereas from the bulk coherent binodal (see Fig.~\ref{fig2}) we would expect single phase equilibrium states for $\conc<0.29$.
Apparently, near free surfaces the solubility limit differs from the bulk coherent prediction and is in between the solubility limit with and without stress effects.
The two phase region near a free interface is then delimited by a new kind of binodal curve, which we denote as the {\em coherent surface binodal}, and which will be determined numerically in the following, and more fundamentally in the next sections.

To find this coherent surface binodal, we proceed as follows.
Starting from the phase separated states, as in Fig.~\ref{fig6}, we reduce the concentration sequentially, until the system reaches a single phase state.
Repeating this procedure for several temperatures allows to extract the low concentration branch of the coherent surface binodal.
Similarly, we start in the high concentration regime, in which the system mainly consists of the hydride and only has a small metallic region.
This entails that in Figs.~\ref{fig5} and \ref{fig6} the role of $\alpha$ and $\beta$ phase are exchanged.
With increasing average concentration the $\alpha$ volume fraction decreases and finally this phase disappears, which delimits the two phase region in the high concentration regime.
Since the model is symmetric with respect to an exchange $\conc\to 1-\conc$, also the phase diagram has this symmetry property.

\begin{figure}
\begin{center}
\includegraphics[width=8.5cm]{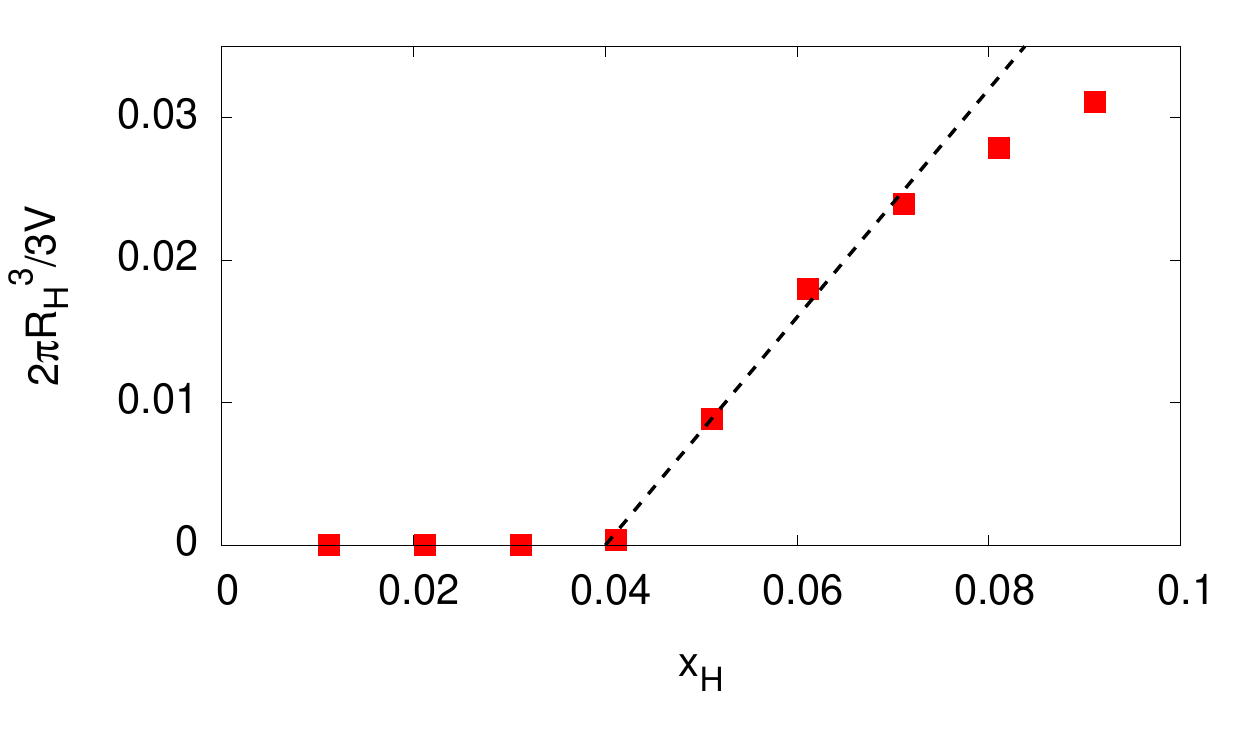}
\caption{(Color online) Hydride volume fraction as function of the average concentration $\conc$ at $T=500\,\mathrm{K}$. For $\conc<0.04$ the system is in a single phase state and the hydride vanishes.
The red points are results of the numerical simulations, the line is a linear fit.}
\label{fig10}
\end{center}
\end{figure}

\begin{figure}
\begin{center}
\includegraphics[width=8.5cm]{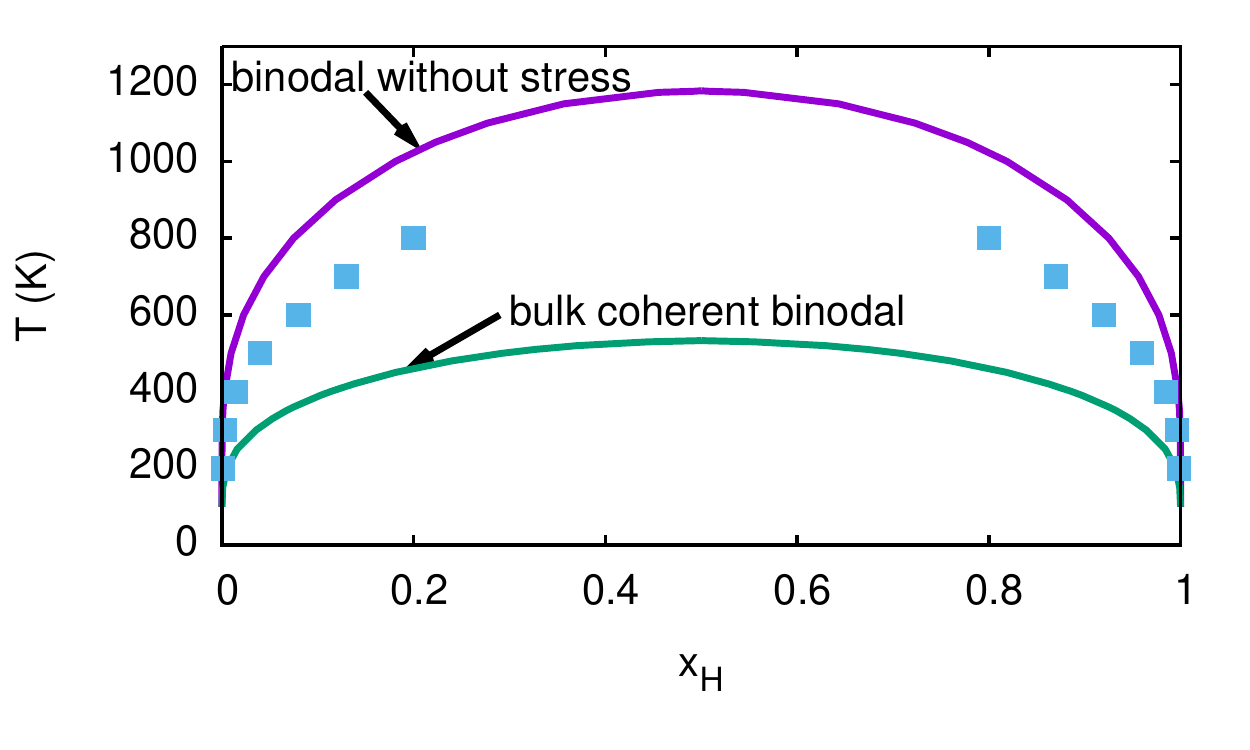}
\caption{(Color online) The phase diagram with the coherent surface binodal, obtained from the continuum model (blue data points).
The focus is here on the low temperature data.
For higher temperatures, closer to the critical point, a reliable determination of the solubility limit becomes numerically difficult, as then the concentrations in the two phases do not differ a lot.
Also, in the very low temperature regime, the solubility limit becomes very small, such that a proper numerical resolution cannot be achieved with sufficient precision. 
Therefore, data below 200 K and above 800 K is not shown.
}
\label{fig11}
\end{center}
\end{figure}

More explicitly, we measure for each simulation in equilibratium the position of the sharp concentration drop along the $z$ axis.
This position $z=\Rhydride$ is defined to be the hydride radius.
An appropriate measure for the (approximate) volume fraction is the expression $v=2\pi \Rhydride^3/3V$, with $V$ being the cylinder volume.
Here we assume a semi-spherical cap shape, as shown in Fig.~\ref{fig6}.
If we plot this volume fraction as function of the average concentration, we can linearly interpolate to the concentration where the two-phase region begins, as shown in Fig.~\ref{fig10}.
Repeating this procedure for different temperatures gives the surface coherent phase diagram, as shown in Fig.~\ref{fig11}.
In the low temperature regime the equations become very stiff, and the present numerical approach is not suitable to predict the low solubility limits with high accuracy.
At this point, analytical predictions, which will be developed in the subsequent section, are particularly useful.

\begin{figure}
\begin{center}
\includegraphics[width=8.5cm]{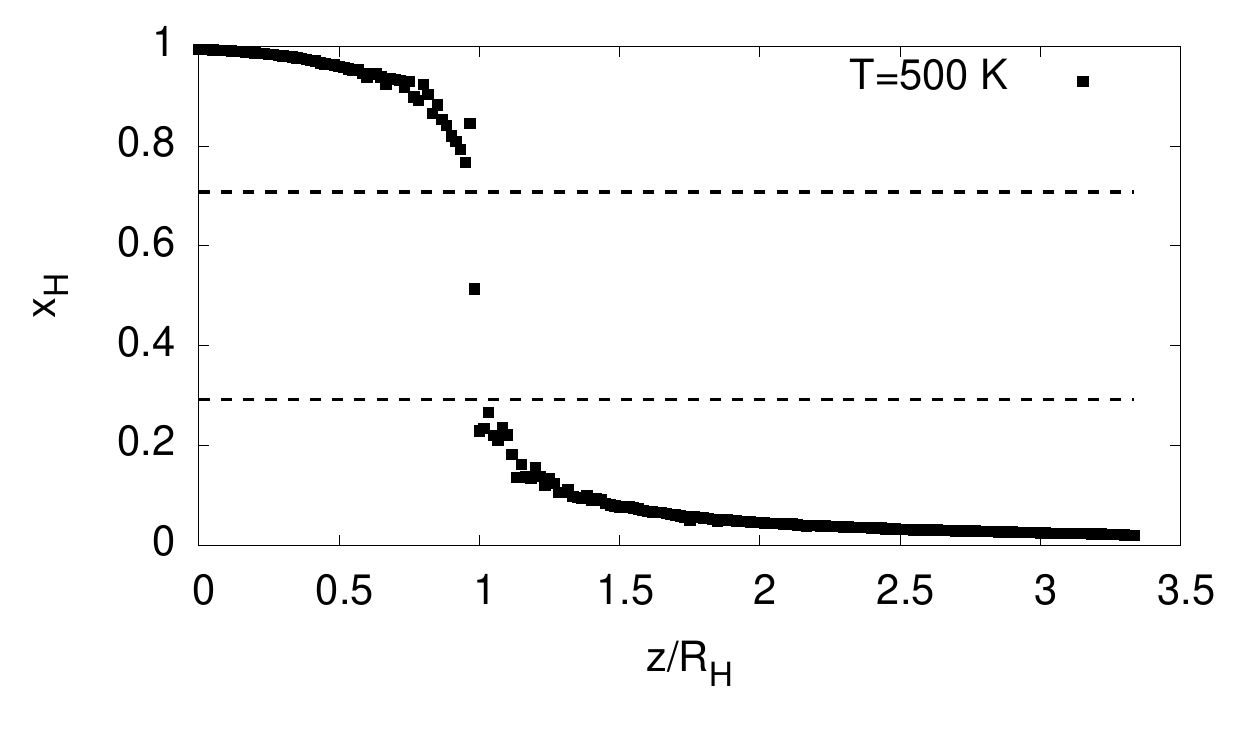}
\includegraphics[width=8.5cm]{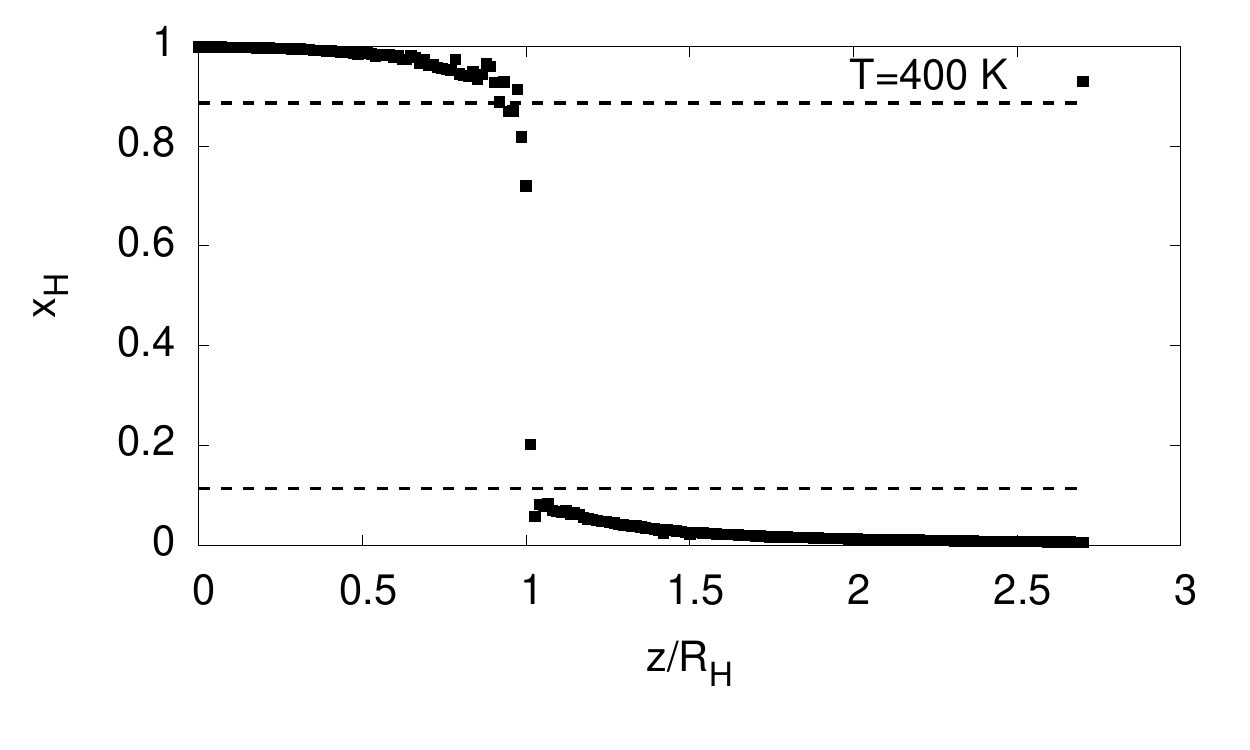}
\includegraphics[width=8.5cm]{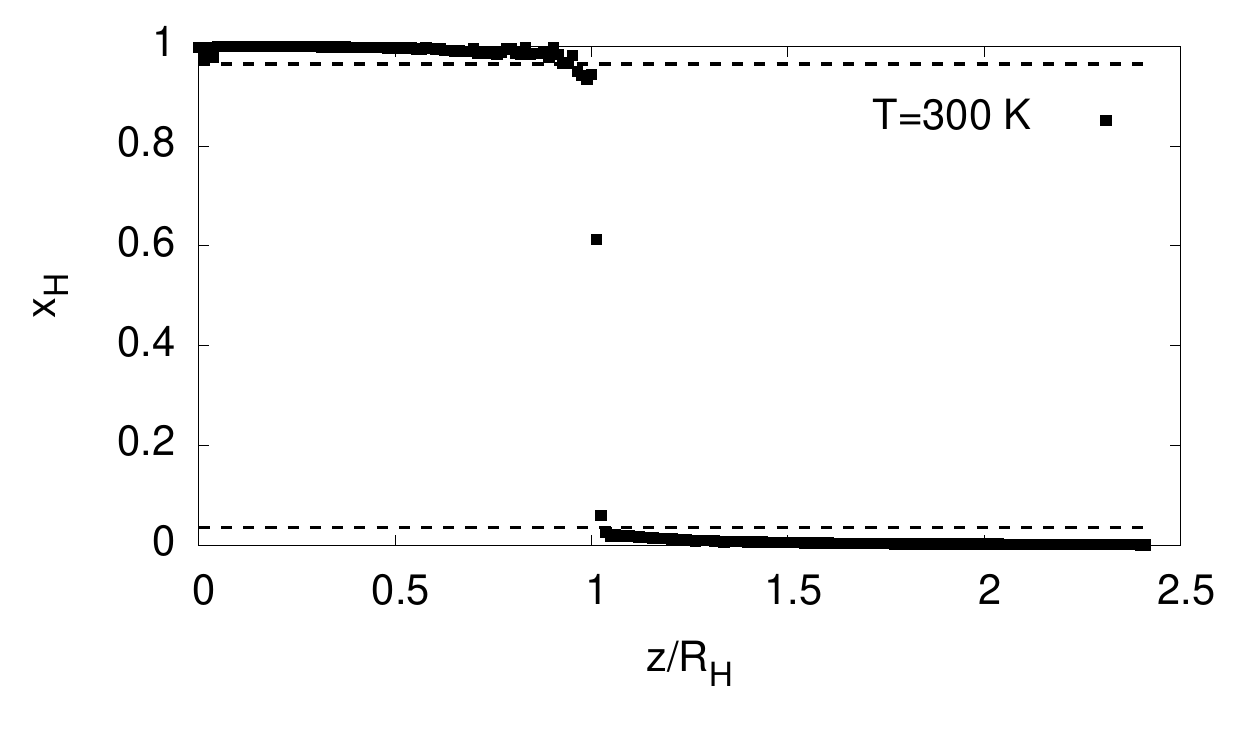}
\caption{The hydrogen concentration along the axis of symmetry, $r=0$, as defined in Fig.~\ref{fig5}. 
At the interface between the phases the local concentrations are given by the bulk coherent phase diagram, as indicated by the dashed horizontal lines.
Top panel: The data stems from an equilibrium configuration for $T=500\,\mathrm{K}$ and average concentration of $\conc=0.061$ (see also Fig.~\ref{fig6}).
In the center and bottom panel the same is shown for $T=400\,\mathrm{K}$ and $T=300\,\mathrm{K}$.
The concentration gradients inside the phases become smaller at lower temperatures.
}
\label{fig8bbbbbb}
\end{center}
\end{figure}

A noticeable outcome of the simulations is that the concentrations are no longer constant inside the phases, and therefore this situation differs from the phase separation in the bulk.
The combination of elasticity and free surfaces changes the thermodynamic situation conceptually, in comparison to phase separation in the bulk.
This is demonstrated exemplarily in Fig.~\ref{fig8bbbbbb} for the concentration profile along the axis of symmetry, $r=0$, as defined in Fig.~\ref{fig5}.
In this plot, the equilibrium concentrations given by the bulk coherent phase diagrams are also shown as dashed lines for comparison.
One can see that the equilibrium concentrations are spatially no longer constant inside each phase.
The simulations indicate that the equilibrium concentrations of the {\em bulk} coherent phase diagram are locally reached exactly at both sides of the interface.
We will later verify this observation analytically.

This result has an important consequence:
Despite the varying concentrations in the phases at finite temperatures, the concentration variations decrease for lower temperatures and finally vanish in the limit $T\to 0$.
The reason is that the coherent bulk phase diagram shows zero hydrogen solubility in the metal and similarly the hydride is free of hydrogen vacancies, see the phase diagram in Fig.~\ref{fig2}. 
%
%
This result of constant phase concentrations in the low temperature limit is essential for further analytical predictions in the following section.

This discussion concludes the phenomenological inspection of the solubility limits near free surfaces using the continuum model.
In the next section, we will support and generalise the findings by analytical considerations for the low temperature limit.


\section{Analytical description}
\label{analytical::section}

In this section we complement the above description of phase separation by analytical descriptions, which allow for a closed understanding of the coherent surface binodal.
It extends the phenomenological results of the continuum model, which were discussed in the previous section, by a more thorough and generic understanding and is a prerequisite for the scale bridging connection to atomic scale inspections in the following section.

The starting point is the coherent nucleation in the bulk, as it is the reference case to which we add surface effects.
First, we derive from the exact phase coexistence conditions the solubility limit in the low temperature regime for the bulk system.
Next, we solve the elastic problem of precipitates near free surfaces and interfaces, and show how this enters into the description of phase coexistence.
From this step, we get an analytical expression for the low temperature solubility limit, comparable to the previous bulk case.
This prediction asymptotically matches the previous numerical results for the surface binodal for low temperatures and concentrations and demonstrates the influence of the different material parameters and boundary conditions.
Apart from minimisation of the total Gibbs energy, we also discuss local equilibrium conditions at the interface between the $\alpha$ and $\beta$ phase.
This inspection explains the  concentration jump observed numerically in Fig.~\ref{fig8bbbbbb}.
We raise the question whether the appearance of a single nucleus is favorable to a breakup into smaller nuclei.
Finally, we compare the solubility limits for bulk and surfaces, beyond which phase separation can occur, to spinodal decomposition.

\subsection{The coherent bulk phase diagram}
\label{coherent::section}

In the isotropic case with equal elastic constants of the two phases the elastic problem of coherent inclusions in an infinite or periodic matrix can be solved analytically.
Only the volume fraction of the secondary phase matters, and the detailed arrangement and shape of the precipitates is irrelevant according to the Bitter-Crum theorem\cite{Fratzl:1999aa}.
As a well known consequence, there is no elastic interaction between two precipitates in this case, as the change of inclusion separation does not change their volume fraction.
We denote the volume fraction of the hydride ($\beta$ phase) by $v$.
In the case that the system of volume $V$ can expand freely (external pressure zero), the elastic energy is given by (for an infinite system, one can look instead at the average energy which is contained in a subvolume $V$)
\begin{equation} \label{BitterCrumEq}
\Gelb = V v(1-v) (\concb-\conca)^2 \chi^2 \frac{E}{1-\nu},
\end{equation}
with Young's modulus $E=G(3\lambda+2G)/(\lambda+G)$ and the Poisson ratio $\nu=\lambda/[2(\lambda+G)]$.
Here it is assumed that the concentrations $\conca$ and $\concb$ are spatially constant inside the individual phases.
For phase separation in the bulk, this statement is exact and is corroborated by the numerical simulations in Fig.~\ref{fig4aaa}.
For the specific case of a concentric spherical inclusion in a spherical matrix (Eshelby problem), the above energy expression can also easily be calculated, see Appendix \ref{Eshelby::section}.
Also, there we show that the assumption of spatially constant concentrations inside each phase is indeed satisfied, as no chemical potential gradients appear.
Expression (\ref{BitterCrumEq}) reflects that for single phase configurations ($v=0$ or $v=1$) no coherency stresses arise.

Whereas in general a common tangent construction is not applicable in the presence of elastic effects, the situation is more fortunate in the case of coherent interfaces, linear and isotropic elasticity with equal elastic constants in both phases and a dilatational eigenstrain.
More general situations can conceptually be considered as perturbations of this ideal situation.

The stress free phases are assumed to be characterised by Gibbs energies (per host atom) $g_\alpha(x_\alpha)$ and $g_\beta(x_\beta)$, including the configurational entropy.
The temperature dependence is suppressed in the notation for brevity.
We emphasize that these Gibbs energies shall not contain long-ranged elastic effects, which are accounted for separately.
It is assumed that the above energies are taken at the equilibrium volume, i.e.~for stress free situations.
The above formulation generalises the model used in Eqs.~(\ref{eq1})-(\ref{gchdef}), for which we used a single expression for the Gibbs energy density for both phases.
If we apply the present, more general, case to the previous situation, we have to use $g_\alpha = g_\beta = (\goh + \gch) \omz/\Nz$.
In contrast to the continuum description in the previous section, here we normalize all intensive quantities to the number of metal atoms and not to the volume, as such a representation will be more useful for the later connection to the discrete atomistic description.
To make this normalization transparent in the notation we do not decorate the symbols by a hat, and we have the conversion rule $g=\gh \omz/\Nz$.

The concentrations in the phases are $x_\alpha = N_\mathrm{H}^\alpha/N_\mathrm{M}^\alpha$ and $x_\beta = N_\mathrm{H}^\beta/N_\mathrm{M}^\beta = (N_\mathrm{H}-N_\mathrm{H}^\alpha)/(N_\mathrm{M}-N_\mathrm{M}^\alpha)$.
Here, $N_\mathrm{M}$ is the total (and fixed) number of host atoms, $N_\mathrm{H}$ the total (and fixed) number of interstitials, as we use a canonical ensemble.
$N_\mathrm{M}^\alpha$ is the total number of host atoms which belong to the $\alpha$ phase, and $N_\mathrm{H}^\alpha$ the number of interstitial atoms in the $\alpha$ phase.
Similar quantities are defined in the $\beta$ phase.
Notice that $N_\mathrm{M}^\alpha$ and $N_\mathrm{H}^\alpha$ are not fixed and degrees of freedom for the Gibbs energy minimisation.
The total Gibbs energy reads (bulk contributions only)
\begin{equation}
\Gibbs=\Gibbs_\alpha + \Gibbs_\beta + \Gelb
\end{equation}
with
\begin{equation}
\Gibbs_\alpha = N_\mathrm{M}^\alpha g_\alpha\left( \frac{N_\mathrm{H}^\alpha}{N_\mathrm{M}^\alpha} \right),
\end{equation}
\begin{equation}
\Gibbs_\beta = (N_\mathrm{M}-N_\mathrm{M}^\alpha) g_\beta\left( \frac{N_\mathrm{H}-N_\mathrm{H}^\alpha}{N_\mathrm{M}-N_\mathrm{M}^\alpha} \right).
\end{equation}
For the elastic energy we use $V= N_\mathrm{M} \omz/\Nz$.
Here $\omz$ is the unit cell volume of the hydrogen free lattice which serves as Lagrangian reference state, and $N_0$ is the number of host atoms per unit cell (e.g.~$N_0=4$ for fcc).
The $\beta$ phase volume fraction reads $v=(N_\mathrm{M}-N_\mathrm{M}^\alpha)/N_\mathrm{M}$.
Hence, the elastic energy for a system, which is free of external stresses, is expressed through the particle numbers as
\begin{eqnarray} 
\Gelb &=& \frac{E}{1-\nu} \chi^2 \frac{N_\mathrm{M}-N_\mathrm{M}^\alpha}{N_\mathrm{M}} \frac{N_\mathrm{M}^\alpha}{N_\mathrm{M}} \times \nonumber \\
&& \times \left( \frac{N_\mathrm{H}-N_\mathrm{H}^\alpha}{N_\mathrm{M}-N_\mathrm{M}^\alpha} - \frac{N_\mathrm{H}^\alpha}{N_\mathrm{M}^\alpha} \right)^2 \frac{N_\mathrm{M} \omz}{\Nz}, \label{bc}
\end{eqnarray}
see Eq.~(\ref{BitterCrumEq}) above.

In thermodynamic equilibrium the Gibbs energy is minimised with respect to the internal degrees of freedom $N_\mathrm{H}^\alpha$ and $N_\mathrm{M}^\alpha$.
The first minimisation condition is
\begin{equation}
\left(\frac{\partial \Gibbs}{\partial N_\mathrm{H}^\alpha}\right)_{N_\mathrm{H}, N_\mathrm{M}, N_\mathrm{M}^\alpha}=0.
\end{equation}
It gives
\begin{eqnarray} 
\left(\frac{\partial \Gibbs}{\partial N_\mathrm{H}^\alpha}\right)_{N_\mathrm{H}, N_\mathrm{M}, N_\mathrm{M}^\alpha} &=& g_\alpha'(x_\alpha) + \frac{2E}{1-\nu} \chi^2  \frac{\omz}{\Nz}  x_\alpha  \nonumber \\
&-& \left( g_\beta'(x_\beta) + \frac{2E}{1-\nu} \chi^2  \frac{\omz}{\Nz}  x_\beta \right) \nonumber \\
&=& 0. \label{cahneq1}
\end{eqnarray}
Without elastic effects, it reduces to the usual equality of chemical potentials $\muH=g'(\conc)$ in the two phases.

The second minimisation condition is
\begin{equation}
\left(\frac{\partial \Gibbs}{\partial N_\mathrm{M}^\alpha}\right)_{N_\mathrm{H}, N_\mathrm{M}, N_\mathrm{H}^\alpha}=0.
\end{equation}
It becomes
\begin{eqnarray} 
&& g_\alpha(x_\alpha) - x_\alpha g_\alpha'(x_\alpha)  - \frac{E}{1-\nu}\chi^2\frac{\omz}{\Nz} x_\alpha^2 \nonumber \\
&-& \left( g_\beta(x_\beta) + x_\beta g_\beta'(x_\beta) - \frac{E}{1-\nu}\chi^2\frac{\omz}{\Nz} x_\beta^2 \right) \nonumber \\
&=&0. \label{cahneq2}
\end{eqnarray}
Without elastic effects, it reduces to the usual equality of grand potentials $\omega=g-\conc\muH$ in the two phases.
By the Gibbs-Duhem relation $g=\mu_\mathrm{M} + \muH \conc$ with the chemical potential $\mu_\mathrm{M}$ of the metal atoms, $\mu_\mathrm{M} = (\partial \Gibbs/\partial N_\mathrm{M})_{N_\mathrm{H}}$, hence $\omega=\mu_\mathrm{M}$.
We can therefore alternatively read Eq.~(\ref{cahneq2}) in the absence of stress effects as equality of chemical potentials of the host metal atoms in the two phases.
Without elasticity, the equations (\ref{cahneq1}) and (\ref{cahneq2}) are the usual common tangent construction for binary two-phase systems.

We can define an auxiliary potential $\bar{g}$ for each phase, as suggested by Cahn \cite{Cahn:1962aa},
\begin{equation} \label{CahnAuxiliaryPotential}
\bar{g}_\alpha(x_{\alpha,\beta}) = g_{\alpha,\beta}(x_{\alpha,\beta}) + \frac{E}{1-\nu}\chi^2\frac{\omz}{\Nz} x_{\alpha,\beta}^2,
\end{equation}
such that the generalised chemical potentials become
\begin{equation}
\bar{\mu}_{\alpha,\beta}(x_{\alpha,\beta}) = \bar{g}_{\alpha,\beta}'(x_{\alpha,\beta})
\end{equation}
and the generalised grand potentials are
\begin{equation}
\bar{\omega}_{\alpha,\beta}(x_{\alpha,\beta}) = \bar{g}_{\alpha,\beta}(x_{\alpha,\beta}) - x_{\alpha,\beta} \bar{\mu}_{\alpha,\beta}.
\end{equation}
The above equilibrium conditions (\ref{cahneq1}) and (\ref{cahneq2}) then read
\begin{equation} \label{CahnsBothConditions}
\bar{\mu}_\alpha(x_\alpha)=\bar{\mu}_\beta(x_\beta),\qquad \bar{\omega}_\alpha(x_\alpha)=\bar{\omega}_\beta(x_\beta),
\end{equation}
which means that they can be interpreted as the result of a common tangent construction using the potentials $\bar{g}_\alpha$ and $\bar{g}_\beta$ instead of the stress free Gibbs energies $g_\alpha$ and $g_\beta$.
We have previously used these results in Section \ref{section::coherentphasediagram}, see Eq.~(\ref{CahnModifiedPotentialJustUsed}).

We can further investigate this generalised common tangent construction in the regime of small concentrations.
The (generalised) Gibbs energies of the pure $\alpha$ and $\beta$ phase per host atom are decomposed into a regular contribution $\bar{g}^0_{\alpha, \beta}$ and a singular configurational entropy contribution $g_\mathrm{c}$,
\begin{equation} \label{singulardecomposition}
\bar{g}_{\alpha, \beta}(x_{\alpha, \beta}, T) = \bar{g}^0_{\alpha, \beta}(x_{\alpha, \beta}) + g_\mathrm{c}(x_{\alpha, \beta}, T).
\end{equation}
Here, singular means that $g_\mathrm{c}'(\conc, T)$ diverges for $\conc\to 0$, where the prime denotes differentiation with respect to concentration.
The dominant contribution to $g_\mathrm{c}$ for the low concentration regime is $g_\mathrm{c}\simeq \kB T \conc \ln(\conc/x_0)$, in agreement with Eq.~(\ref{gchdef}).
If a two-phase mixture forms, the averaged Gibbs energy per host atom of the heterogeneous system is
\begin{equation} \label{genGhet}
\bar{g}_\mathrm{het}(\conc, T) = \bar{g}_\alpha(x_\alpha, T) (1-v) + \bar{g}_\beta(x_\beta, T) v.
\end{equation}
In the following we distinguish between the concentrations $x_\alpha$ and $x_\beta$ in the two phases and the average concentration $\conc$.
We consider phase diagrams with zero solubility in the $\alpha$ phase for $T\to 0$, i.e.~$x_\alpha\to x_\alpha^0=0$.
In the same limit, the $\beta$ phase solubility tends to $x_\beta^0$.
Hence by the lever rule $\conc=x_\alpha (1-v) + x_\beta v$ we obtain $v\simeq\conc/x_\beta^0$ for $T\to 0$.

Equality of chemical potentials at the touching point $\conc=x_\alpha$ of the common tangent construction demands
\begin{equation}
\left( \frac{\partial \bar{g}_\mathrm{het}(\conc, T)}{\partial \conc}\right)_{T, x_\alpha, x_\beta, \conc=x_\alpha} =\left(\frac{\partial \bar{g}_\alpha(x_\alpha, T)}{\partial x_\alpha}\right)_{T}, \label{touchingcondition}
\end{equation}
which expresses that the common tangent has the same slope as the Gibbs energy $\bar{g}_\alpha$ at the touching point $\conc=x_\alpha$.
Since the configurational contribution is regular (and vanishingly small) in the Gibbs energy and singular only in the chemical potential near $x_\alpha\approx 0$, we get from Eqs.~(\ref{genGhet}) and (\ref{touchingcondition}) in the limit $T\to 0$
\begin{equation}
\left[ -\bar{g}_\alpha^0(0) + \bar{g}_\beta^0(x_\beta^0) \right] \frac{1}{x_\beta^0}\simeq {\overline{g}_\alpha^0}'(0) + \kB T \ln(\conc/\xn),
\end{equation}
where only the singular contribution is retained in the last term.
This derivation gives for the solubility limit
\begin{equation} \label{sol::eq1}
\conc \simeq \xn \exp\left(- \frac{\Delta \bar{\Gibbs}}{\kB T} \right)
\end{equation}
with
\begin{equation} \label{sol::eq2}
\Delta \bar{\Gibbs} := [-\bar{g}_\beta^0(x_\beta^0)+\bar{g}_\alpha^0(0)]/x_\beta^0+{\overline{g}_\alpha^0}'(0).
\end{equation}
In the special case without elasticity we have $g=\bar{g}$, and then we recover the standard Arrhenius expression for the solubility limit in the low concentration and temperature regime.
Hence, we can interpret the energy
\begin{equation}
\Delta \Gibbs := [-g_\beta^0(x_\beta^0)+g_\alpha^0(0)]/x_\beta^0+{g_\alpha^0}'(0),
\end{equation}
as formation energy, where we omitted the elastic contribution.
We will later confirm this interpretation.

\begin{figure}
\begin{center}
\includegraphics[width=8.5cm]{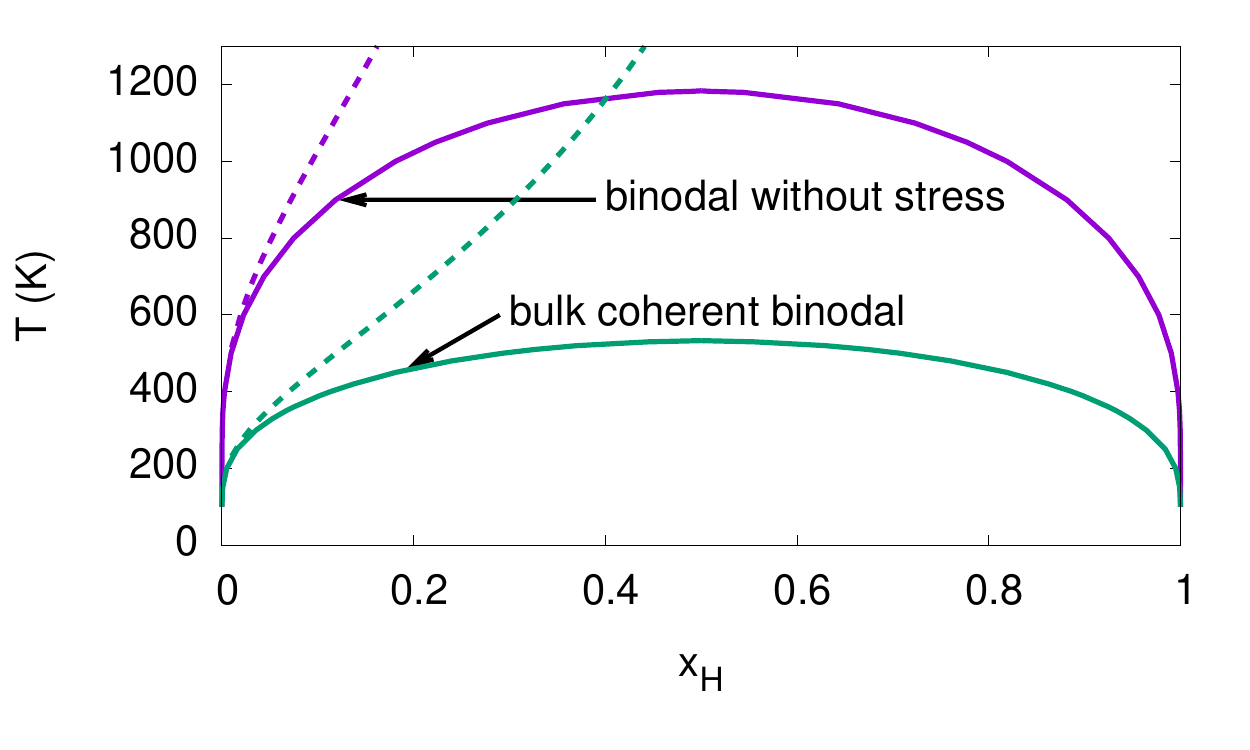}
\caption{(Color online) Comparison of the full phase diagram with and without elasticity (solid curves) with the corresponding asymptotic expressions (\ref{sol::eq1}) in the low concentration regime (dashed curves).}
\label{fig7_5f}
\end{center}
\end{figure}

With elastic effects we get from Eq.~(\ref{CahnAuxiliaryPotential}) the new formation energy
\begin{equation} \label{TotalEffectiveFormationEnergy}
\Delta \bar{\Gibbs} = \Delta \Gibbs + \Delta \Gelb 
\end{equation}
with
\begin{equation} \label{TotalEffectiveFormationEnerg2}
\Delta\Gelb = - \frac{E}{1-\nu} \chi^2 \frac{\omz}{\Nz} \concb^0.
\end{equation}
Since the elastic term reduces the formation energy, the solubility limit is increased.
The agreement of the asymptotic expression (\ref{sol::eq1}) with and without stress and the numerically calculated phase diagram in the low concentration regime is shown in Fig.~\ref{fig7_5f}. 
\subsection{Solubility limits near surfaces and interfaces}
\label{SurfacesInterfaces::section}

The central objective of the present paper is that whereas phase separation in the bulk is only possible below the bulk coherent binodal (green curve in Figs.~\ref{fig11} and \ref{fig7_5f}) and thus for rather high concentrations, it can occur already at significantly lower concentrations at or near free surfaces.
If elastic stresses can relax completely, phase separation can occur already at lower concentrations, given by the binodal without stress (purple curve in Figs.~\ref{fig11} and \ref{fig7_5f}).
We have seen in Section \ref{continuum::section} that phase separation near surfaces typically starts at concentrations in between these two curves (blue points in Fig.~\ref{fig11}).
Conceptually, the near surface phase separation behavior is in accordance with the appearance of surface spinodal modes, which are also ultimately limited by the chemical instead of the coherent spinodal near free surfaces \cite{Tang:2012aa}.
It is the purpose of the present section to predict and understand the {\em coherent surface binodal} analytically in the low temperature regime.

A central point is the solution of the mechanical problem of a coherent inclusion near a free surface or an interface.
For that, we have employed both analytical methods as well as finite element simulations (see Appendix \ref{section::FEM} for details).
All calculations are performed under the same conditions as before, i.e.~linear isotropic elasticity, Vegard's law with isotropic lattice expansion, equal elastic constants in both phases and coherency at the interfaces.
Then, the elastic problem does not have an intrinsic lengthscale, hence the only relevant lengthscale is the size of the precipitate (provided that the sample is large).
In the proximity of a free surface we therefore expect to see deviations from the Bitter-Crum value (\ref{BitterCrumEq}) of the elastic energy, if the distance $h$ of the precipitate from the surface becomes comparable or smaller than the characteristic size $\Rhydride$ of the precipitate.
We will therefore obtain universal curves for the elastic energy modification near surfaces, which depend only on $h/\Rhydride$ for a given geometry, the set of elastic boundary conditions and the Poisson ratio.
Here it should be mentioned that in contrast to the bulk case, where only the volume fraction $v$ of the precipitates matters, the geometry here plays a role and leads to different energies for differently shaped precipitates.
This shape dependence will be elucidated in the following.
We point out that the energy modification appears as a {\em bulk} effect if the distance from the surface is of the order of the precipitate radius $\Rhydride$, and is therefore conceptually different from {\em surface} effects, which affect only the topmost surface layers.
This distinction is supported experimentally\cite{Northemann:2011aa,Burlaka:2015aa} e.g.~for coherent hydrides with a size of 30-40 nm in thin films of niobium, for which the ratio of the number of bulk to surface atoms is of the order $100$.
In the present analysis, we focus on bulk effects and leave out surface contributions.

\begin{figure}
\begin{center}
\includegraphics[width=9cm]{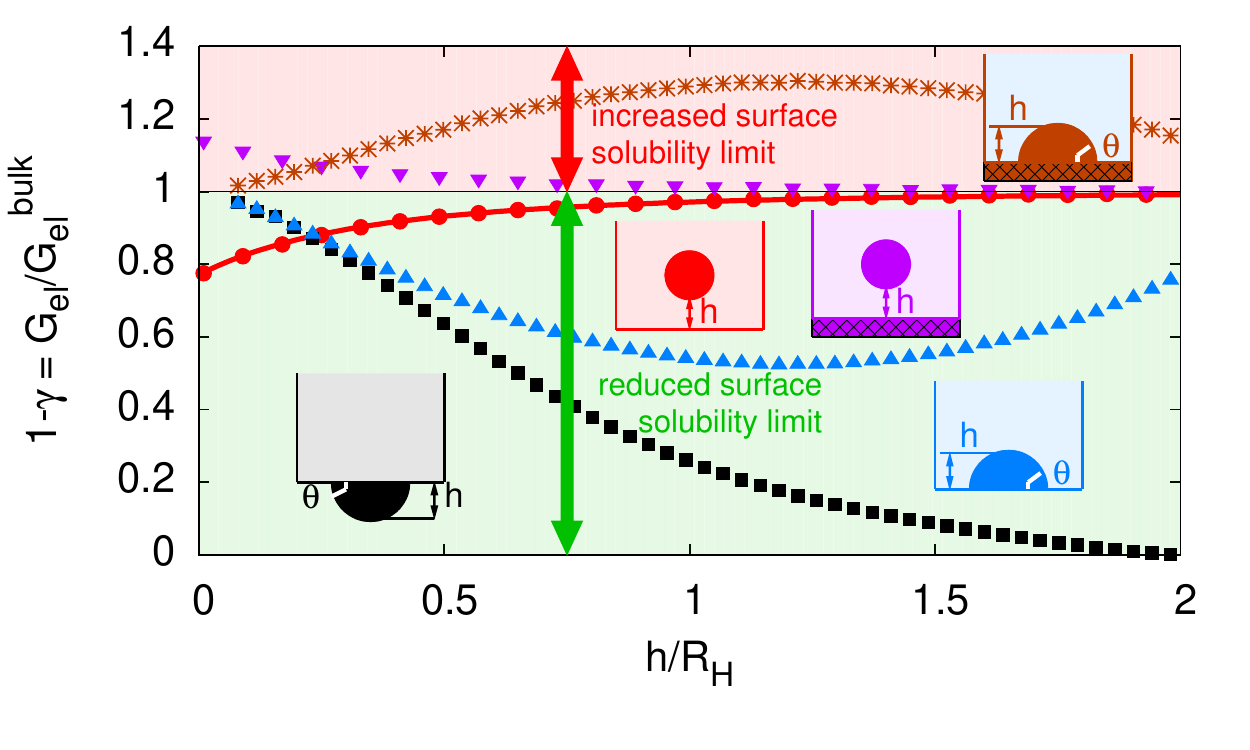}
\caption{(Color online) Elastic energy as function of the position and shape of a three-dimensional homogenous or heterogeneous nucleus, as obtained with the parameters in Table \ref{table1}. 
The small sketches illustrate the geometry for the elastic energy curves in the same color.
In all cases, the lower boundary of the simulation volume is either a free surface with zero tractions or a rigid substrate with fixed displacement boundary conditions.
If the nucleus is deep inside the material, the elastic energy depends only on the volume fraction in isotropic approximation and is given by the Bitter-Crum expression (\ref{BitterCrumEq}). 
In all cases the characteristic scale $\Rhydride$ is defined via the fixed nucleus volume $V_\beta$ as $V_\beta=4\pi \Rhydride^3/3$.
When a spherical inclusion (red) is approaching a free surface, the elastic energy decays according to a power law (solid red curve).
The limit (\ref{BitterCrumEq}) is also recovered if the precipitate is flat and fully wets the free surface ($h=0$ for the blue and black curve).
For a heterogeneous nucleus, which forms a spherical cap inside the sample on the free surface (blue), the elastic energy attains a minimum energy for a contact angle $\theta\approx90^\circ$.
For a spherical cap forming outside (black), the elastic energy decays to zero for a spherical droplet ($h/\Rhydride=2$).
In contrast, nucleation near a rigid substrate increases the elastic energy (purple, brown).
}
\label{fig3}
\end{center}
\end{figure}

For a spherical inclusion near a planar free surface of a semi-infinite (three-dimensional) material, the elastic problem is solved analytically (see Appendix \ref{inclusion::appendix} and Ref.~\onlinecite{Balluffi:2012aa} for details), and the distance dependent part of the elastic energy scales as $[\Rhydride/(h+\Rhydride)]^3$.
This dependence is shown in Fig.~\ref{fig3} as red solid curve.
In the plot we show the actual elastic energy $\Gel$, relative to the bulk value $\Gelb$, which is given by Eq.~(\ref{BitterCrumEq}).
The results are in agreement with finite element simulations (red dots), as explained in Appendix \ref{section::FEM}.
This shows that the elastic energy $\Gel$ is indeed lower for a misfitting inclusion near a free surface than in the bulk, $\Gelb$, because the stresses can partially relax.
In particular, the relevant scale for the decay of the surface-inclusion interaction is the precipitate radius $\Rhydride$.
For $h\gg \Rhydride$ we restore the bulk value (\ref{BitterCrumEq}) for the elastic energy.
We define $\gamma$ as the elastic energy in the near surface region in relation to the bulk value, $\gamma := 1-\Gel/\Gelb$.
A positive (negative) value corresponds to a reduction (an increase) of the elastic energy near a surface, compared to the bulk.
This way, the formation of a homogenous spherical nucleus can lead to a reduction of the energy down to $\gamma=(1+\nu)/6$, when touching the free surface ($h=0$).

Further energy reduction can be obtained if the nucleus forms a spherical cap inside the metallic matrix (blue curve).
This and all following data are obtained by the finite element simulations.
Variation of the cap height at fixed nucleus volume leads to an elastic energy minimisation for a contact angle of $\theta\approx 90^\circ$. 
Notice that the inclusion radius is defined here as the radius of a full sphere, which has the same volume as the cap.
For complete ``wetting'' of the free surface by the nucleus (the limit $h/\Rhydride\to 0$) the bulk Bitter-Crum elastic energy (\ref{BitterCrumEq}) is recovered.
This result is in agreement with the Eshelby problem (Appendix \ref{Eshelby::section}).
In both cases we obtain the same elastic energy as for a precipitate in the bulk.
One can obtain the result by taking into account that for the thin wetting film for $h/\Rhydride=0$ the hydrogen free matrix is fully relaxed, whereas the hydride layer acquires the same lattice constant as the substrate.
From this, the elastic energy can easily be calculated and gives the bulk value $\Gelb$.

First, we find an influence of the nucleus shape on the elastic energy in the isotropic elastic approximation in the presence of surfaces, in contrast to the bulk.
This dependence is the reason for the shape change of an initially spherical hydride seed when it is attracted by a free surface in the continuum simulations, see Fig.~\ref{fig6}.

Second, (heterogeneous) phase separation at free surfaces is favored over (homogeneous) phase separation in the bulk based already on elasticity alone.
The reduction of elastic energy near surfaces can therefore support precipitate formation.

Third, for a complete picture of precipitation thermodynamics, additionally the interfacial energy has to be considered, and the balance of bulk and interfacial terms determines the size of the critical nucleus.
If both are taken into account, the heterogeneous nucleation energy barrier is related to the homogeneous one by a catalytic potency factor $f(\theta)$ via $\Delta \Gibbs_\mathrm{het}= f(\theta) \Delta \Gibbs_\mathrm{hom}$, and $f(\theta)$ depends only on the contact angle \cite{Turnbull:1950aa}.
The latter function is considered as function of interfacial properties only.
The above results suggest that the heterogeneous nucleation barrier and the contact angles are influenced additionally by elasticity, which favours independently of the interfacial energies a wetting angle of $\theta\approx 90^\circ$.

If corrugations of the outer surface are permitted (black curve in Fig.~\ref{fig3}), the elastic energy reduces to zero for a heterogeneous nucleus which is sitting spherically on the substrate.
Trends for such morphologies have recently been confirmed experimentally in Nb-H film using scanning tunnelling microscopy \cite{Northemann:2011aa}.
In this case we obtain a complete elastic relaxation ($\gamma=1$) and consequently a shift from the coherent bulk solubility limit in Fig.~\ref{fig7_5f} to the stress free surface solubility limit.

More generally, the reduction of the elastic energy leads to the replacement of $\Gelb$ by $\Gel$ in the entire derivation in section \ref{coherent::section}.
As a result, we have to replace
\begin{equation}
\Delta\Gelb \to (1-\gamma)\Delta\Gelb
\end{equation}
in Eqs.~(\ref{sol::eq1}) and (\ref{TotalEffectiveFormationEnergy}), using the expression (\ref{TotalEffectiveFormationEnerg2}) for $\Delta\Gelb$.
Hence in the low temperature regime the {\em coherent surface binodal} is given by
\begin{equation} \label{MasterFormula}
\conc \simeq \xn \exp\left(- \frac{\Delta \Gibbs + (1-\gamma)\Gelb}{\kB T} \right),
\end{equation}
which is the central result of the present paper.
The geometrical effect of the proximity of a nucleus to a surface, including shape dependencies, is fully contained in the dimensionless number $\gamma$, which is positive near a free surface.
For values $0<\gamma<1$ we arrive at a solubility limit which is in between the bulk coherent binodal and the solubility limit without consideration of elasticity.

\begin{figure}
\begin{center}
\includegraphics[width=8.5cm]{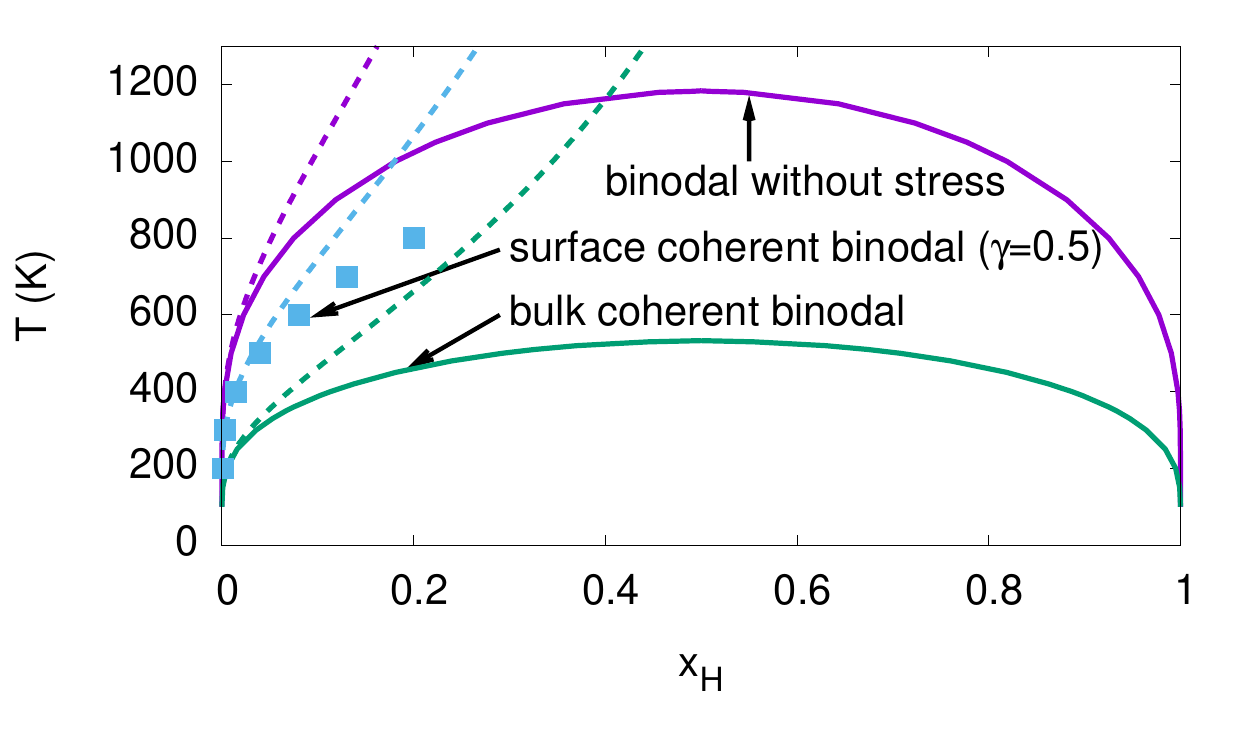}
\caption{(Color online) The same as Fig.~\ref{fig11}, where we have added the asymptotic expression for the solubility limit near a free surface according to Eq.~(\ref{MasterFormula}) as dashed blue line. 
This prediction is in agreement with the Cahn-Hilliard simulation results (blue squares).
}
\label{figsurfacecahnhilliardcomparison}
\end{center}
\end{figure}

\begin{figure}
\begin{center}
\includegraphics[width=8.5cm]{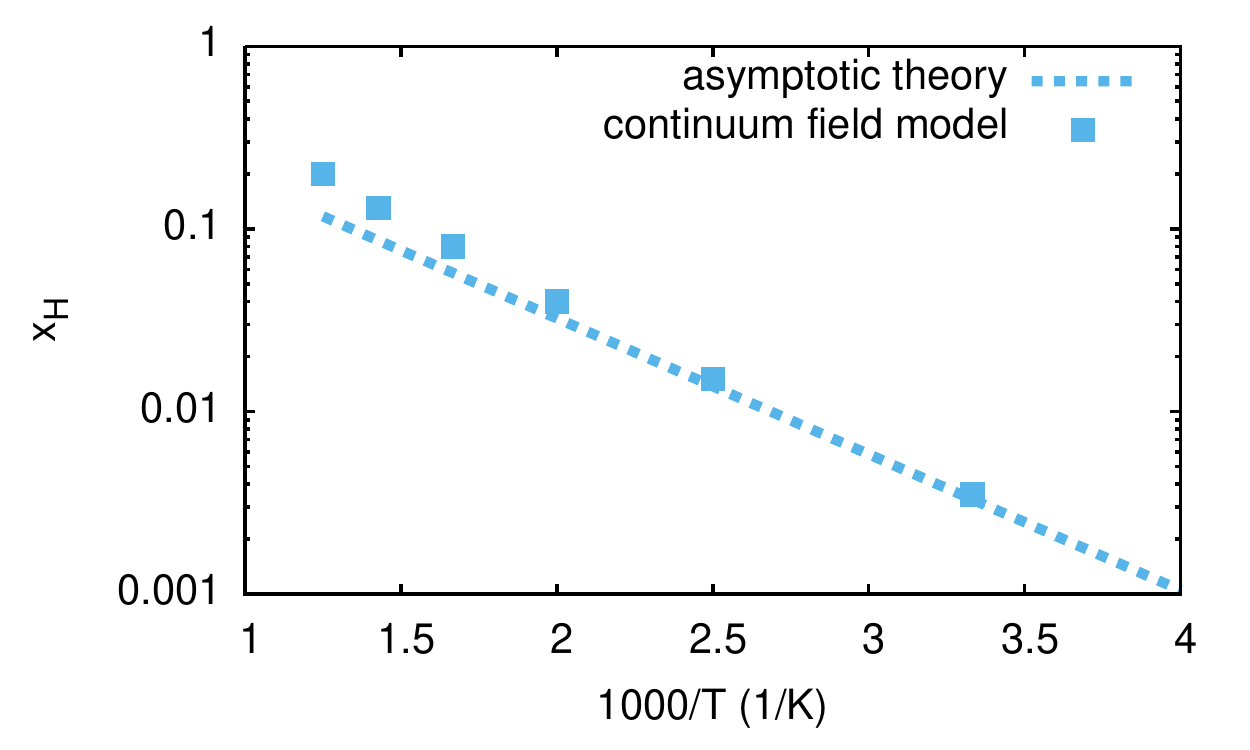}
\caption{(Color online) The same as Fig.~\ref{figsurfacecahnhilliardcomparison} with only the surface solubility limit, as predicted from the asymptotic theory (dashed line) and the continuum model (squares).
}
\label{figsurfacecahnhilliardcomparisonlog}
\end{center}
\end{figure}

We can compare this analytical prediction with the numerically solved continuum model with a free surface, which has been done in Section \ref{CahnHilliardSurface::section}.
There we have found that the hydride prefers to form an almost spherical cap at the free surface, see Fig.~\ref{fig6}.
According to the blue curve in Fig.~\ref{fig3} this corresponds to a value $\gamma\approx 0.5$ at $h/R\approx 1.2$.
From this we obtain Fig.~\ref{figsurfacecahnhilliardcomparison}, which shows the agreement between the theoretical expression (\ref{MasterFormula}) and the numerical simulations for low temperatures.
This agreement is further confirmed in an Arrhenius plot for the low concentration regime, see Fig.~\ref{figsurfacecahnhilliardcomparisonlog}.
In the limit $T\to 0$ the numerical results converge to the analytical prediction.

We note that Eq.~(\ref{MasterFormula}) is a highly useful formula to include long-ranged elastic coherency effects e.g.~in {\em ab initio} calculations of phase diagrams. 
It allows to include mechanical effects, which otherwise would be impossible to describe in state-of-the art quantum mechanical simulations due to the restricted number of atoms.
Applications of this approach will be demonstrated in Section \ref{abinitio::section}.

We also point out that the elastic deformation state can be fairly nontrivial, and in particular both phases are typically deformed.
In contrast, an approximation, where one phase accommodates to the lattice constant of the other, overestimates the elastic energy.

Whereas the precise prediction of the solubility limits requires knowledge of $\Delta\Gibbs$ with an accuracy in the meV range, this quantity drops out when we calculate the ratio of the solubility limits near a surface, $\conc^\mathrm{surface}$, and in the bulk, $\conc^\mathrm{bulk}$.
The first quantity follows from Eq.~(\ref{MasterFormula}), the latter can be obtained from the same expression with $\gamma=0$.
Their ratio defines the generic near-surface phase diagram shift by the {\em solubility modification factor} 
\begin{equation} \label{solmodfac}
s := \frac{\conc^\mathrm{surface}}{\conc^\mathrm{bulk}} = \exp\left( \frac{\gamma\Delta \Gelb}{\kB T}\right),
\end{equation}
which is independent of details of the phase diagram, i.e.~the value of $\Delta\Gibbs$.
A numerical evaluation of this solubility modification factor will be done in Section \ref{abinitio::section}, where we use in particular {\em ab initio} determined parameters to predict the solubility limit change near free surfaces and interfaces in comparison to the bulk.

\subsection{Local equilibrium conditions}

The above expressions for the solubility limits rely on the assumption that the concentrations are homogeneous in the individual phases in the low temperature regime.
We have seen before in the equilibrium results of the Cahn-Hilliard model that this is indeed the case for $T\to 0$.
Therefore, in the derivation of the asymptotic expressions of the solubility limits it is legitimate to use these spatially constant concentrations $\conca^0(=0)$ and $\concb^0$.
Here we investigate in more detail the concentration gradients, which exist inside the phases, see also Fig.~\ref{fig8bbbbbb}.

In this section we derive local equilibrium conditions at the phase boundary between the lattice gas phase ($\alpha$) and the hydride ($\beta$).
Two phase coexistence conditions, the equality of (generalized) hydrogen chemical potentials and grand potentials, are needed to uniquely determine $\conca$ and $\concb$.
In Section \ref{analytical::section} we have derived them for bulk and near surface regions using a minimisation of the {\em total} Gibbs energy.
This evaluation requires in particular the knowledge of the integrated elastic energy, which in general cannot be determined analytically in the presence of surfaces and interfaces.
Here we show that these conditions can be derived without solving the global elastic problem explicitly.
We need to determine the fields only locally, which can be done analytically.
The results in this section are therefore independent of the previous arguments, and in particular they do not require the numerical solution of the equilibrium fields via the continuum approach, which was developed in  Section \ref{continuum::section}.
In the following, all quantities are evaluated directly either on the $\alpha$ or $\beta$ side of a coherent interface between these phases.
For brevity of notation, we do not explicitly write this for the individual quantities in the following.
As a result, we find that the equilibrium concentrations right at the interface are given by the bulk coherent phase diagram, irrespective of potential concentration gradients inside the phases for phase separation near free surfaces.

\begin{figure}
\begin{center}
\includegraphics[trim=0cm 9cm 0cm 0cm, clip=true, width=8.5cm]{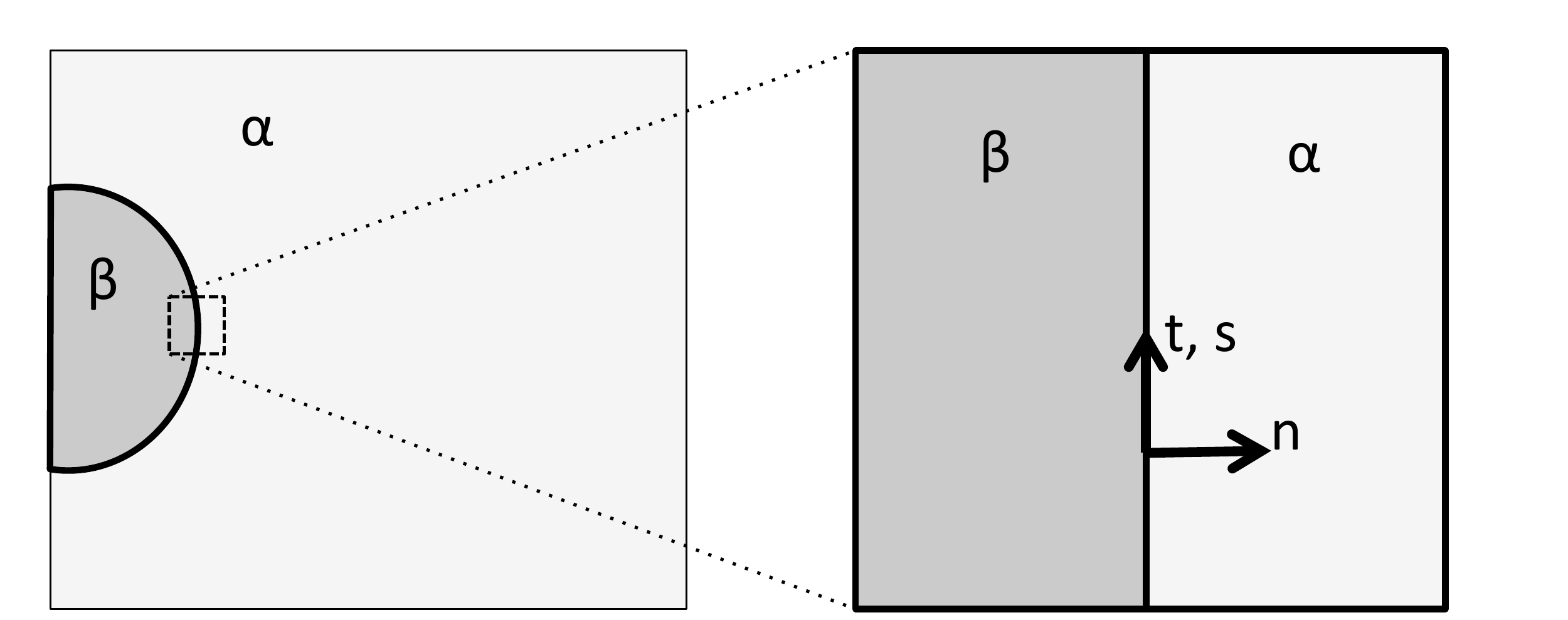}
\caption{Sketch of the geometry for the local equilibrium consideration.}
\label{fig17}
\end{center}
\end{figure}

The geometry is sketched (in a two-dimensional projection) in Fig.~\ref{fig17}.
We consider the energetics in a small subsystem, which contains part of the interface, as magnified in the right panel.
We use $n$ as normal and $t$ and $s$ as local tangential directions to the interface.
Since in the following we will work with the (local) Helmholtz free energy as thermodynamic potential, we assume that $T$, as well as the average concentration $\conc$, are fixed in the considered infinitesimally small sub-system.
In the local picture, the phase concentrations $\conca$ and $\concb$, as well as all strain components in each phase are spatially constant, as we choose the sub-system as in the right panel of Fig.~\ref{fig17} to be sufficiently small.
Moreover, the local volume is kept fixed, which means that average strains $\bar{\epsilon}_{ij}$ are prescribed.
In the end, these values do not enter into the equilibrium conditions. 

The strain components, which involve a tangential direction, have to obey the no-slip coherency constraint
\begin{eqnarray}
\epsilon_{tt}^{(\alpha)} &=& \epsilon_{tt}^{(\beta)} = \bar{\epsilon}_{tt}=const, \\
\epsilon_{ss}^{(\alpha)} &=& \epsilon_{ss}^{(\beta)} = \bar{\epsilon}_{ss}=const, \\
\epsilon_{st}^{(\alpha)} &=& \epsilon_{st}^{(\beta)} = \bar{\epsilon}_{st}=const.
\end{eqnarray}
For the other strain components we invoke the stress continuity at the interface and use the fact that the elastic constants are equal in both phases.
Therefore
\begin{eqnarray}
\sigma_{nt}^{(\alpha)} = \sigma_{nt}^{(\beta)} & \Rightarrow & \epsilon_{nt}^{(\alpha)} = \epsilon_{nt}^{(\beta)}= \bar{\epsilon}_{nt} = const, \\
\sigma_{ns}^{(\alpha)} = \sigma_{ns}^{(\beta)} & \Rightarrow & \epsilon_{ns}^{(\alpha)} = \epsilon_{ns}^{(\beta)} =  \bar{\epsilon}_{ns}=const, \\
\sigma_{nn}^{(\alpha)} = \sigma_{nn}^{(\beta)} & \Rightarrow & \epsilon_{nn}^{(\beta)} = \epsilon_{nn}^{(\alpha)} + \frac{1+\nu}{1-\nu} \chi (\concb-\conca).
\end{eqnarray}
Furthermore, with $v$ being the local volume fraction of the $\beta$ phase we have from the volume constraint
\begin{equation} \label{eq18}
\epsilon_{nn}^{(\beta)} v + \epsilon_{nn}^{(\alpha)} (1-v) = \bar{\epsilon}_{nn}=const.
\end{equation}
Hence the latter two relations determine the normal strain in both phases uniquely.
In particular,
\begin{equation}
\epsilon_{nn}^{(\alpha)} = \bar{\epsilon}_{nn} - v \frac{1+\nu}{1-\nu} \chi (\concb - \conca).
\end{equation}
The strain component $\epsilon_{nn}^{(\beta)}$ is determined via relation (\ref{eq18}).
The conservation law for the concentration reads
\begin{equation}
v \concb + (1-v)\conca = \conc = const.
\end{equation}

We split the free energy densities into contributions which do not depend on strain and an elastic term, $\hat{f} = \hat{f}_0(\conc, T) + \hat{f}_{el}(\conc, \epsilon_{ij})$.
In the continuum model the first term corresponds therefore to $\hat{g}_o+\hat{g}_c$.
For the elastic term we use isotropic linear elasticity, whereas we do not further specify $\hat{f}_0$, which  can include the case of separate functions for the two phases.
The elastic free energy density therefore reads in our local coordinate system
\begin{eqnarray}
\hat{f}_{el}(\conc, \epsilon_{ij}) &=& \mu \Big[ (\epsilon_{nn}-\chi \conc)^2 +  (\epsilon_{tt}-\chi \conc)^2 \nonumber \\
&&+  (\epsilon_{ss}-\chi \conc)^2 + 2\epsilon_{nt}^2 + 2 \epsilon_{ns}^2 + 2\epsilon_{st}^2 \Big] + \nonumber \\
&&+ \lambda (\epsilon_{nn} + \epsilon_{tt} + \epsilon_{ss} -3 \chi \conc)^2
\end{eqnarray}
for each phase (using the appropriate indices $\alpha$ and $\beta$).
The energy density of the entire sub-system becomes then
\begin{eqnarray}
\hat{f}(\conca, v) &=& (1-v)\left[ \hat{f}_0(\conca, T) + \hat{f}_{el}(\conca, \epsilon_{ij}^{(\alpha)})\right] \nonumber \\
&&+ v \left[ \hat{f}_0(\concb, T) + \hat{f}_{el}(\concb, \epsilon_{ij}^{(\beta)})\right].
\end{eqnarray}
By the above conservation laws we can eliminate the strains in the individual phases as well as the concentration in the $\beta$ phase, and consequently the only undetermined parameters are $\conca$ and $v$.
Local equilibrium demands minimisation with respect to these parameters.
From the minimisation with respect to $\conca$ (at fixed volume fraction $v$) we obtain the continuity of the chemical potentials,
\begin{equation}
\hat{\mu}_0(\conca, T) - \hat{\mu}_0(\concb, T) + 2\frac{E\chi^2}{1-\nu}(\conca-\concb)=0,
\end{equation}
with
\begin{equation}
\hat{\mu}_0(\conc, T) = \left(\frac{\partial \hat{f}_0}{\conc}\right)_T.
\end{equation}
This condition motivates to {\em define} a chemical potential for each phase according to
\begin{equation}
\hat{\bar{\mu}}(\conc, T) = \hat{\mu}_0(\conc, T) + 2\frac{E\chi^2}{1-\nu} \conc,
\end{equation}
such that the equilibrium condition reads
\begin{equation} \label{eq19}
\hat{\bar{\mu}}_\alpha(\conca, T) = \hat{\bar{\mu}}_\beta(\concb, T), 
\end{equation}
in agreement with Cahn's definition, cf.~Eqs.~(\ref{CahnAuxiliaryPotential})-(\ref{CahnsBothConditions}).
We note that this condition is the same as we naturally get from the global thermodynamic description, but here we used a purely local analysis.

From the second minimisation condition
\begin{equation}
\left( \frac{\partial \hat{f}}{\partial v}\right)_{\conca}=0
\end{equation}
we get the local equality of grand potential densities,
\begin{equation} \label{eq20}
\hat{\bar{\omega}}(\conca, T) = \hat{\bar{\omega}}(\concb, T),
\end{equation}
being defined as
\begin{equation}
\hat{\bar{\omega}}(\conc, T) = \hat{f}_0(\conc, T) - \conc \hat{\mu}_0(\conc, T) - \frac{E\chi^2\conc^2}{1-\nu}.
\end{equation}
These conditions are the same as the ones which we found before from the global energy approach.
We point out that we do not have to specify the external boundary conditions to arrive at these results, hence they equally hold for the infinite (or periodic) system as well as the finite system with free (or confined) surfaces.
Hence the concentrations at the $\alpha-\beta$ interface are exactly those determined by the coherent phase diagram.
We have previously found this result in the continuum model (Figs.~\ref{fig8bbbbbb}), and this observation is explained by the present analysis.

Returning to an ensemble with given stress instead of strain, the above potentials $\bar{\mu}$ and $\bar{\omega}$ (written here per particle, not per volume) can be obtained from a Gibbs energy per matrix atom
\begin{equation}
\bar{g} = g_0(x, T) + \frac{E\chi^2\conc^2 \Omega_0}{N_0(1-\nu)},
\end{equation}
where $g_0$ is the stress independent part of the Gibbs energy.
From that we get
\begin{equation}
\bar{\mu}_\mathrm{H} = \frac{\partial \bar{g}}{\partial \conc}, \qquad \bar{\omega} = \bar{g} - \bar{\mu}_\mathrm{H} \conc
\end{equation}
as determining quantities for a local common tangent construction.
We note that the constant background strain $\bar{\epsilon}_{ij}$ does not appear in the phase equilibrium conditions. 

\begin{figure}
\begin{center}
\includegraphics[width=8.5cm]{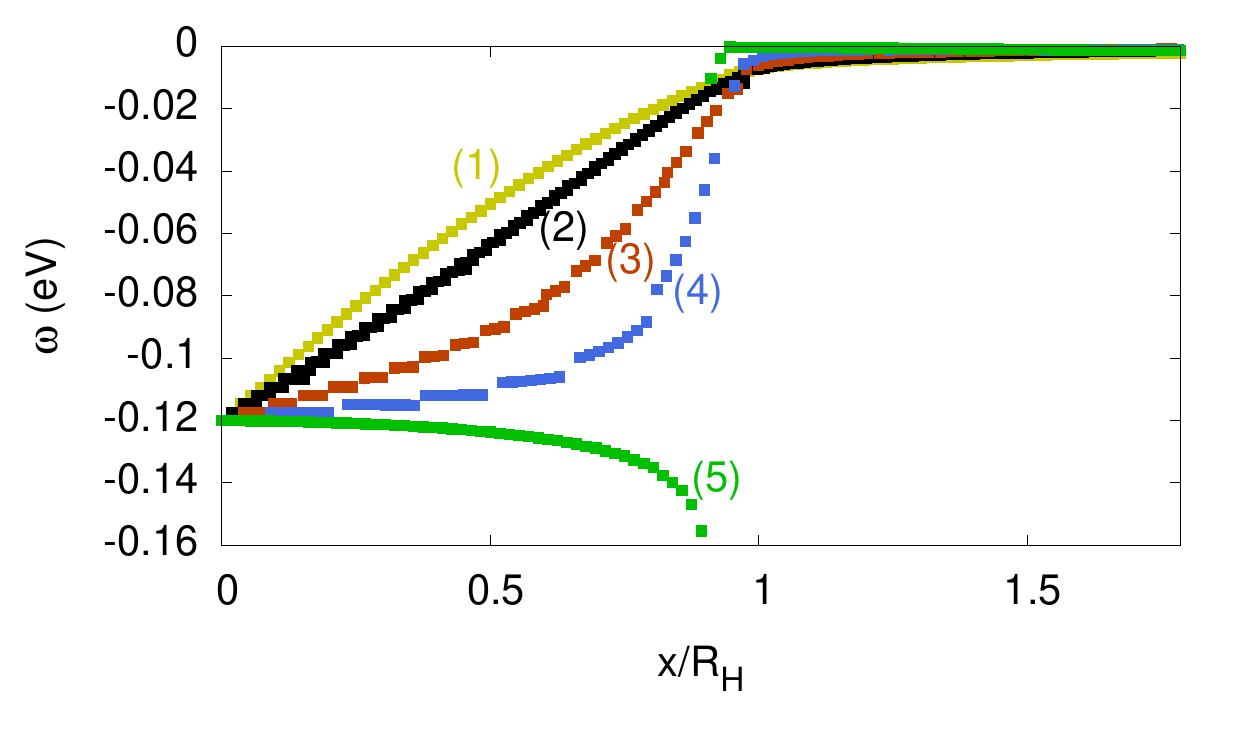}
\caption{(Color online) Grand potential $\bar{\omega} = \hat{\bar{\omega}} \omz/\Nz$ along different paths (shown in Fig.~\ref{fig16}) through the interface at $T=500\,\mathrm{K}$.
The position is normalized to the hydride radius $\Rhydride$.
At the interface, $\bar{\omega}$ exhibits a kink but is continuous, as predicted by condition (\ref{eq20}).
Only for the path marked by number (5), which is directly along the free surface of the sample, the condition (\ref{eq20}) is not fulfilled as the analysis does not apply in that case.
}
\label{fig18}
\end{center}
\end{figure}

\begin{figure}
\begin{center}
\includegraphics[width=6cm]{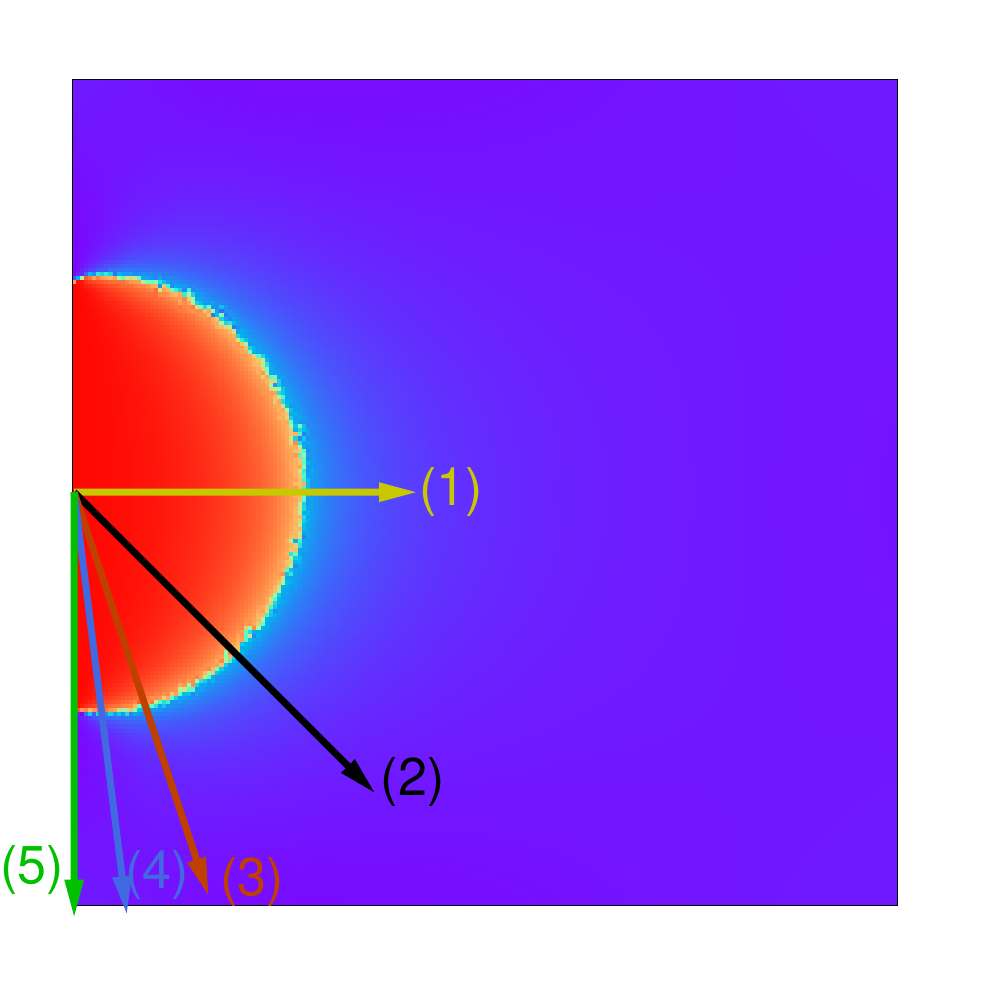}
\caption{(Color online) Concentration profiles along certain directions. 
The different paths mark the directions along which the grand potential densities $\hat{\bar{\omega}}$ are plotted in Fig.~\ref{fig18}.
The equilibrium nucleus at a free surface at the left boundary is found from the continuum simulations at $T=500\,\mathrm{K}$.
Red regions correspond to the hydride, blue to the lattice gas phase.
}
\label{fig16}
\end{center}
\end{figure}

Numerically, the continuity of the grand potential $\hat{\bar{\omega}}$ through the interface is demonstrated as one-dimensional cuts in Fig.~\ref{fig18}, along the different paths defined in Fig.~\ref{fig16}.
Indeed, in the numerical simulations based on the continuum model we find a continuous value of $\hat{\bar{\omega}}$ at the interface positions, which confirms the analytical predictions.
Only directly at the free surface (i.e.~for the green path marked by (5) in Fig.~\ref{fig16}) the grand potential becomes singular and exhibits a jump, because the local analysis does not apply due to the neglect of the free boundary conditions.

\subsection{Surface-induced precipitate interaction}

One of the well known consequences of the elastic isotropy, that we assume in this article, is that in an infinite or periodic system no interaction between precipitates exists.
As mentioned before, this non-interaction is due to the fact that the elastic energy (\ref{BitterCrumEq}) depends only on the total volume fraction of the precipitates and not on their shape or fragmentation.
Hence, separation of two inclusions does not change the energy.
We mention that deviations from the isotropy assumptions lead to mutual interactions.
Here we point out that even in the absence of anisotropy the proximity of surfaces can induce an attractive or repulsive interaction.

This effect also raises another point, namely whether the appearance of a single nucleus is --- from point of view of the elastic energy --- the most favorable configuration, or whether, due to the effective surface induced nucleus-nucleus interaction, a breakup into smaller precipitates may lead to a lower energy, which might affect the solubility limit.

We have performed a series of elastic energy calculations using finite element methods, to understand the energetics of multi-precipitate situations.
We simulate here two-dimensional situations, as the computational effort is lower.
We expect that the results will only change slightly in three-dimensional situations.
More precisely, in the used plane strain setup the system is translation invariant in the third direction, hence the precipitates are cylindrical instead of spherical.
The concentration $\conc$ is assumed to be spatially homogeneous in each phase.

\begin{figure}
\begin{center}
\includegraphics[width=7cm]{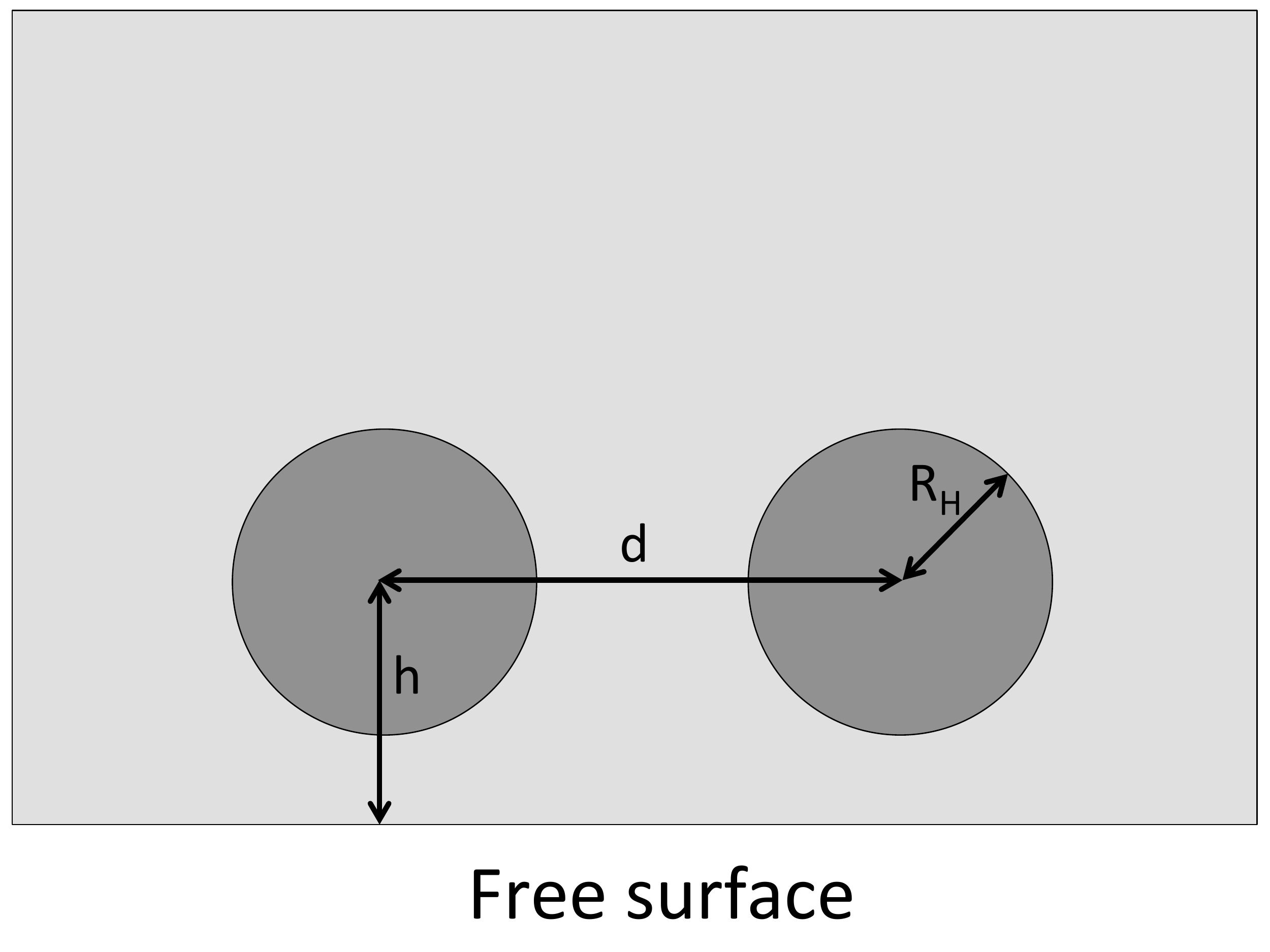}
\caption{Two spherical inclusions near a free surface.}
\label{fig26}
\end{center}
\end{figure}

\begin{figure}
\begin{center}
\includegraphics[width=8.5cm]{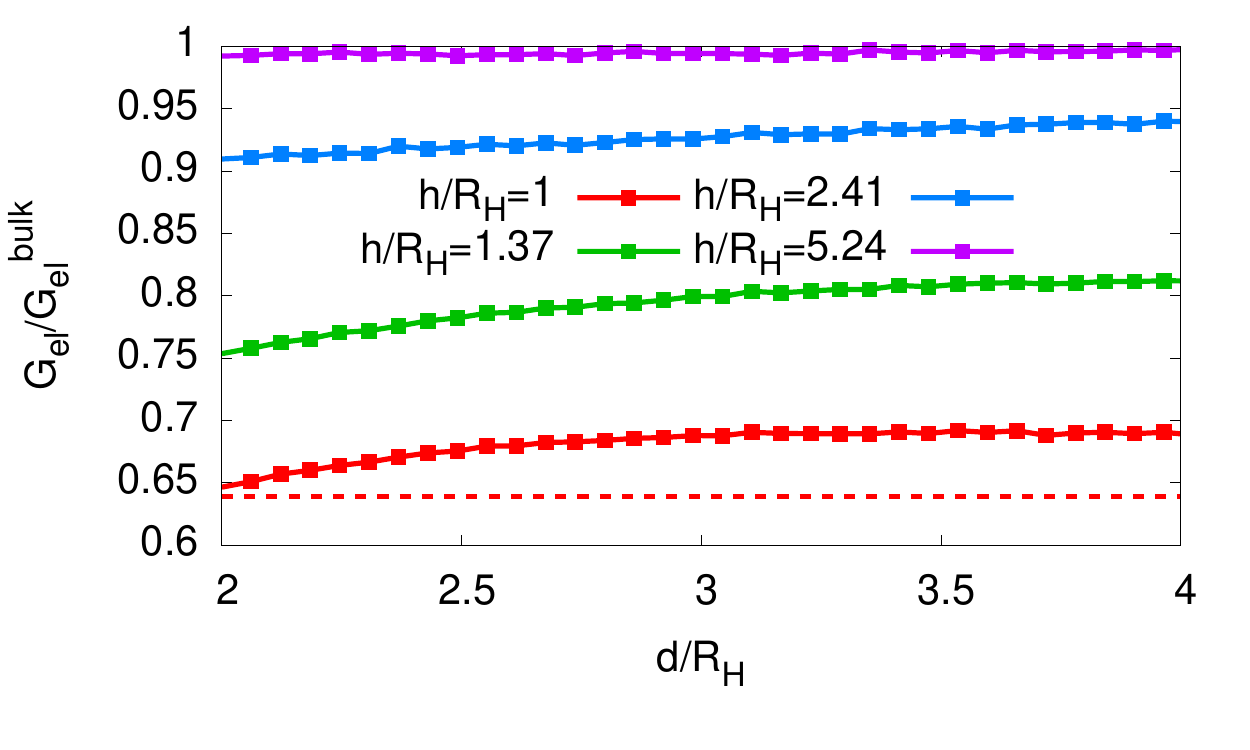}
\caption{(Color online) Energy of two spherical inclusions near a free surface.
$\Gel^\mathrm{bulk}$ is the energy for the bulk state (where the shape dependence disappears), analogous to expression (\ref{BitterCrumEq}).
For comparison, the elastic energy of a single surface touching sphere, which has the same area as the two separate spheres together, is shown as horizontal, dashed red curve.
}
\label{fig27}
\end{center}
\end{figure}

First, two spheres of radius $\Rhydride$ are placed at a mutual distance $d$ and at a distance $h$ from a free surface, see Fig.~\ref{fig26}.
The calculated elastic energy, relative to the energy of two isolated spheres in an infinite medium, is plotted in Fig.~\ref{fig27} as function of the separation $d/\Rhydride$ for different distances $h/\Rhydride$.
If the spheres are far away from the free surface, $h/\Rhydride\gg 1$, they do not experience a mutual interaction, in agreement with the Bitter-Crum theorem.
However, if they are closer to the free surface, an effective attraction appears.
First we note that this interaction is weak compared to the distance dependence of the interaction with the free surface.
The energy is lowest for the case $h/R=1$, when the spheres touch the surface.
We have compared the two-sphere configuration with that of a single sphere, which also touches the surface and which is shown for comparison as horizontal dashed red line in Fig.~\ref{fig27}.
We note that the volume of the single sphere is the same as the sum of the volumes of the two smaller spheres.
We see that the energy of the single sphere is lower that for two separate spheres.
Hence we conclude that it is energetically favorable to have a single nucleus instead of a breakup into smaller spheres.

\begin{figure}
\begin{center}
\includegraphics[trim=0cm 1cm 0cm 0cm, clip=true, width=7cm]{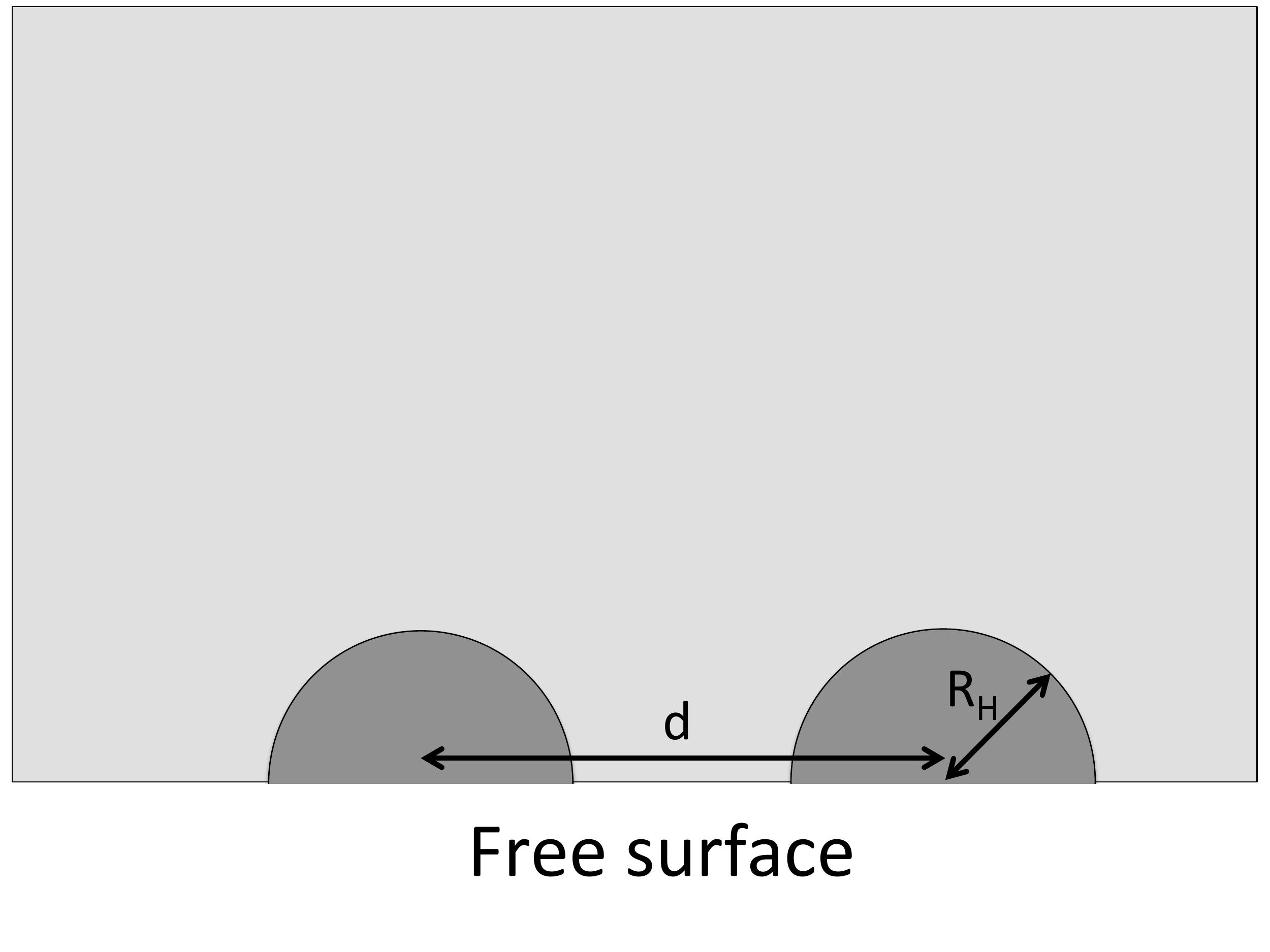}
\caption{Two spherical caps near a free surface.}
\label{fig28}
\end{center}
\end{figure}

\begin{figure}
\begin{center}
\includegraphics[width=8.5cm]{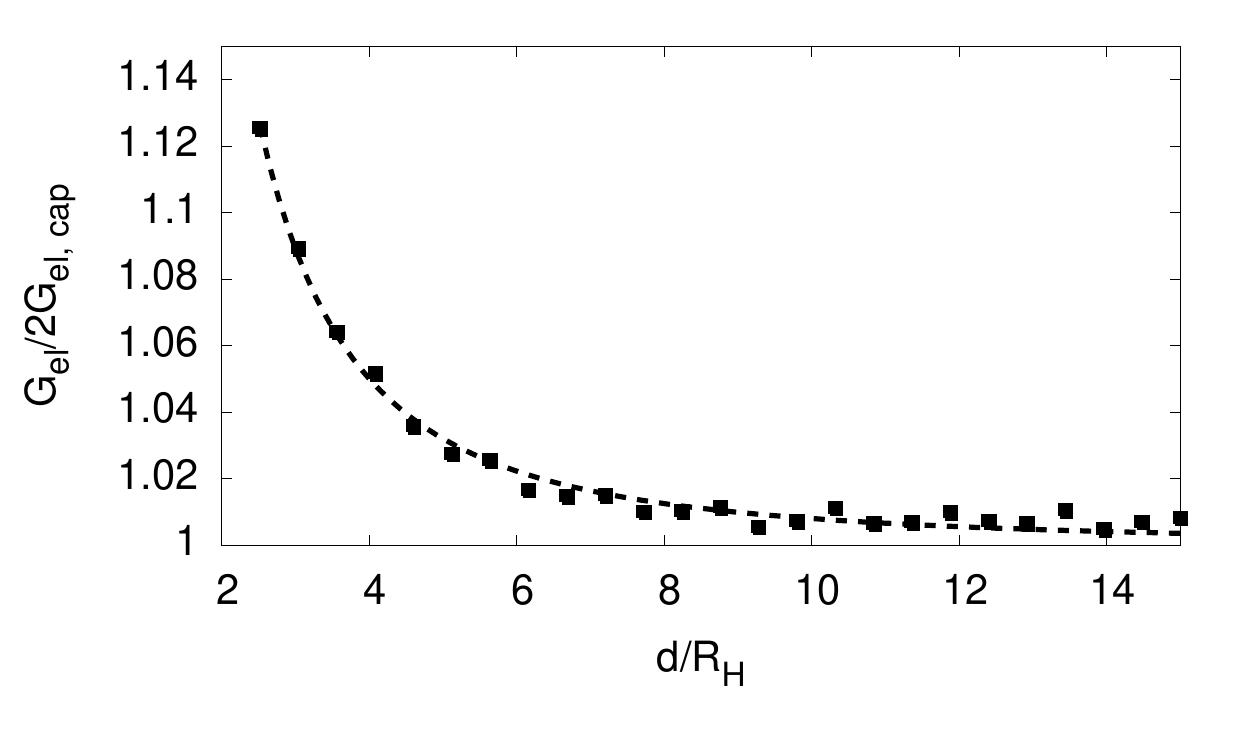}
\caption{Interaction of the caps: total elastic energy as function of the cap separation $d/\Rhydride$, normalized to twice the elastic energy of a single cap at a free surface. 
The dashed curve is a fit of the interaction energy as $\Gel - 2 {\cal G}_\mathrm{el, cap} \sim 1/d^2$, hence $k=2$ in Eq.~(\ref{distanceScaling}).}
\label{fig29}
\end{center}
\end{figure}

The situation is different if we use a spherical cap instead of full spheres (see Fig.~\ref{fig28}).
Fig.~\ref{fig29} shows the interaction energy, normalized to twice the elastic energy of a single cap at a free surface.
In contrast to the spherical inclusions the interaction is repulsive.

\begin{figure}
\begin{center}
\includegraphics[width=8.5cm]{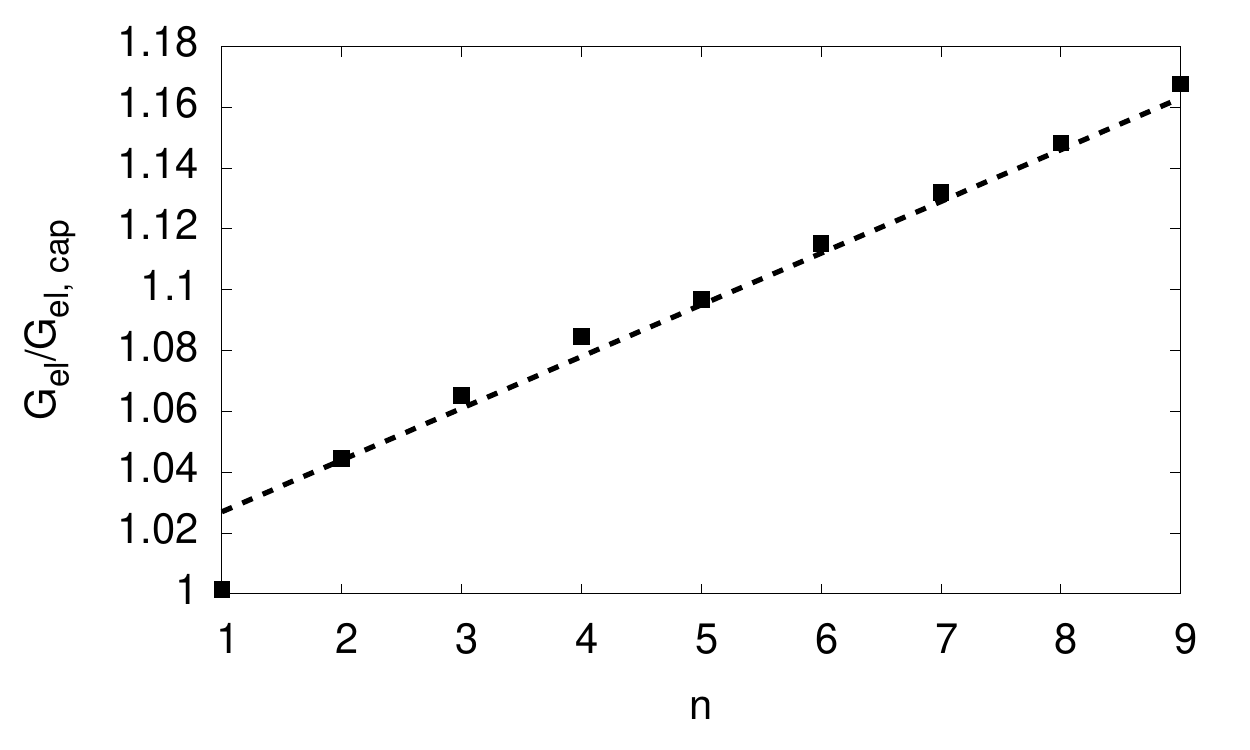}
\caption{Total energy as number of caps in a system with periodic boundary conditions in lateral direction. 
${\cal G}_\mathrm{el, cap}$ is the energy of a single cap ($n=1$). 
The dashed curve is a linear fit, which agrees with $k=2$ in Eq.~(\ref{interactionscalingmanycaps}).}
\label{fig30}
\end{center}
\end{figure}

This result may suggest that a splitting of a single cap into two (or more) is energetically favorable. 
This is however an improper conclusion.
Two competing effects play a role here.
First, due to the volume reduction of each precipitate the surface induced interaction gets weaker.
Second, due to the inter-precipitate distance reduction the strength of the interaction is increased.
To study this competition numerically we split the single cap into $n$ smaller ones, which in sum have the same volume as the original one. 
Fig.~\ref{fig30} shows the energy as function of the number of caps $n$, which are equally spaced due to the mutual repulsion.
%
We see that the energy increases linearly with the number of caps. 
A splitting into several caps is therefore energetically unfavourable according to elasticity.

We can also understand this behavior analytically.
The elastic interaction energy $\Gibbs_\mathrm{el, int} = \Gel - 2 {\cal G}_\mathrm{el, cap}$ per unit length $L$ in out-of-plane direction for two nearby caps with distance $d$ scales as
\begin{equation} \label{distanceScaling}
\Gibbs_\mathrm{el, int}/L \sim E\epsilon_0^2 \frac{\Rhydride^{k+2}}{d^k}.
\end{equation}
The elastic constant $E$ has to appear in order to get the dimension energy, and the expression has to be quadratic in the misfit $\epsilon_0=\chi(x_\beta-x_\alpha)$, because the elastic energy has to be positive, irrespective which phase has the larger lattice constant.
From the spatial integration of the elastic Green's function for the calculation of the total energy we expect a dependence on the cap's area, hence $k=2$.
Altogether, the expression has to have the dimension energy per length.
Hence a scaling with $d^{-k}$ has to appear.
This value of the exponent is indeed found in Fig.~\ref{fig29}, and the above considerations explain the $1/d^2$ interaction.

With the total area of the caps being $A$ we have $n\Rhydride^2 \sim A$ in our two dimensional setup.
Hence
\begin{equation}
\Rhydride \sim \left( \frac{A}{n} \right)^{1/2}.
\end{equation}
The lateral distance between neighbouring caps is 
\begin{equation}
d \sim R/n,
\end{equation}
where we assume that the horizontal system length is $R$.
The interaction from the nearest neighbours matters most for the total elastic energy.
Hence, apart from the purely volume-dependent part without the interaction, we get
\begin{equation}
\Delta \Gel \sim n \Gibbs_\mathrm{el, int}.
\end{equation}
This equation gives for the dependence on the number of caps $n$
\begin{equation} \label{interactionscalingmanycaps}
\Delta \Gel/L  \sim n^{k/2} E \epsilon_0^2 A^{1+k/2} R^{-k} \sim n^{k/2}.
\end{equation}
For $k=2$ -- as argued and confirmed above -- we therefore obtain a scaling $\Delta \Gel  \sim n$, which we observe in the numerical results in Fig.~\ref{fig30}.

We can therefore conclude that the formation of isolated spherical caps at free surfaces is the energetically most favorable configuration.



\subsection{Spinodal decomposition}
\label{section::spinodaldecomposition}

Spinodal decomposition is not in the focus of the present work but briefly discussed for completeness.
As obvious from Fig.~\ref{fig2}, typically rather high hydrogen concentrations are required to trigger this nucleation-free phase separation mechanism.
Hence we believe that it is not relevant at least for many metal-hydrogen systems which experience hydrogen embrittlement at already much lower hydrogen concentrations.
To obtain low temperature predictions of the spinodal curves, we perform an asymptotic analysis, to get expressions analogous to the solubility limits (\ref{sol::eq1}) with and without bulk elastic effects.

\begin{figure}
\begin{center}
\includegraphics[width=8.5cm]{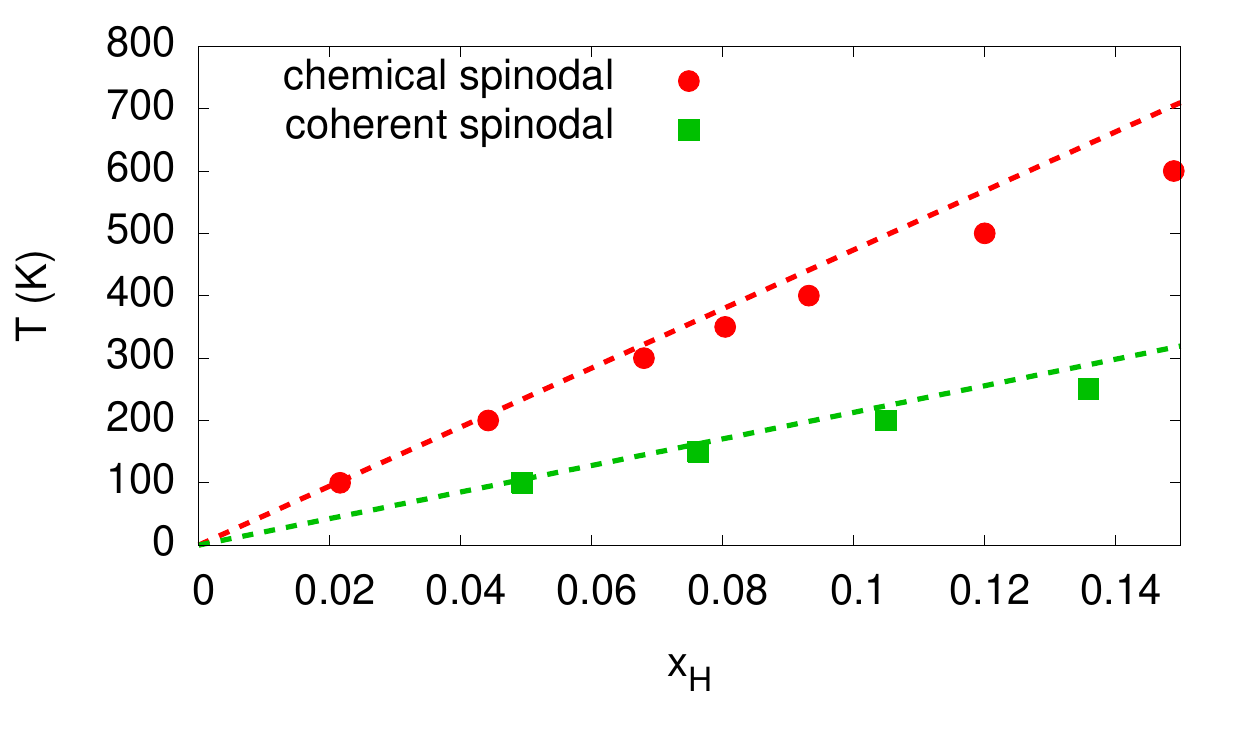}
\caption{(Color online) Asymptotic limit of the chemical (red points) and coherent (green squares) spinodal from Fig.~\ref{fig2}, and the convergence to the analytical solution (dashed curves).}
\label{fig20}
\end{center}
\end{figure}

According to the seminal work by Cahn and Hilliard\cite{Cahn:1962aa,Cahn:1958aa} the chemical spinodal is determined by the condition $[\mu_c(\conc)+\mu_o(\conc)]'=0$.
This condition reads explicitly for the Gibbs energy function defined in Section \ref{continuum::section}
\begin{equation}
\kB T \left(\frac{1}{\conc} + \frac{1}{1-\conc} \right ) - 2\alpha = 0.
\end{equation}
Asymptotically for $T\to 0$ and for the low concentration branch
\begin{equation}
\conc \simeq \frac{\kB T}{2\alpha}.
\end{equation}
The convergence of the full numerical data from Fig.~\ref{fig2} to the asymptotic solution is shown in Fig.~\ref{fig20}.

The coherent spinodal is determined by the condition\cite{Cahn:1962aa}
\begin{equation}
\mu_c'(\conc) + \mu_o'(\conc) + 2\frac{E\chi^2 \Omega_0}{(1-\nu)N_0} = 0,
\end{equation}
which is the same as $\bar{\mu}'(x)=0$.
We obtain
\begin{equation}
\kB T \left(\frac{1}{\conc} + \frac{1}{1-\conc} \right ) - 2\alpha + 2\frac{E\chi^2 \Omega_0}{(1-\nu)N_0} = 0.
\end{equation}
For the entire concentration and temperature regime the solutions of this equation are shown in Fig.~\ref{fig2}.
In the low temperature limit for the low concentration branch we get
\begin{equation}
\conc \simeq \kB T \left( 2\alpha - 2\frac{E\chi^2\Omega_0}{(1-\nu) N_0} \right)^{-1}.
\end{equation}
Again, this prediction is confirmed in Fig.~\ref{fig20}.

Surface spinodal modes have been discussed \cite{Tang:2012aa}, and they can appear in between the coherent and the surface spinodal in the vicinity of a free surface, which is comparable to our findings concerning the solubility limit near free surfaces.

\section{Atomistic perspective and applications}
\label{abinitio::section}

In this section we inspect the above results, which were obtained from a continuum description of the thermodynamic system, on a discrete atomistic level.
We first compare in Section \ref{conversioncontinuumdiscrete::subsection} the predicted bulk solubility limit without elasticity to descriptions used e.g.~in {\em ab initio} descriptions and confirm the consistency of the approaches.
The continuum results allow to incorporate the elastic effects relevant for the bulk coherent solubility limit to the atomistic description.
This continuum approach allows to take into account long-ranged effects, which would otherwise be difficult to consider in {\em ab initio} simulations alone, as the reachable system sizes are too small.
In the following subsections, the general results for solubility limits in the bulk and near surfaces are applied to Ni-H, Fe-H and Nb-H, using {\em ab initio} or experimental data.
We find that for coherent nucleation the solubility limits near surfaces can be drastically different from the bulk.

\subsection{Linking continuum and atomistic descriptions}
\label{conversioncontinuumdiscrete::subsection}

To link the continuum description to an atomistic and {\em ab initio} perspective we recapitulate the central expression (\ref{sol::eq1}) for the bulk solubility limit of the $\alpha$ phase in the low temperature regime,
\begin{equation} \label{atomistic::eq1}
\conc \simeq \xn \exp\left(- \frac{\Delta \bar{\Gibbs}}{\kB T} \right),
\end{equation}
and remind that $x_0$ is the ratio of available interstitial sites relative to the number of metal atoms.
The derived expression for $\Delta \bar{\Gibbs}$ reads
\begin{equation}
\Delta \bar{\Gibbs} := [-\bar{g}_\beta^0(x_\beta^0)+\bar{g}_\alpha^0(0)]/x_\beta^0+{\overline{g}_\alpha^0}'(0),
\end{equation}
see Eqs.~(\ref{sol::eq1}) and (\ref{sol::eq2}).
Here $\bar{g}$ are the (modified) Gibbs energies per metal atom for the different phases as function of the concentration of the interstitial species $x = N_\mathrm{H}/N_\mathrm{M}$ of the considered phase $\alpha$ or $\beta$.
The Gibbs energies are evaluated at $T=0$, hence they are equal to enthalpies.
Notice that they are also evaluated for zero external pressure, therefore they also coincide with total energies, for which we use the letter $\Energy$.
This is important as the long ranged elastic effects are taken into account separately and lead to the replacement $g\to\bar{g}$.
The (Gibbs) energies $g$ are evaluated for the zero pressure equilibrium situation, as otherwise we would double-count the elastic energy.
For an extended discussion of this issue we refer to Appendix \ref{micromesoelastic::appendix}.

Let us first consider the case without elastic effects, hence $\bar{g}\to g$ are the conventional Gibbs energies per metal atom, and similarly $\bar{\Gibbs}\to\Gibbs$.
In an atomistic simulation we set up sufficiently large supercells which consist of a number $\NM$ of metal and $\NH$ interstitial hydrogen atoms, and perform a full electronic and ionic relaxation with external pressure $P=0$.
As will become more clear below, we need for the necessary atomistic or {\em ab initio} simulations only information about the hydrogen free and the fully saturated hydride, which are both stress free, as well as a single interstitial atom in a metallic matrix.
Since the presence of isolated interstitial elements lead to local deformations, we have to ensure that the metal system is large enough, $\NM\gg \NH$, such that no interaction with the system's boundaries or periodic images of the impurities (in case of periodic boundary conditions) play a role.
For practical {\em ab initio} simulations such a limit cannot be reached, and therefore it has to be ensured that the interactions with the periodic images are sufficiently low, in order not to influence the results.
The obtained energies are in the following written as $\Energy(\NM; \NH)$ for a system which consist of $\NM$ metal  and $\NH$ hydrogen atoms.
We therefore identify $g_\alpha^0(0)=\Energy_\alpha(\NM; 0)/\NM$ and $g_\beta^0(x_\beta^0)=\Energy_\alpha(\NM; x_\beta^0 \NM)/\NM$ and remind of the notation introduced in Section \ref{coherent::section} (see in particular Eq.~(\ref{singulardecomposition}) there).
We also remind that the Gibbs energies $g_{\alpha, \beta}^0$ are given per host metal atom.
Then $\Delta \Gibbs$ becomes
\begin{equation}
\Delta \Gibbs = \frac{1}{x_\beta^0} \left[ - \frac{\Energy_\beta(\NM;x_\beta^0\NM)}{\NM} + \frac{\Energy_\alpha(\NM;0)}{\NM}\right] + {g_\alpha^0}'(0).
\end{equation}
The derivative term is written as finite difference,
\begin{equation}
{g_\alpha^0}'(0) = \frac{\Energy_\alpha(\NM; 1)/\NM-\Energy_\alpha(\NM; 0)/\NM}{\Delta x}
\end{equation}
with the concentration difference $\Delta x=1/\NM$, i.e.~$N_\mathrm{M}$ should be large. 
Since the energy is extensive, the following property holds for arbitrary scaling factors $M$
\begin{equation}
\Energy(\NM;\NH)/M = \Energy(\NM/M; \NH/M).
\end{equation}
Also, we have the additivity
\begin{equation}
\Energy(N_1; 0) + \Energy(N_2; 0) = \Energy(N_1+N_2; 0).
\end{equation}
Therefore we get
\begin{equation}
\Delta \Gibbs = \Energy_\alpha(\NM; 1) - \left[ \Energy_\beta(1/x_\beta^0; 1) + \Energy_\alpha(\NM-1/x_\beta^0;0) \right].
\end{equation}
This expression is exactly what we expect\cite{Freysoldt:2014aa}:
The first term is the energy of the metallic matrix with one isolated hydrogen atom inside.
It is compared to the energy term in square brackets, which represents the phase separated state consisting of the $\alpha$ phase, which is free of hydrogen and has $\NM-1/x_\beta^0$ metal atoms (the supercell for an {\em ab initio} simulation has to be scaled such that all atom numbers are integers), and of the $\beta$ phase with the remaining atoms, such that the concentration of this phase is $x_\beta^0$.

We can further rewrite this expression as
\begin{equation}
\Delta \Gibbs = \Energy^f_\alpha - \Energy^f_\beta
\end{equation}
with the formation energy of the ``defect'' in the dilute $\alpha$ phase,
\begin{equation} \label{form1}
\Energy^f_\alpha = \Energy_\alpha(N_M; 1) - \Energy_\alpha(N_M; 0) - \Energy_\mathrm{H}^\mathrm{ref}(0; 1)
\end{equation}
with an arbitrary reference potential $\Energy_\mathrm{H}^\mathrm{ref}(0; 1)$ for an isolated hydrogen atom.
Often, this will be expressed through the $T=0$ chemical potential for a hydrogen molecule at zero pressure\cite{Aydin:2012aa}.
Hence, the formation energy is the energy cost if the isolated reference constituents, a perfect bulk $\alpha$ metal and a single hydrogen atom, form the $\alpha$ ``alloy''. 
Similarly, for the $\beta$ phase
\begin{equation} \label{form2}
\Energy^f_\beta = \Energy_\beta(1/x_\beta^0; 1) - \Energy_\alpha(1/x_\beta^0; 0) - \Energy_\mathrm{H}^\mathrm{ref}(0;1)
\end{equation}
expresses the energy change for formation of the hydride phase with concentration $x_\beta^0$ from the reference bulk phases of the pure $\alpha$ metal and the hydrogen atom.
The choice of $\Energy_\mathrm{H}^\mathrm{ref}$ is arbitrary, as it drops out from the expression for $\Delta \Gibbs$.

For the {\em ab initio} predicted low temperature phase diagram the $T=0$ energies are computed.
This is valid as long as lattice vibrations are not yet excited (appreciably below the Debye temperature), such that only zero point energies of the vibrational modes have to be added.
This procedure will be illustrated in the following subsections.
The theoretical predictions for the formation enthalpies have been shown to be in good agreement with experimental values for various elements, see Ref.~\onlinecite{Aydin:2012aa} for details.

Next, let us include the effect of bulk elasticity for coherent interfaces.
As worked out before in section \ref{coherent::section} we have
\begin{equation}
\Delta \bar{\Gibbs} = \Delta \Gibbs + \Delta \Gelb
\end{equation}
with
\begin{equation}
\Delta\Gelb = - \frac{E}{1-\nu} \chi^2 \frac{\omz}{\Nz} \concb^0.
\end{equation}
The bulk solubility limit is
\begin{equation}
\conc \simeq \xn \exp\left(- \frac{\Delta \Gibbs}{\kB T} \right) \exp \left( \frac{E \chi^2 \omz \concb^0}{(1-\nu)\Nz \kB T}\right),
\end{equation}
which is increased in comparison to the stress free case (\ref{atomistic::eq1}).
With the first factor being calculated from {\em ab initio} determined formation enthalpies, the second exponential additionally takes into account the long-ranged elastic coherency effects, as calculated from a continuum perspective.

The coherent solubility limit near free surfaces or interfaces contains the additional geometrical factor $1-\gamma$, hence the solubility limit becomes then
\begin{equation}
\conc \simeq \xn \exp\left(- \frac{\Delta \Gibbs}{\kB T} \right) \exp \left( \frac{E \chi^2 \omz \concb^0(1-\gamma)}{(1-\nu)\Nz \kB T}\right),
\end{equation}
which is in general between the bulk coherent and stress free solubility limit for free surfaces and higher than the bulk solubility limit near rigid substrates.

The important result is therefore the identification of the proper energy terms from an atomistic description.
The additional consideration of long ranged elastic effects, which are otherwise difficult to take into account in an atomistic description alone, become feasible through the present combination with continuum theory.
Through the additive decomposition of a ``chemical'' and the mechanical contribution to the Gibbs energy $\bar{\Gibbs}$ the expression for the solubility limit factorizes.
This means that even without the knowledge of the chemical contribution $\Delta \Gibbs$ still the elastic contribution can be calculated as a modification factor of the solubility limit.

\subsection{Solubility limit prediction}

In the following we predict the influence of long-ranged elastic effects in the presence of surfaces on the solubility limits of Nb-H, Ni-H and Fe-H.
For that, we use experimental and {\em ab initio} calculated data.

\subsubsection{Nb-H}

For the Nb-H and Nb-D system the solubility of the bulk bcc $\alpha$ phase in the low temperature regime is experimentally well described by
\begin{equation} \label{Nb1}
\conc^\mathrm{bulk}=\xn \exp(-\Delta {\cal H}_p/\kB T)
\end{equation}
with $\Delta {\cal H}_p= 0.12\,\mathrm{eV}$ and $\xn=5.35$, which is in agreement with the occupation of tetrahedral sites in the bcc lattice (see Ref.~\onlinecite{Schober:1978aa} and references therein), hence also $N_0=2$.
The $\beta$ phase is an ordered interstitial solid solution of hydrogen with concentrations about $x_\beta^0\approx 0.7$, with a face centred orthorhombic structure ($a/c\approx 1.4$).
From experimental observations\cite{Northemann:2008uq} it has been concluded that the hydride precipitates remain coherent in thin films up to a size of about 30 nm.
Since the tetragonal distortion is not taken into account in our isotropic approximation, we mainly aim at a qualitative understanding.

Experimental values for material parameters are summarised in Ref.~\onlinecite{Pick:1976aa}.
For the lattice constant of pure $\alpha$ Nb we use $a_\alpha=3.3\,\mathrm{\AA}$, which is determined both experimentally and through {\em ab initio} calculations.
For the elastic constants we use $E=105\,\mathrm{GPa}$ and $\nu=0.4$ for pure Nb.
In isotropic approximation the Vegard coefficient is $\chi=0.058$, see Ref.~\onlinecite{Northemann:2008uq} and references therein.

\begin{figure}
\begin{center}
\includegraphics[trim=0.0cm 0cm 0cm 0cm, clip=true, width=8.8cm]{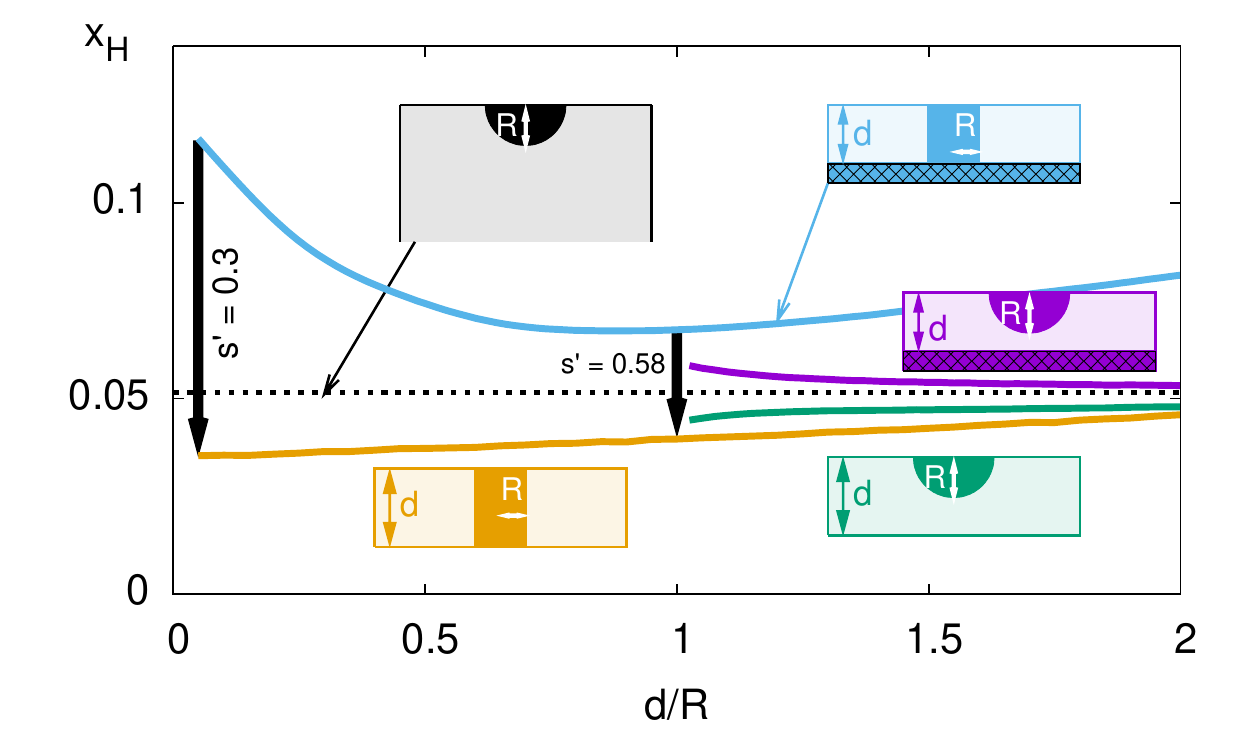}
\includegraphics[trim=0.0cm 0cm 0cm 0cm, clip=true, width=8.8cm]{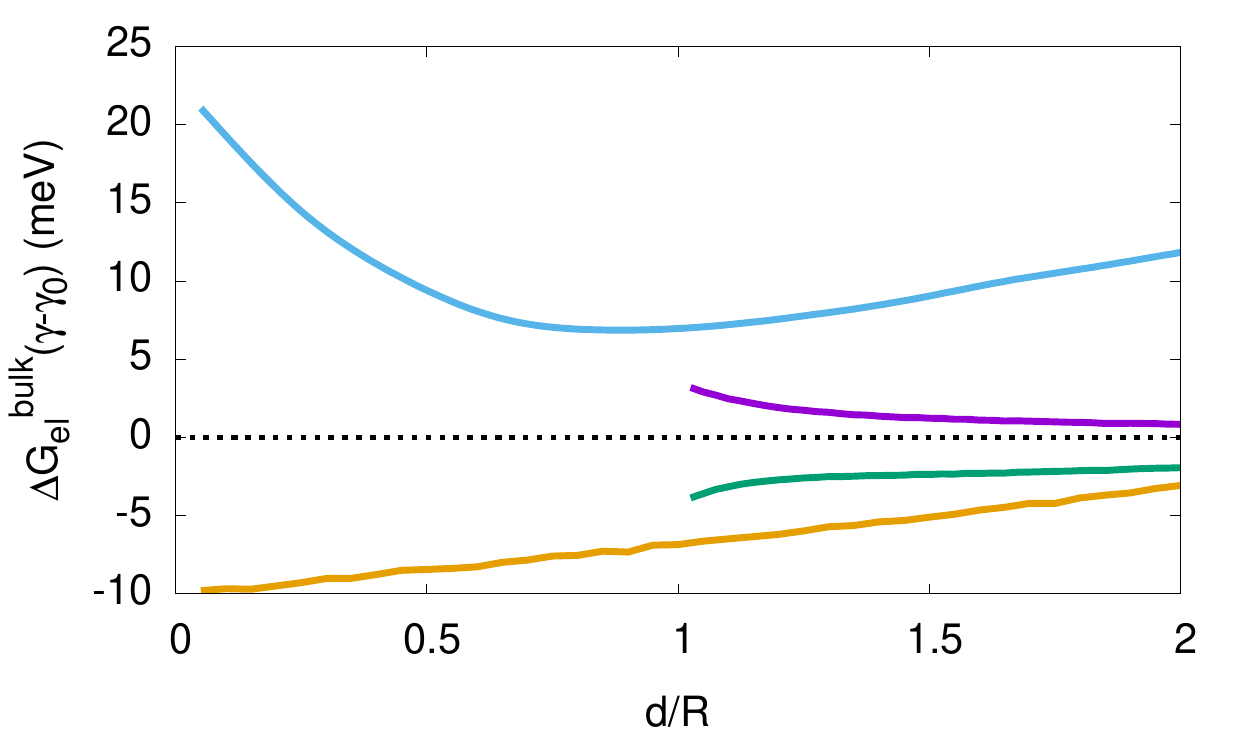}
\caption{(Color online) 
{\em Top}: Predicted solubility limit of the $\alpha$ phase in Nb-H at $T=300\,\mathrm{K}$ in thin films with thickness $d$.
The small sketches show cuts through the film with a cylindrical symmetry.
Starting from thick films, where nucleation is assumed to start from free surfaces (black), cap-like spherical inclusions (radius $R$) can exist for $d/R>1$ in films with finite thickness (green, purple).
The solubility limit in thin films with a precipitate opposite to a free surface (green) or a rigid substrate (purple) hardly changes and asymptotically approach the black curve for $d/R\to\infty$.
For cylindrical nuclei on a rigid substrate (blue) the solubility limit increases with smaller ratio $d/R$.
Delamination from the substrate (orange) reduces the solubility limit by the factor $s'$, which is defined as the ratio of solubility limits of a free standing film to that of a film with a rigid substrate.
The value of $s'$ depends on the aspect ratio  $d/R$.
{\em Bottom}: Elastic contribution $\Delta \Gibbs_\mathrm{el}^\mathrm{bulk}(\gamma-\gamma_0)$ to the formation energy according to Eq.~(\ref{onemoreeqlabel}).
The color coding is the same as in the top panel.
}
\label{fig4}
\end{center}
\end{figure}

To allow for a comparison with experimental results on Nb-H by Pundt et al.\cite{Northemann:2008uq,Pundt:2006aa,Northemann:2011aa}, we consider thin film geometries additionally to the free or clamped surfaces of large systems, as discussed in Fig.~\ref{fig3}.
For the calculation of the elastic energy in film geometries with either free or clamped surfaces ($u_i=0$ there, i.e.~an infinite large contrast in the elastic constants, see top panel of Fig.~\ref{fig4}) we use the finite element method.
This allows to compute $\gamma$ as function of the aspect ratio $d/R$ of the film thickness to the hydride radius for different anticipated geometries (bottom panel of Fig.~\ref{fig4}).

For the bulk behavior, the experimental result (\ref{Nb1}) applies.
If it reflected the true coherent nucleation deep in the bulk in the absence of any free surfaces, it would imply that nucleation near surfaces should appear already at lower concentrations, which is not in agreement with the experimental observation that the solubility limit is close to the bulk value in free standing thin films \cite{Pundt:2006aa}.
This argument suggests that the experimentally observed ``bulk'' solubility limit corresponds to the appearance of precipitates near the free surface of a macroscopic sample.
Therefore, the expression (\ref{Nb1}) should correspond to near-surface nucleation.
The hydride precipitate at the surface will assume the most favourable configuration, which is a spherical cap inside the metal (blue curve in Fig.~\ref{fig3}) with $h\approx R$, hence $\gamma=0.48=:\gamma_0$ for $\nu=0.4$, shown as black dotted line in Fig.~\ref{fig4}.
If we compare this to the case of a cylindrical precipitate in a free standing film, for which the solubility limit is given as orange curve in Fig.~\ref{fig4} (with $\gamma=0.4$) -- which will be discussed more explicitly below -- we indeed see that this value is close to the value of the surface solubility limit for the spherical cap at the surface of a bulk sample, as observed experimentally\cite{Pundt:2006aa}.
Hence we identify
\begin{equation}
\exp(-\Delta {\cal H}_p/\kB T) = \exp \left( - \frac{\Delta \Gibbs + (1-\gamma_0)\Delta \Gibbs_\mathrm{el}^\mathrm{bulk}}{\kB T}\right).
\end{equation}
For other situations with a different value of $\gamma$ the solubility is then given by
\begin{equation}
\conc = \xn \exp \left( - \frac{\Delta \Gibbs + (1-\gamma)\Delta \Gibbs_\mathrm{el}^\mathrm{bulk}}{\kB T}\right).
\end{equation}
Hence we get for the ratio
\begin{equation} \label{onemoreeqlabel}
\frac{\conc}{\conc^\mathrm{bulk}} = \exp\left( \frac{\Delta \Gibbs_\mathrm{el}^\mathrm{bulk}(\gamma-\gamma_0)}{\kB T} \right).
\end{equation}
For fixed temperature and varying $\gamma=\gamma(d/R)$ we can therefore write
\begin{equation}
\conc = \bar{x}_0 \exp \left( \frac{\Delta \Gibbs_\mathrm{el}^\mathrm{bulk}\gamma}{\kB T} \right),
\end{equation}
with the reference concentration
\begin{equation}
\bar{x}_0 = \conc^\mathrm{bulk} \exp \left( -\frac{\Delta \Gibbs_\mathrm{el}^\mathrm{bulk}\gamma_0}{\kB T} \right).
\end{equation}
The latter expression is just a constant for a given temperature.
Consequently, the behavior of $\conc(d/R)$ depends only on the geometry and is phase-diagram insensitive, as all related parameters combine to the single prefactor $\bar{x}_0$.
These expressions are used to predict the solubility limit curves in Fig.~\ref{fig4}.

Apart from free standing thin films, we have inspected also the situation of hydride nucleation near a rigid substrate, where the displacement at the surface is fixed, see Fig.~\ref{fig4}.
In this case the elastic energy even increases in comparison to the bulk, and therefore $\gamma$ becomes negative.
We note that this effect is attributed to the stiffness contrast between the film and the rigid substrate, not due to mismatch stresses.
Consequently, the solubility limit in the $\alpha$ phase is higher in these regions as compared to the bulk.

As discussed above, thin film experiments show that free standing Nb-H films have essentially the same solubility limit as bulk material \cite{Pundt:2006aa}.
This is an agreement with the current predictions, see the horizontal dotted curve in Fig.~\ref{fig4} for the ``bulk'' solubility limit (near a free surface) and the orange and green curves (for differently shaped precipitates in a thin free standing film).
There is only a small remaining solubility limit depression below the bulk value. 
Experimentally, it is presumably shadowed by microstructural effects \cite{Pundt:2006aa}.
Our analysis suggests that phase separation starts from the surfaces, and then progresses into the bulk.

Two scenarios with cap like precipitates in thin film films (green and purple) lead only to small differences in the solubility limits, despite the presence of a rigid surface for the second case.
As expected, the solubility limit is slightly higher in this case as the elastic energy cannot be relaxed as much as for the free standing film.
Still the difference is not large and vanishes in the limit $d/R\to\infty$, as then the presence of the remote rigid substrate is not ``visible'' to the precipitate.
This result is similar to experimental findings for $d>200\,\mathrm{nm}$ on a stiff substrate, where the ``bulk'' value $\conc=0.06$ is found as solubility limit\cite{Pundt:2006aa}.

For $d/R<1$ the cap does not fit into the film anymore, and instead a cylindrical nucleus forms, as confirmed experimentally \cite{Northemann:2008uq}.
This geometry significantly increases the elastic energy for films on a rigid substrate and therefore raises the solubility limit (blue curve), as found experimentally, where a solubility limit of about $\conc=0.2$ for a film thickness of $d=30\,\mathrm{nm}$ is found (see Fig.~17 in Ref.~\onlinecite{Pundt:2006aa}).
Delamination from the substrate (orange curve in Fig.~\ref{fig4}) reduces the solubility limit by a factor $s'\approx 0.3-0.6$, well in the range of the experimental drop $s'\approx 0.06/0.2 = 0.3$, see Ref.~\onlinecite{Pundt:2006aa}.

It is interesting to compare the present predictions for the formation of the {\em high} concentration phase at the surfaces, in comparison to the work on Nb-H by Zabl and Peisl \cite{Zabel:1979aa, Zabel:1980aa}.
They find that spinodal decomposition leads to macroscopic modes with a wavelength of the order of the sample size, and the {\em low} concentration $\alpha$ phase of Nb-H forms more frequently at the free surfaces of the samples, which seems to contradict the predictions in the present work at a first glance.
However, here one should take into account that the experiments \cite{Zabel:1979aa, Zabel:1980aa} have been conducted in the vicinity of the critical point of the $\alpha-\alpha'$ decomposition, where the two phases appear essentially ``symmetrically''.
In the present study, in contrast, we always considered the low concentration regime and therefore the emerging hydride is the minority phase, which breaks the symmetry.
Close to the critical point small artifacts like hydrogen leakage through a protective oxide layer may favour the formation of the low concentration phase at the free surfaces.
The experiments showed that also the high concentration phase can form near the free surfaces.
Thin disk shaped samples split and then also high concentrations of hydrogen were found at free surfaces (see Fig.~1b in Ref.~\onlinecite{Zabel:1979aa}).


\subsubsection{Ni-H}

%

Instead of basing the analysis on experimental data, nowadays also predictions of phase diagrams via {\em ab initio} techniques are feasible, and the inclusion of elastic effects is demonstrated here.
For the technical details of the {\em ab initio } based computations we refer to section II.C in Ref.~\onlinecite{Aydin:2012aa}, including the calculation of the quantum mechanical zero-point energy of H and magnetism.
From high pressure experiments \cite{Fukai:2005aa} and simulations \cite{Korbmacher:2014aa} it is known that a hydride with full occupation of the octahedral sites forms at $x_\beta^0=1$, therefore also $x_0=1$ and $N_0=4$.
We obtain from the {\em ab initio} simulations $a_\alpha=3.52\,\mathrm{\AA}$, $E=200\,\mathrm{GPa}$, $\nu=0.31$ in isotropic approximation.
In agreement with the previous notation we denote the hydrogen poor phase by $\alpha$ and the hydride by $\beta$.
The Vegard coefficient is $\chi=0.058$.
The bare $T=0$ total energies are computed using VASP for a fully relaxed lattice, i.e.~it is both macroscopically stress free and also the internal atom positions are relaxed. 
Appreciably below the Debye temperature $T_\mathrm{D}\approx 450\,\mathrm{K}$ vibrational excitations are negligible, thus $T=0$ energies are used for $\Delta \Gibbs$. 
All energies contain an arbitrary reference energy, which drops out in the final expression, as obvious from the preceding equations.
From these values we can calculate the formation energy difference $\Delta \Gibbs=\Energy^f_\alpha-\Energy^f_\beta$ with the definitions given in Eqs.~(\ref{form1}) and (\ref{form2}).
We use half of the $P=0$ energy of an isolated $\mathrm{H}_2$ molecule as reference state for the hydrogen. 
Its energy is $G_\mathrm{H}^\mathrm{ref}:=\Energy_\mathrm{tot}(\mathrm{H}_{2})/2 = (-3.359 + 0.271)\,\mathrm{eV}$ (the added number is the contribution from the zero point lattice vibrations).
From this, we obtain for the formation energies of the $\alpha$ phase $\Energy^f_\alpha = 0.07\,\mathrm{eV}$. 
The value for $\Energy^f_\alpha$ is close to zero and in good agreement with the experimental value of $\Energy^{f, \mathrm{exp}}_\alpha = 0.124\,\mathrm{eV}$, see Ref.~\onlinecite{Aydin:2012aa} for a more general discussion.
For the hydride ($x_\beta^0=1$) we get similarly $\Energy_\beta^f = \Energy_\mathrm{tot}(\mathrm{Ni}_{32}\mathrm{H}_{32})/32 - \Energy_\mathrm{tot}(\mathrm{Ni}_{32})/32 - \Energy_\mathrm{tot}(\mathrm{H}_{2})/2 = -0.134\,\mathrm{eV}$ and consequently $\Delta \Gibbs=0.2039\,\mathrm{eV}$, where the value of the reference for the hydrogen drops out.



%
\begin{figure}
\begin{center}
\includegraphics[width=8.5cm]{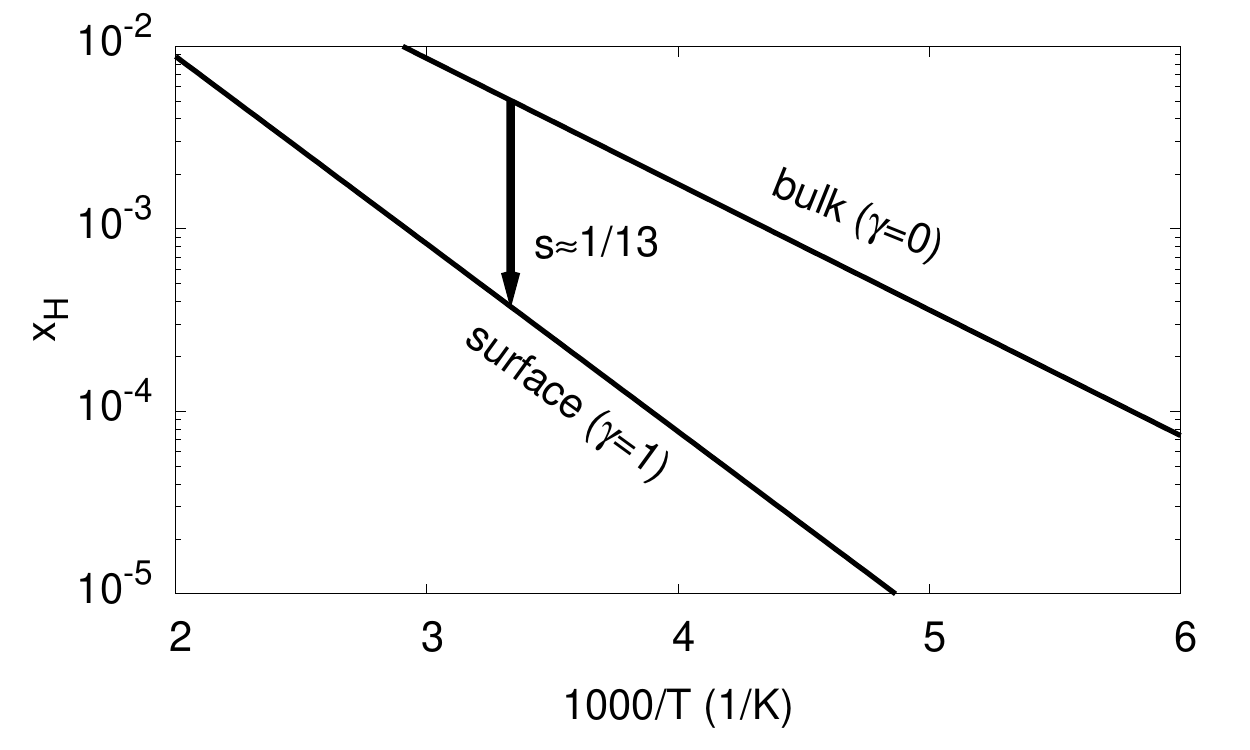}
\caption{
Solubility limits of the $\alpha$ phase of Ni-H for coherent nucleation in the bulk ($\gamma=0$) and for heterogeneous nucleation at free surfaces ($\gamma=1$), based on {\em ab initio} calculated parameters for $\Delta \Gibbs$ and $\Delta \Gibbs_\mathrm{el}^\mathrm{b}$. 
}
\label{fig4b}
\end{center}
\end{figure}
%
The predictions are demonstrated in Fig.~\ref{fig4b} for Ni-H. 
Whereas in the bulk the solubility limit is about $\conc^\mathrm{bulk}\approx 5.1\cdot 10^{-3}$ in the room temperature regime, it is only around $\conc^\mathrm{surface}\approx 3.8\cdot 10^{-4}$ for surface nucleation ($\gamma=1$). 
At room temperature, this leads to a solubility modification factor $s\approx 1/13$ for {\em ab initio} based descriptions of Ni-H for complete elastic relaxation near surfaces ($\gamma=1$) in comparison to the bulk.

\subsubsection{Fe-H (nonmagnetic, fcc)}

In the low temperature regime the bcc iron phase is stable, but nevertheless e.g.~in austenitic steels iron can be stabilised as fcc phase by alloying.
All calculations have been performed with VASP at $T=0$.
Again, hydrogen occupies the octahedral sites of the lattice, with $N_0=4$, and $x_\beta^0=1$. 

For the consideration of elastic effects the elastic constants of pure fcc iron are extracted from {\em ab initio} simulations.
We get $C_{11}=427.6\,\mathrm{GPa}$, $C_{12}=217.6\,\mathrm{GPa}$ and $C_{44}=239.8\,\mathrm{GPa}$.
In isotropic approximation the Lam\'e coefficient is $\lambda=(C_{11}+2C_{12})/3$ and the shear modulus $G=(C_{11}-C_{12})/2$.
Young's modulus and Poisson ratio are then $E=G(3\lambda+2G)/(\lambda+G)\approx 286\,\mathrm{GPa}$ and $\nu=\lambda/[2(\lambda+G)]\approx 0.37$.
We obtain from {\em ab initio} calculations a lattice expansion per hydrogen atom $\Omega=2.03\,\mathrm{\AA}^3$, which is in good agreement with the experimental value $\Omega=1.9\,\mathrm{\AA}^3$ in the fcc iron lattice \cite{Fukai:2005aa}.
The Vegard coefficient $\chi=0.0623$ follows from the lattice constant of iron $a_\alpha=3.45\,\mathrm{\AA}$ and $\Omega$.

From these values we get $s\approx 1/77$ for $\gamma$-Fe for complete elastic relaxation near surfaces ($\gamma=1$) in the room temperature regime. 
Such a drastic reduction of the solubility limit near free surfaces may trigger hydrogen embrittlement by hydride formation at free surfaces and cracks even in the absence of external stresses.
If this brittle phase has a lower fracture toughness, it can enhance a tendency for {\em delayed} fracture.
The time delay may result from the necessity to nucleate hydrides.
Additional tensile stress concentrations lower the chemical potential of the hydrogen in the crack tip region and therefore favour hydride formation even more, leading to hydrogen enhanced decohesion \cite{Lynch:2012aa}. 

In the analysis we have assumed that the material behaves elastically.
Large strains can cause plastic deformation and release stresses.
This relaxation effectively brings the coherent binodal closer towards the stress free binodal and therefore decreases the preference for free surface nucleation.
In turn, materials with a higher flow stress can sustain larger stresses and should therefore exhibit a stronger relative tendency for surface nucleation of hydrides in comparison to the bulk.
This argument is in line with the general trend that higher strength steels exhibit a stronger susceptibility to hydrogen embrittlement.


\section{Summary and conclusions}

In this paper we discussed the role of coherent elasticity in the presence of surfaces on phase separation.
We inspected these processes on multiple scales, ranging from atomistic to macroscopic descriptions. 
On the atomistic level, the formation energies for the relevant phases were calculated.
The long ranged elastic effects, which arise due to coherency stresses between the phases and stress relaxation through the presence of sample surfaces, are considered from larger scale approaches.
In particular, geometric information on the precipitate shapes can be gained from continuum simulations, which are used as input for finite element computations of the solubility modification factor, that expresses the influence of nearby free or confined surfaces.
As a central result, we obtain the closed and generic formula (\ref{MasterFormula}) for the low concentration solubility limit.
This equation connects all the aforementioned approaches and scales.
As a consequence, we predict the equilibrium formation of secondary phases, which have a misfit strain with the matrix phase, to appear primarily at free surfaces of a sample in the low concentration regime. 
As demonstrated in various applications on metal-hydrogen systems, the appearance of hydrides in the presence of free surfaces can start at concentrations up to two orders of magnitude lower than in the bulk.
Experimental benchmarking has in particular been done for the Nb-H system, for which results of thin film experiments can be understood in the framework of the present theory.

Although we put the focus on metal-hydrogen systems, the theory is fully applicable also to other systems.
A particular case are lithium ion batteries\cite{Bai:2011aa, Li:2014aa}, which will be investigated in the future.
Phase separation in LiFePO$_4$ nanoparticles as storage material is influenced by surface effects\cite{Ferguson:2014aa} in combination with elasticity\cite{Cogswell:2013aa, Cogswell:2012aa, Bazant:2013aa}.
The thermodynamics in combination with reaction dynamics leads to a complex behaviour with potential severe influence on capacity and currents during charging and discharging.

In the present article we focused on {\em bulk} effects and neglected the size of the samples.
We emphasise that also the considered elastic effects, which are influenced by the presence of surfaces, are still a bulk effect, as (linear) elasticity does not have an intrinsic lengthscale.
Consequently, the elastic interaction of precipitates with surfaces decays on the scale of the precipitate size.
For a complete description surface effects shall additionally be considered.
A particular challenge will be the connection of the present scale-free description to a microscopic model\cite{Tersoff:1995ab,Tersoff:1995aa}.

\begin{acknowledgements}
This work has been supported by the DFG Collaborative Research Center 761 {\em Steel ab initio}.
The authors gratefully acknowledge the computing time granted on the supercomputer JURECA at the J\"ulich Supercomputing Centre (JSC).


\end{acknowledgements}

\appendix

%

\section{Finite element simulations}
\label{section::FEM}

For the finite element simulations we represent the elastic equilibrium equations in their weak form.
We assume that the concentrations are spatially constant inside each phase, which is appropriate in the low temperature limit.
For the Cahn-Hilliard simulations we use a general concentration field, which follows its own evolution equation.
In both cases, the concentration couples to the stress via the stress free eigenstrain in Eq.~(\ref{isostress}) using Vegard's law.
To solve the elastic problems most efficiently in three dimensions, we anticipate a cylindrical symmetry (Fig.~\ref{fig5}), where the top and bottom surfaces are the stress free surfaces of interest.
The cylinder mantle is assumed to be far away, i.e.~$R\gg R_H$ and the hydride is located on the symmetry axis $r=0$.
In the cylindrical symmetry we only have displacement components $u_r(r, z)$ and $u_z(r, z)$.
Then the non-vanishing strain components are
\begin{eqnarray}
\epsilon_{rr} &=& \frac{\partial u_r}{\partial r}, \\
\epsilon_{zz} &=& \frac{\partial u_z}{\partial z}, \\
\epsilon_{\phi\phi} &=& \frac{u_r}{r}, \\
\epsilon_{rz} &=& \frac{1}{2} \left( \frac{\partial u_r}{\partial z} + \frac{\partial u_z}{\partial r} \right).
\end{eqnarray}
Newton's stationary equations are
\begin{eqnarray}
\frac{\partial \sigma_{rr}}{\partial r} + \frac{\partial \sigma_{rz}}{\partial z} + \frac{1}{r} (\sigma_{rr} - \sigma_{\phi\phi}) &=& 0, \\
\frac{\partial \sigma_{rz}}{\partial r} + \frac{\partial\sigma_{zz}}{\partial z} + \frac{1}{r}\sigma_{rz} &=& 0.
\end{eqnarray}
For the weak form, we multiply them by test functions $v$ and $w$, integrate over the volume and add them.
We treat here both equations successively.
The first equation gives
\begin{equation}
2\pi \int\limits_{r=0}^R \int\limits_{z=0}^H r \left( \frac{\partial \sigma_{rr}}{\partial r} + \frac{\partial \sigma_{rz}}{\partial z} + \frac{1}{r} (\sigma_{rr} - \sigma_{\phi\phi}) \right) v dr dz =0.
\end{equation}
By the product rule we get
\begin{eqnarray}
&&-2\pi \int\int dr dz \left( r \sigma_{rr} \frac{\partial v}{\partial r} + \sigma_{\phi\phi} v + r \sigma_{rz}\frac{\partial v}{\partial z} \right)  \nonumber \\
&&+ 2\pi \int\int \left[ \frac{\partial}{\partial r} \left(r\sigma_{rr} v \right) + \frac{\partial}{\partial z} \left(r\sigma_{rz} v\right) \right] dr\,dz= 0.
\end{eqnarray}
After application of Gauss' theorem the second integral vanishes for stress free boundaries, i.e.~$\sigma_{rr}=0$ at $r=R$ and $\sigma_{rz}=0$ at $z=0$ and $z=H$.
At the inner boundary $r=0$ this term also does not contribute for finite stresses due to the factor $r$ ($=0$).
Hence we get
\begin{equation}
-2\pi \int\int dr dz \left( r \sigma_{rr} \frac{\partial v}{\partial r} + \sigma_{\phi\phi} v + r \sigma_{rz}\frac{\partial v}{\partial z} \right)  = 0
\end{equation}
as first contribution to the weak form.

Similarly for the second equation
\begin{equation}
2\pi \int \int dr\, dz\, r \left( \frac{\partial \sigma_{rz}}{\partial r} + \frac{\partial\sigma_{zz}}{\partial z} + \frac{1}{r}\sigma_{rz} \right) w =0.
\end{equation}
We then get
\begin{eqnarray}
&&-2\pi \int \int dr\, dz\, r\left( \sigma_{zz} \frac{\partial w}{\partial z} + \sigma_{rz} \frac{\partial w}{\partial r} \right) \nonumber \\
&&+ 2\pi \int\int dr\, dz \left[ \frac{\partial}{\partial z} \left( r\sigma_{zz} w \right) + \frac{\partial}{\partial r} \left( r\sigma_{rz} w \right) \right] \nonumber \\
&=& 0.
\end{eqnarray}
Again the second integral vanishes according to Gauss' theorem, and provided that $\sigma_{rz}=0$ on the mantle and $\sigma_{zz}=0$ at the upper and lower surface of the cylinder.
Hence we arrive at the following weak form for the entire elastic problem
\begin{eqnarray}
 -2\pi \int\int dr dz &\Bigg(& r \sigma_{rr} \frac{\partial v}{\partial r} + \sigma_{\phi\phi} v + r \sigma_{rz}\frac{\partial v}{\partial z} +  r \sigma_{zz} \frac{\partial w}{\partial z} \nonumber \\
&& + r \sigma_{rz} \frac{\partial w}{\partial r}\Bigg)  = 0
\end{eqnarray}
for traction free boundaries.
This equation in solved using the open source finite element library FreeFEM\cite{Hecht:2012aa}.
We use a non-uniform mesh with a higher density in the surrounding of the hydride to properly resolve field gradients.
We checked the convergence of the solutions for different mesh resolutions.

The finite element implementation is also used to determine the concentration field.
In order to find its equilibrium distribution, different approaches can be used.
For diffusional transport as a physical description for the hydrogen redistribution in the metal we use
\begin{equation}
\frac{\partial c}{\partial t} = \nabla\cdot \left( D \nabla \muHh \right)
\end{equation}
with a diffusion coefficient $D$ and the chemical potential
\begin{equation} \label{chempotdefpervolume}
\muHh = \frac{\delta \Gibbs}{\delta \conc}.
\end{equation}
Computationally more efficient is the use of nonlocal dynamics according to
\begin{equation} \label{eq10}
\frac{\partial \conc}{\partial t} = -K (\muHh - \lagrangeh)
\end{equation}
with a kinetic coefficient $K$.
The Lagrange multiplier is added to guarantee global mass conservation,
\begin{equation}
\frac{d}{dt} \int_V \conc d\rv = 0.
\end{equation}
Hence we use
\begin{equation}
\lagrangeh = \frac{1}{V} \int_V \muHh d\rv.
\end{equation}
We note that both equations reduce the energy of the system in the course of time, $d\Gibbs/dt \leq 0$, and become equivalent in equilibrium.
The choice of the kinetic parameters $D$ and $K$ is arbitrary for equilibrium investigations and only set the timescale for the relaxation dynamics.

\section{The Eshelby problem}
\label{Eshelby::section}

\begin{figure}
\begin{center}
\includegraphics[trim=0cm 3cm 10cm 1cm, clip=true, width=6cm]{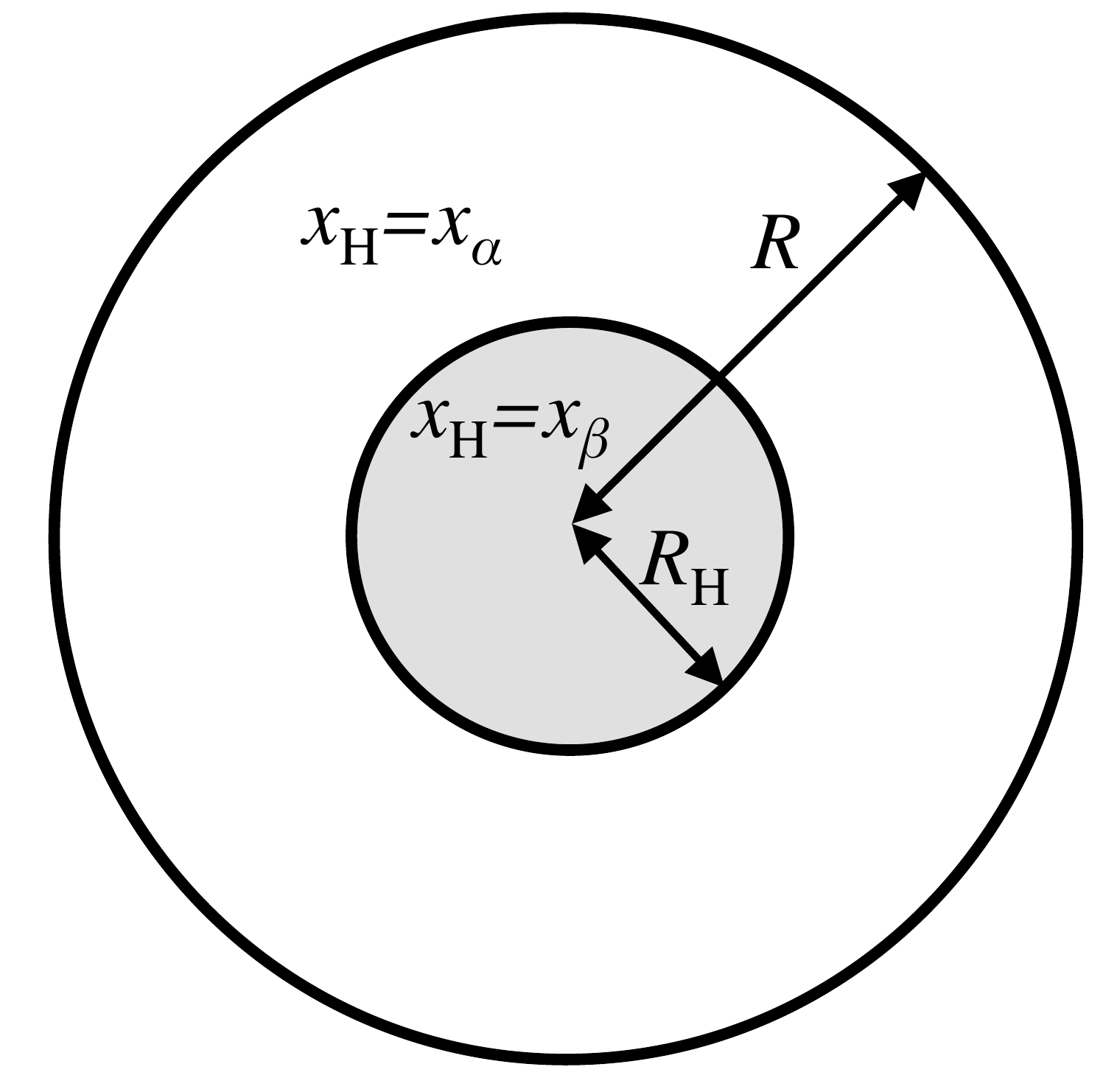}
\caption{Sketch of the geometry for the Eshelby problem.
The figure shows a cut through the three-dimensional system.
The outer shell is the hydrogen poor lattice gas phase, the core the hydride.
}
\label{fig8}
\end{center}
\end{figure}

The problem of a spherical coherent inclusion inside a spherical matrix, as sketched in Fig.~\ref{fig8}, is a special case of Eshelby's more general solution for elliptical inclusions.
It has the advantage that the solution is analytical and turns out to be equivalent to bulk phase separation.

In detail, the system is a three dimensional sphere with radius $R$.
The outer shell is assumed to be the lattice gas phase.
Inside it contains a concentric spherical hydride inclusion of radius $\Rhydride$.
The interface between the two phases is coherent, and the system is stress free at the outer boundary.
Concerning the concentrations, we assume that they are spatially constant within each phase.
They have values $\conca$ and $\concb$ ($\alpha$ and $\beta$ correspond to the lattice gas and the hydride phase).
Later we will see that the assumption of homogeneous concentrations in the individual phases is actually correct in equilibrium.
Under these conditions the displacement field is rotational invariant and has only a radial component.
In the hydride it is
\begin{equation}
u_r^\beta = c_\beta r,
\end{equation}
and in the lattice gas phase
\begin{equation}
u_r^\alpha = c_\alpha r + \frac{b_\alpha}{r^2},
\end{equation}
with constants $b_\alpha, c_\alpha, c_\beta$.
These solutions satisfy the elastic force balance, $\partial_i\sigma_{ij}=0$.
The coefficients are determined by the conditions $u_r^\alpha(\Rhydride)=u_r^\beta(\Rhydride)$, $\sigma_{rr}^\alpha(R)=0$, $\sigma_{rr}^\alpha(\Rhydride) = \sigma_{rr}^\beta(\Rhydride)$.
They correspond to coherency at the interface and a stress free outer surface.
We obtain
\begin{eqnarray}
c_\beta &=& \chi \big[ 2(1-2\nu)\Rhydride^3 (\concb-\conca) + 2R^3 (1-2\nu)\conca \nonumber \\
&& + R^3(1+\nu)\concb \big] /[3(1-\nu) R^3],\\
c_\alpha &=& \chi \frac{ 3(1-\nu)R^3 \conca + 2(1-2\nu) \Rhydride^3(\concb-\conca) }{3(1-\nu)R^3}, \\
b_\alpha &=& \chi \frac{(1+\nu)\Rhydride^3(\concb-\conca)}{3(1-\nu)}.
\end{eqnarray}
From these equations, we can derive the stresses in the two phases.
They are constant in the hydride, and in spherical coordinates given by ($r<\Rhydride$)
\begin{equation}
\sigma_{rr}^\beta = \sigma_{\theta\theta}^\beta = \sigma_{\phi\phi}^\beta = -\frac{2}{3} \chi (\concb-\conca) \frac{E}{1-\nu} \left( 1- \frac{\Rhydride^3}{R^3} \right),
\end{equation}
thus the stress state is hydrostatic and the shear stresses vanish there.
In the surrounding  $\alpha$ phase the non-vanishing stress components are ($\Rhydride\leq r \leq R$)
\begin{eqnarray}
\sigma_{rr}^\alpha(r) &=& -\frac{2}{3} \chi (\concb-\conca) \frac{E}{1-\nu} \left( \frac{\Rhydride^3}{r^3}- \frac{\Rhydride^3}{R^3} \right), \\
\sigma_{\theta\theta}^\alpha(r) &=& \sigma_{\phi\phi}^\alpha(r) \nonumber \\
&=& \frac{1}{3}\chi(\concb-\conca)\frac{E}{1-\nu} \left(  \frac{2\Rhydride^3}{R^3} + \frac{\Rhydride^3}{r^3} \right),
\end{eqnarray}
which have a $1/r^3$ dependence and are not hydrostatic.
For the trace of the stress tensor we have
\begin{eqnarray}
\sigma_{kk}^\alpha(r) &=& 2\chi(\concb-\conca)\frac{E}{1-\nu} \frac{\Rhydride^3}{R^3}, \\
\sigma_{kk}^\beta(r) &=& -2\chi (\concb-\conca)\frac{E}{1-\nu} \left( 1-\frac{\Rhydride^3}{R^3} \right),
\end{eqnarray}
which is a constant also in the $\alpha$ phase.

For thermodynamic considerations we consider the different contributions to the (hydrogen) chemical potentials, which is defined as the energy change if the number of hydrogen atoms is changed.
Apart from the definition as chemical potential $\muHh$ per unit volume, as defined in Eq.~(\ref{chempotdefpervolume}), we use the more common form of energy per particle,
\begin{equation} \label{mudef}
\mu = \left(\frac{\partial {\cal G}}{\partial \NH}\right)_{p, T} = \frac{\omz}{\Nz} \frac{\delta G}{\delta \conc},
\end{equation}
where the first expression is here a shorthand thermodynamical notation, which applies for spatially homogeneous fields and an isotopic pressure $P$.
The more precise local definition of the chemical potential is given by the variational derivative, which is normalised by the prefactor as energy per hydrogen atom.

The homogeneous hydrostatic stress implies that the elastic contribution to the chemical potential,
\begin{equation} \label{mueldef}
\muel = - \frac{\chi \omz}{\Nz} \tr \sigma
\end{equation}
is spatially constant in both phases.
Equation (\ref{mueldef}) follows readily from the definition (\ref{mudef}) using Eqs.~(\ref{eq3}) and (\ref{isostress}).
Explicitly we get
\begin{eqnarray}
\muela &=& - \frac{2\chi^2E\omz \Rhydride^3(\concb-\conca)}{\Nz(1-\nu)R^3}, \\
\muelb &=& \frac{2\chi^2E\omz (R^3-\Rhydride^3)(\concb-\conca)}{\Nz(1-\nu)R^3}.
\end{eqnarray}
Hence, the difference of the chemical potentials does not depend on the hydride volume fraction $v=\Rhydride^3/R^3$,
\begin{equation} \label{eq2}
\muelb-\muela = \frac{2\chi^2 E \omz (\concb-\conca)}{\Nz (1-\nu)}.
\end{equation}
This is the well-known effect of the elastic hysteresis \cite{Brener:2007aa, Spatschek:2013aa}.

The elastic energy density (per volume) in each phase is given by Eq.~(\ref{eq3}).
Integrating over the system gives the total elastic energy
\begin{equation} \label{eq4}
\Gel = \frac{4\chi^2 E \pi \Rhydride^3 (R^3-\Rhydride^3)(\concb-\conca)^2}{3(1-\nu) R^3}.
\end{equation}

To fully describe phase coexistence, we need additionally a description of the stress free contributions to the Gibbs energy, which we choose here in accordance with the continuum model.
Per metal atom we have in both phases the configurational contribution
\begin{equation}
\gc(\conc, T) = \kB T \left[ \conc \log\conc + (1-\conc)\log(1-\conc) \right]
\end{equation}
and the H-H interaction term
\begin{equation}
\go(\conc) = \alpha(\conc- \conc^2),
\end{equation}
in analogy to the Gibbs energy densities $\gch$ and $\goh$ in Eqs.~(\ref{gchdef}) and (\ref{gohdef}).
Since the concentrations are constant in each phase, the integrated energy contributions are
\begin{equation} \label{eq5}
\Gc = \frac{4\pi R^3 \Nz}{3 \omz} \left[ \gc(\concb, T) v + \gc(\conca, T) (1-v) \right]
\end{equation}
and
\begin{equation} \label{eq6}
\Go = \frac{4\pi R^3 \Nz}{3 \omz} \left[ \go(\concb) v + \go(\conca) (1-v) \right]
\end{equation}
with the hydride volume fraction $v=\Rhydride^3/R^3$.
The prefactor $4\pi R^3\Nz/3\omz$ equals the number of metal atoms $\NM$.
The total Gibbs energy of a two phase state is
\begin{equation} \label{eq7}
\Gibbs = \Gel + \Gc + \Go.
\end{equation}
To determine the phase equilibrium conditions we miminimze $\Gibbs$ with respect to the degrees of freedom $\conca$ and $\concb$, while maintaining the conservation law
\begin{equation} \label{eq8}
\concb v + \conca (1-v) = \conc = const.
\end{equation}
This is in accordance with the concept that we use a $(T, p, \NM, \NH)$ ensemble with external pressure $P=0$.
After straightforward algebraic manipulations we can rewrite these conditions as
\begin{equation}
\muhat(\conca) = \muhat(\concb)
\end{equation}
for the first condition, using
\begin{equation}
\muhat(\conc) := g'(\conc) + \frac{2\chi^2 E\omz \conc}{\Nz (1-\nu)}
\end{equation}
with $g(\conc)=\go+\gc$ for each phase.
The second condition is
\begin{equation}
\omegahat(\conca) = \omegahat(\concb)
\end{equation}
with the ``grand potential'' 
\begin{equation}
\omegahat(\conc) := g(\conc) + \frac{E\chi^2 \conc^2 \omz}{\Nz(1-\nu)} - \muhat(\conc) \conc.
\end{equation}
The above equilibrium conditions are identical to Cahn's expressions for the coherent bulk binodal, as discussed in section \ref{coherent::section}.
A central conclusion is therefore that the spherical inclusion inside a pressure free sphere is the same as a bulk system. 
This is a somewhat surprising result, as one might have expected that surface effects play a role here.
The calculation therefore shows explicitly, that this example should be taken with care if interpreted in the spirit of free surface effects.

In view of the results from the continuum model one may expect that it might be energetically favourable to form the hydride at the outer surface of the spherical system instead in its centre.
However, exchanging the role of the lattice gas phase $\alpha$ and the hydride $\beta$ leaves the elastic energy (\ref{eq4}) invariant, as the energy is quadratic in the concentration difference $\concb-\conca$.
Formation of a thin covering hydride layer is therefore not energetically different than confining it inside the matrix phase.
This is in agreement with the elastic energy calculations in Fig.~\ref{fig3}, which show that a thin film of the hydride phase covering a free surface ($h/r=0$ for the black and blue curves) is energetically equivalent to bulk phase separation.
This argument again shows that the Eshelby problem is equivalent to bulk phase separation.

\section{Spherical inclusions near free surfaces}
\label{inclusion::appendix}


%
\begin{figure}
\begin{center}
\includegraphics[width=6cm]{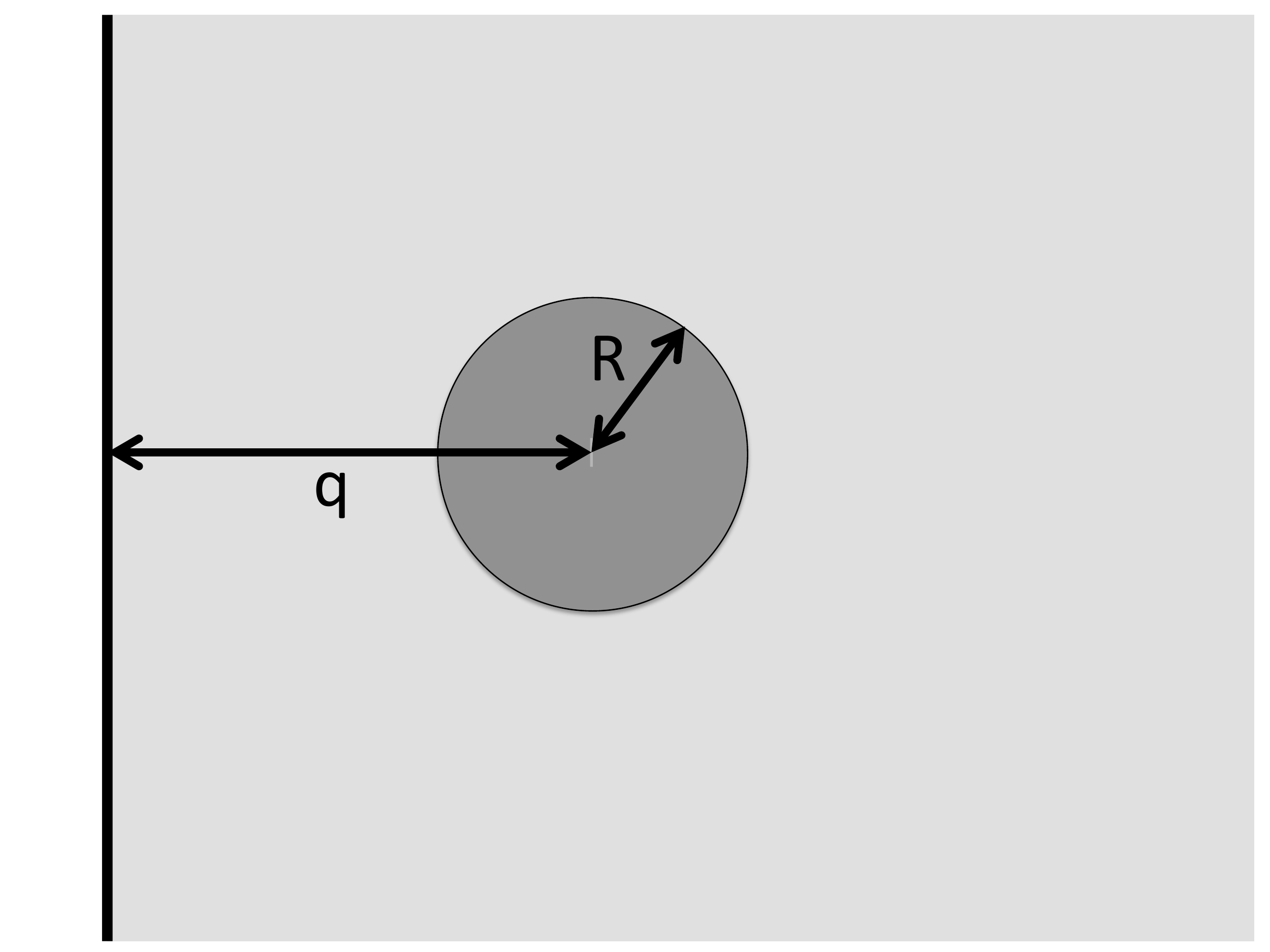}
\caption{Spherical hydride inclusion near a planar free surface. The system is semi-infinite and three-dimensional.}
\label{figSphereNearFreeSurface}
\end{center}
\end{figure}
If a spherical inclusion of radius $R$ comes close to a planar free surface (see Fig.~\ref{figSphereNearFreeSurface}), the energy drops from the bulk value (\ref{BitterCrumEq}) according to $\Gel=\Gelb+\Delta {\cal G}_\mathrm{el}^\mathrm{surface}(q)$ with
\begin{equation}
\Delta {\cal G}_\mathrm{el}^\mathrm{surface}(q) = -\frac{4\pi G(1-\nu) c^2}{q^3},
\end{equation}
with $q$ being the distance of the sphere center from the surface and the shear modulus $G$, see Ref.~\onlinecite{Balluffi:2012aa}.
The constant $c$ is given by
\begin{equation}
c = \frac{(1+\nu) \chi (x_\beta-x_\alpha) V_\beta}{4 \pi (1-\nu)}
\end{equation}
with the sphere volume $V_\beta=4\pi R^3/3$.
Hence we get the scaling $\Delta {\cal G}_\mathrm{el}^\mathrm{surface}/{\cal G}_\mathrm{el}^\mathrm{bulk}\sim V_\beta/q^3$ for a precipitate volume fraction $v\ll 1$.
This expression leads to the red curve in Fig.~\ref{fig3}.


\section{Separation of microscopic \& mesoscopic elasticity}
\label{micromesoelastic::appendix}

We derived the elastic contribution to the free energy assuming coherency between the phases.
The other bulk contributions to the free energy, as contained in $\goh$ and $\gch$, are assumed not to contain (long-ranged) elastic effects, as they describe the homogeneous single phase states.
From a microscopic perspective, however, the situation is somewhat more complicated, as e.g.~in the dilute lattice gas phase we have far separated interstitial hydrogen atoms, and each of them deforms locally the surrounding lattice.
In the {\em ab initio} calculations in Section \ref{abinitio::section} we accounted for relaxation of the lattice, such that the net force on each atom is zero.
At a first glance it therefore seems that we take into account elasticity twice, both one the microscopic and mesoscopic level.
Here we show that we do not double-count elastic effects.

For that, let us first discuss the different roles of the elastic deformations from a conceptual and intuitive level.

On the microscopic level, the lattice is locally deformed around each hydrogen atom in the dilute $\alpha$ phase.
The range of these distortions is short, of the order of a few lattice units.
For low hydrogen concentrations at low temperatures, the contribution to the mutual interaction of hydrogen atoms is therefore small, as long as they are sufficiently far away from each other.
This mutual interaction is captured by $\goh$ on the continuum level (in general, it contains also potential other sources for H-H interactions, which are not necessarily of elastic origin).
This function starts with a linear contribution for $\conc\ll 1$ according to $\goh\approx \alpha \Nz/\omz \conc$, which is related to the chemical potential of the hydrogen atoms (see Section \ref{abinitio::section} for details).
The interaction is contained in the quadratic contribution $-\alpha \conc^2 \Nz/\omz$, which lowers (for $\alpha>0$) the energy relative to the leading, linear term, and therefore describes an attractive interaction.
On the mesoscopic level, isolated hydrogen atoms appears as a homogeneous phase with a coarse-grained hydrogen concentration $\conc$.
As long as this concentration is homogeneous and the phase can expand freely, the material remains stress free and therefore the mesoscopic elastic energy is zero.
We can therefore conclude that the microscopic, short-ranged elastic effects are completely contained in $\goh$.
This argument is reflected by the decisive parameter $\alpha$, which is obtained from {\em ab initio} simulations, which take into account the local relaxation as well the global expansion of the lattice.
Similarly, the hydride phase can be considered.
If it is fully saturated, i.e.~all octahedral sites are filled, the lattice will only expand globally (as mimicked by the Vegard coefficient on the mesoscopic level), but locally it will fully preserve the crystalline symmetries.
If hydrogen vacancies appear here at higher temperatures, the lattice will again relax locally.
The situation is therefore similar to the dilute lattice gas phase, where now vacancies instead of the interstitial hydrogen atoms appear as defects.
Since we focus on the solubility limit of dilute phase, we do not introduce a secondary parameter analogous to $\alpha$, which characterizes the hydrogen vacancy interaction in the hydride phase.
The extension is however straightforward and requires only proper adaption of the free energy contribution $\goh$ on the mesoscopic level.

The mesoscopic elastic term $\gelh$ becomes relevant as soon as concentration gradients, external constraints and, most importantly, (coherent) two-phase situations appear.
The range of these elastic distortions is much longer, of the order of the precipitate size (which we assume to be much larger than the atomic spacing).
This assumption is consistent with a thermodynamic equilibrium consideration, i.e.~the size of the fully equilibrated phases scales with the system size.
As demonstrated explicitly for the Eshelby problem, the coherency leads to a long-ranged deformation of both phases.
On the microscopic level e.g.~the individual hydrogen atom in the dilute phase is therefore placed in a homogeneously distorted environment.
The homogeneity results from the fact that the mesoscopic elastic deformation varies on a larger scale than the lattice unit.
This deformation changes the chemical potential of the hydrogen atoms.
For a local compression the interstitial sites become smaller and therefore it is less favourable to insert hydrogen atoms there.
Hence the chemical potential depends on the mesoscopic stress, and this is reflected by the expression for the chemical potential
\begin{equation} \label{elasticchempotappendix}
\mu_\mathrm{el} = \frac{\omz}{\Nz} \frac{\delta \Gel}{\delta \conc} = - \frac{\omz}{\Nz} \chi \mathrm{tr}\,\sigma,
\end{equation}
which follows readily from (\ref{eq3}) and (\ref{isostress}).
Hence, a compressive stress with $\mathrm{tr}\,\sigma<0$ raises the chemical potential of the hydrogen.

To validate the underlying expression (\ref{eq3}) for the elastic energy density we compare the above results to the discussion in Ref.~\onlinecite{Aydin:2012aa}.
This study uses the thermodynamic relation
\begin{equation} \label{Aydin1}
\frac{\partial \Delta {\cal H}(P)}{\partial P} = V_\mathrm{H}(P)
\end{equation}
with the enthalpy difference
\begin{equation} \label{Aydin0}
\Delta {\cal H}= {\cal H}_\mathrm{MH}(P)-{\cal H}_\mathrm{M}(P) - \frac{1}{2} {\cal H}_{H_2}(P=0),
\end{equation}
involving the enthalpy of the (dilute) metal-hydrogen system ${\cal H}_\mathrm{MH}(P)$ at a given pressure $P$, the same for the pure metal, ${\cal H}_\mathrm{M}(P)$ and a reference chemical potential of an isolated hydrogen molecule at zero pressure, $\frac{1}{2} {\cal H}_{H_2}(P=0)= \mu_\mathrm{H}(P=0)$.
On the right hand side $V_\mathrm{H}(P)$ is the excess volume created by insertion of a hydrogen atom into the metallic matrix at pressure $P$.
With the bulk modulus
\begin{equation} \label{Aydin2}
B = - V \frac{\partial P}{\partial V}
\end{equation}
the above equation (\ref{Aydin1}) can equivalently be rewritten as \cite{Griessen:1985aa}
\begin{equation} \label{Aydin3}
\frac{\partial \Delta {\cal H}(P)}{\partial\ln \omz} = -B V_\mathrm{H}(P).
\end{equation}
This relation was verified against {\em ab initio} simulations with different hydrostatic deformations for a large number of metals, see Fig.~3 in Ref.~\onlinecite{Aydin:2012aa}.

We can compare this equation to our formulation, noting that at $T=0$ (as used for the {\em ab initio} simulations in Ref.~\onlinecite{Aydin:2012aa}) the relation ${\cal H}={\cal G}$ holds.
In continuum approximation (see also Section \ref{conversioncontinuumdiscrete::subsection}), the enthalpy difference $\Delta {\cal H}$ becomes according to Eq.~(\ref{Aydin0})
\begin{eqnarray}
\Delta {\cal H} &=& \Delta \NH \left( \frac{\partial {\cal G}}{\partial \NH} \right)_{P, \NH=0}  -  \frac{1}{2} {\cal H}_{H_2}(P=0) \nonumber \\
&=& \muH -  \frac{1}{2} {\cal H}_{H_2}(P=0),
\end{eqnarray}
with the change of number of hydrogen atoms being $\Delta\NH=1$, using the definition of the chemical potential of hydrogen.
Furthermore, with the bulk modulus $B=\lambda + 2G/3$ and an isotropic deformation $\epsilon_{ij}=\epsilon\delta_{ij}$ we readily get from Eq.~(\ref{elasticchempotappendix})
\begin{equation}
\frac{\partial \Delta {\cal H}}{\partial P} = \frac{3\chi \omz}{\Nz},
\end{equation}
where we note that only $\muel$ is stress dependent and used $3P=-\mathrm{tr}\,\sigma$.
From the stress free strain $\epsilon_{ij}^0=\chi \conc\delta_{ij}$ we can further identify the volume per hydrogen
\begin{equation}
V_\mathrm{H} = \frac{\omz}{\Nz} \mathrm{tr}\epsilon_{ij}^0 = \frac{3\chi \omz}{\Nz},
\end{equation}
since $\conc=1$ corresponds to fully occupied octahedral sites, hence $\NH=\Nz$.
Consequently, Eq.~(\ref{Aydin1}) is reproduced.
We can therefore conclude that the formulation of the continuum model, as given by Eqs.~(\ref{eq1})-(\ref{eq3}), is consistent with the {\em ab initio} perspective.

The other aspect of the separation of elastic effects into microscopic and mesoscopic contributions is discussed in the following.
In practise, we first perform the {\em ab initio} calculations of the single defect allowing for full lattice relaxation, using free volume expansion (i.e.~the ``microscopically'' stress free system) and calculate the energy of this configuration.
Next we calculate the long-ranged elastic field in the continuum description, where the previous effects enter only via an eigenstrain $\epsilon_{ij}^0$, which expresses the widening of the lattice by the hydrogen.
Again, we calculate the elastic energy, but this time using only the long-ranged field.
The elastic field is a superposition of these two fields, which we denote in the following e.g.~by $u_i$ for the short-scale displacements and $u_i^\infty$ for the long-ranged contribution.
Hence the total field reads $u_i^{tot}=u_i + u_i^\infty$.
This decomposition is similar for the total stress and strain fields in linear elasticity.
The elastic energy, however, is a quadratic expression,
\begin{equation}
{\cal G}_{el}^{tot} = \frac{1}{2} C_{ijkl} \int (\epsilon_{ij}^{tot}-\epsilon_{ij}^0) (\epsilon_{kl}^{tot}-\epsilon_{kl}^0) d\rv.
\end{equation}
Using the above decomposition, it therefore seems that apart from the ``self-energy contributions'' of the two scales,
\begin{equation}
{\cal G}_{el}^\infty = \frac{1}{2} C_{ijkl} \int (\epsilon_{ij}^{\infty}-\epsilon_{ij}^0) (\epsilon_{kl}^{\infty}-\epsilon_{kl}^0) d\rv,
\end{equation}
on the mesoscale and 
\begin{equation}
{\cal G}_{el} =  \frac{1}{2} C_{ijkl} \int \epsilon_{ij} \epsilon_{kl}d\rv,
\end{equation}
on the microscale, which is effectively encapsulated in $\go$, also a cross term between the two fields should show up.

To further shed light on this, we discuss for simplicity the dilute lattice gas phase for low temperatures, for which the eigenstrain vanishes. 
On the scale of the short-ranged deformations the long-ranged strain is spatially constant, which expresses the scale separation.
This means that this field reads
\begin{equation}
u_i^\infty = a_{ij} x_j
\end{equation}
with constants $a_{ij}$.
On the other hand, the short scale field is given by the dilatational centre solution
\begin{equation}
u_r = \frac{C}{r^2}
\end{equation}
in spherical coordinates, and all other components vanish (identical to the Eshelby solution in Appendix \ref{Eshelby::section} if the outer radius tends to infinity).
$C$ is a constant which combines all material parameters.

We express all fields in spherical coordinates,
\begin{eqnarray}
x &=& r \sin\theta \cos\phi, \\
y &=& r \sin\theta \sin\phi, \\
z &=& r \cos\theta.
\end{eqnarray}
We further use the unit vectors of the spherical coordinate system
\begin{eqnarray}
\hat{\mathbf{r}} &=& \sin\theta \cos\phi\, \hat{\mathbf{x}} + \sin\theta\sin\phi\, \hat{\mathbf{y}} + \cos\theta\, \hat{\mathbf{z}}, \\
\hat{\mathbf{\theta}} &=& \cos\theta \cos\phi\, \hat{\mathbf{x}} + \cos\theta\sin\phi\, \hat{\mathbf{y}} - \sin\theta\, \hat{\mathbf{z}}, \\\
\hat{\mathbf{\phi}} &=& -\sin\phi\, \hat{\mathbf{x}} + \cos\phi\, \hat{\mathbf{y}}
\end{eqnarray}
with the Cartesian unit vectors $\hat{\mathbf{x}}$, $\hat{\mathbf{y}}$, $\hat{\mathbf{z}}$.
We have 
\begin{equation}
\mathbf{u} = u_x\, \hat{\mathbf{x}} + u_y\, \hat{\mathbf{y}} + u_z\, \hat{\mathbf{z}}.
\end{equation}
Then the spherical components of the displacement vector are given by $u_r = \mathbf{u}\cdot \hat{\mathbf{r}}$ and similarly for the other components.
The strains follow from
\begin{eqnarray}
\epsilon_{rr} &=& \frac{\partial u_r}{\partial r}, \\
\epsilon_{\theta\theta} &=& \frac{1}{r} \left( \frac{\partial u_\theta}{\partial\theta} + u_r \right), \\
\epsilon_{\phi\phi} &=& \frac{1}{r\sin\theta} \left( \frac{\partial u_\phi}{\partial\phi} + u_r\sin\theta +u_\theta\cos\theta\right), \\
\epsilon_{r\theta} &=& \frac{1}{2} \left( \frac{1}{r} \frac{\partial u_r}{\partial\theta} + \frac{\partial u_\theta}{\partial r} - \frac{u_\theta}{r} \right), \\
\epsilon_{\theta \phi} &=& \frac{1}{2r} \left( \frac{1}{\sin\theta} \frac{\partial u_\theta}{\partial\phi} + \frac{\partial u_\phi}{\partial \theta} - u_\theta \cot\theta \right), \\
\epsilon_{r\theta} &=& \frac{1}{2} \left( \frac{1}{r\sin\theta} \frac{\partial u_r}{\partial \phi} + \frac{\partial u_\phi}{\partial r} - \frac{u_\phi}{r}  \right).
\end{eqnarray}
From these equations we calculate the elastic energy density as [see Eq.~(\ref{eq3})]
\begin{equation}
g_{el} = G \epsilon_{ij}^2 +\frac{1}{2}\lambda (\epsilon_{kk})^2,
\end{equation}
and integrate it according to
\begin{equation}
{\cal G}_{el} = \int_{R_0}^R \int_{\phi=0}^{2\pi}\int_{\theta=0}^\pi dr\, d\phi\, d\theta\, r^2\sin\theta g_{el}.
\end{equation}
Here, $R_0$ is a cutoff, below which the continuum description should be replaced by a discrete formulation.
We get for the total elastic energy
\begin{equation}
{\cal G}_{el}^{tot} = {\cal G}_{el} + {\cal G}_{el}^\infty
\end{equation}
with
\begin{equation}
{\cal G}_{el} = 8 C^2 G \pi\left( \frac{1}{R_0^3} - \frac{1}{R^3}\right).
\end{equation}
For the homogeneous part we have
\begin{eqnarray}
&& {\cal G}_{el}^\infty = \frac{2\pi (R^3-R_0^3)}{3} \Bigg[ 2(a_{kk})^2 \lambda + \Big\{ 2(a_{11}^2 + a_{22}^2 + a_{33}^2) \nonumber \\
&&+ (a_{12}+a_{21})^2 + (a_{13}+a_{31})^2 + (a_{23}+a_{32})^2\Big\} G \Bigg].
\end{eqnarray}
We therefore see that no cross term appears, as such a term would involve products of the form $C a_{ij}$.


\bibliography{references}

\end{document}